\documentclass[11pt,a4paper]{article}
\ifdefined\pdfminorversion\pdfminorversion=7\fi

\usepackage{jheppub}

\usepackage{array}
\usepackage{mathrsfs}
\usepackage{colonequals}
\usepackage{amscd}
\usepackage{relsize}
\usepackage{suffix}
\usepackage{mathtools}
\usepackage{cancel}
\usepackage{tikz-cd}
\usepackage{tikz}
\usepackage[english]{babel}
\usepackage{csquotes}
\usepackage{comment}
\usepackage{physics}
\usetikzlibrary{arrows.meta,decorations.markings}
\usepackage{tabularx,array,booktabs}

\newcolumntype{Y}{>{\raggedright\arraybackslash\hyphenpenalty=10000\exhyphenpenalty=10000}X}

\title{\boldmath \Large
Effective Field Theory for Holographic Compact Objects
 }

\author[a]{Miguel Correia,}
\author[b]{Vasco Gon\c{c}alves,}
\author[b]{Filipe Serrano}

\affiliation[a]{Department of Physics, McGill University, 3600 Rue University, Montreal, H3A 2T8, QC, Canada}
\affiliation[b]{Centro de F\'isica do Porto e Departamento de F\'isica e Astronomia,
Faculdade de Ci\^encias da Universidade do Porto, Porto 4169-007, Portugal}

\emailAdd{miguel.ribeirocorreia@mcgill.ca}
\emailAdd{vasco.dfg@gmail.com}
\emailAdd{filipeserranophysics@gmail.com}

\abstract{
We study heavy--heavy--light--light correlators in holographic CFTs when the heavy operator is dual to a compact object localized on scales much smaller than the AdS radius. In the long-wavelength regime, the internal structure of the heavy state is described by a worldline effective field theory, whose local operators encode finite-size response coefficients such as tidal Love numbers. Focusing on weakly self-gravitating objects, we show that the Witten diagrams generated by local heavy--light interactions reorganize into the Born series of an effective wave equation for the light bulk field. We reproduce the same correlator by solving this wave equation with localized interactions and identify how the worldline Wilson coefficients are encoded in the CFT data. Our results provide a direct relation between the finite-size response of compact objects in AdS and the dimensions and OPE coefficients of heavy--light and light--light composite operators in the dual CFT.
}

\begin{document}
\maketitle
\flushbottom

\newpage

\section{Introduction}\label{sec:Introduction}

The AdS/CFT correspondence relates quantum gravity in asymptotically
anti-de Sitter spacetime to a conformal field theory living on its boundary
\cite{Maldacena:1997re,Gubser:1998bc,Witten:1998qj}. When the bulk
admits a weakly coupled description, boundary correlation functions can be
computed perturbatively as sums of Witten diagrams. In this dictionary,
single-trace primary operators create bulk particles, while their dimensions and
correlation functions determine the masses and interactions of the
corresponding bulk fields.

Correlation functions involving operators with parametrically different
dimensions provide an interesting probe of this dictionary. We consider
heavy--heavy--light--light (HHLL) correlators of the form
\begin{equation}
    G(P_i)
    \equiv
    \left\langle
        \mathcal O_H(P_1)
        \mathcal O_L(P_2)
        \mathcal O_L(P_3)
        \mathcal O_H(P_4)
    \right\rangle,
    \label{eq:HHLL-correlator}
\end{equation}
where the heavy operator has conformal dimension $\Delta_H$ much larger than
the dimension $\Delta_L\sim \mathcal O(1)$ of the light operator,
\begin{equation}
    \Delta_H\gg\Delta_L.
\end{equation}
By the state-operator correspondence, the normalized four-point function can
be interpreted as a two-point function of $\mathcal O_L$ in the state
$|\mathcal O_H\rangle$ created by $\mathcal O_H$. From the bulk perspective,
the light field therefore acts as a probe of the state associated with the
heavy operator.

The nature of this bulk state depends crucially on how $\Delta_H$ scales
relative to the central charge $c$ (defined here as the coefficient of the stress-tensor two-point function, $c\propto L_{\rm AdS}^{d-1}/G_N$). Heavy operators with $\Delta_H\sim c$
can generate geometries whose characteristic scales are comparable to the
AdS radius. In particular, sufficiently energetic states may be dual to
AdS-sized black holes, and their HHLL correlators are closely related to
thermal two-point functions. This regime has been studied extensively through
semiclassical propagation in black-hole geometries, multi-stress-tensor
exchange, and the spectrum of heavy--light composite operators
\cite{Kulaxizi:2018dxo,Karlsson:2019dbd,
Dodelson:2022eiz,Huang:2024wbq}. The analytic structure of holographic
thermal correlators has also been investigated directly from both the bulk
and boundary perspectives
\cite{Alday:2020eua,
Dodelson:2023vrw,
Dodelson:2023nnr}. In two-dimensional CFTs,
analogous semiclassical heavy--light correlators admit descriptions in terms
of light probes propagating in conical-defect and BTZ geometries
\cite{Hijano:2015rla,Fitzpatrick:2015zha,
Kraus:2017kyl}. 
 
\begin{figure}
    \centering
    \includegraphics[width=0.54\linewidth]{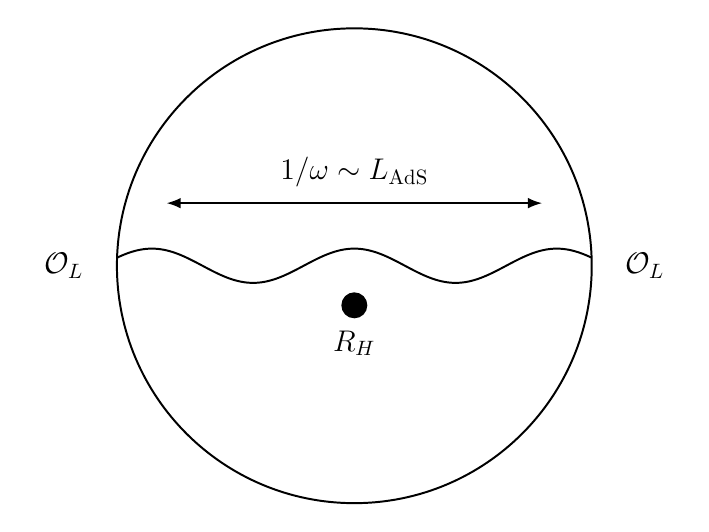}
     \caption{Long-wavelength probe of a heavy state dual to a compact object in AdS.}
    \label{fig:longprobe}
\end{figure}

Here, we instead consider the parametrically different regime
\begin{equation}
    1\ll\Delta_H\ll c.
    \label{eq:subplanckian-heavy-regime}
\end{equation}
The heavy state is energetic compared with the light probe, while its
gravitational backreaction remains perturbative on scales of order the AdS
radius $L_{\rm AdS}$. Within this regime, we further restrict attention to
heavy states whose bulk duals are localized \emph{compact objects} of
characteristic size $R_H$. If $\omega$ denotes the frequency of the light
probe in global AdS, we study the long-wavelength regime
\begin{equation}
    \omega R_H\ll1,
    \label{eq:long-wavelength-regime}
\end{equation}
illustrated in Figure~\ref{fig:longprobe}. In this limit, the probe cannot resolve the
internal structure of the heavy object \cite{Goldberger:2004jt,Kol:2011vg}, which can therefore be described by
a worldline effective theory, introduced below.

Note that for low-lying AdS modes, for which
$\omega\sim L_{\rm AdS}^{-1}$, the condition
Eq.~\eqref{eq:long-wavelength-regime} follows from the compact object being much smaller than AdS,
$R_H\ll L_{\rm AdS}$. Moreover, provided that no object can be more compact than a black hole, this hierarchy also guarantees the upper bound Eq.~\eqref{eq:subplanckian-heavy-regime} on the dimension of the heavy operator.\footnote{For $d>2$, let $R_s$ denote the Schwarzschild radius
associated with the mass $M$ of the heavy state. The standard parametric relations $R_s^{d-2}\sim G_N M$, $\Delta_H\sim M L_{\rm AdS}$, and $c\sim\frac{L_{\rm AdS}^{d-1}}{G_N}$ imply that $\frac{\Delta_H}{c} \sim  \left(\frac{R_s}{L_{\rm AdS}}\right)^{d-2}$. For an object satisfying $R_s\lesssim R_H$, one therefore finds $\frac{\Delta_H}{c}
    \lesssim
    \left(\frac{R_H}{L_{\rm AdS}}\right)^{d-2}
    \ll 1$.}

When Eq.~\eqref{eq:long-wavelength-regime} holds, the heavy state can be
regarded as a point-like compact object in AdS. Its microscopic structure is confined
to a region much smaller than the curvature radius, while its center of mass
follows a worldline extending along global AdS time. Examples may include
massive particles, compact stars, small black holes, and sufficiently short
closed strings. This description does not apply directly to extended branes
or macroscopic field configurations whose spatial profiles remain resolved
on scales comparable to $L_{\rm AdS}$ \cite{Giusto:2018ovt,
Giusto:2019pxc,
Giusto:2023awo,McGreevy:2000cw,
Corley:2001zk,
Lin:2004nb,
Giusto:2024trt,Turton:2024afd,Aprile:2024lwy}.

The long-distance gravitational field of the heavy object is controlled by
the dimensionless parameter
\begin{equation}
    \mu
    \equiv
    \frac{\Delta_H}{c}
    \sim
    \frac{G_N M}{L_{\rm AdS}^{d-2}}.
\end{equation}
For heavy states dual to AdS black holes, correlators of light operators can
be computed semiclassically by propagating the corresponding bulk fields in
the black-hole geometry. Their expansion at small $\mu$ is organized on the
CFT side by the exchange of operators constructed from increasing numbers of
stress tensors
\cite{Fitzpatrick:2019zqz,
Kulaxizi:2018dxo,
Karlsson:2019dbd,
Li:2019zba,
Li:2020dqm,
Parnachev:2020fna, Dodelson:2022yvn,
Huang:2024wbq}, together with light double-twist operators in the $t$-channel.
This multi-stress-tensor sector captures the universal long-distance
gravitational interaction between the heavy state and the light probe. See also 
\cite{Ceplak:2024bja,
Buric:2025anb,
Buric:2025fye,
Barrat:2025nvu,
Barrat:2025twb,
Afkhami-Jeddi:2025wra,
Giombi:2026kdz,
Arnaudo:2026der,
Jia:2026ryl} for recent work.

At large conformal spin, heavy--light double-twist operators correspond semiclassically to bulk configurations in which the light probe remains far from the heavy object. Their CFT data are then organized by the lightcone bootstrap as an asymptotic expansion in inverse conformal spin, whose leading terms are controlled by low-twist crossed-channel exchanges and reproduce the universal long-distance gravitational interaction described above
\cite{Fitzpatrick:2012yx,
Komargodski:2012ek,
Fitzpatrick:2014vua,
Caron-Huot:2017vep,
Li:2019zba,
Li:2020dqm}.

By contrast, interactions localized near the heavy object contribute through
finite-spin data or through terms that are nonperturbative in the inverse-spin
expansion and hence suppressed at large conformal spin
\cite{Heemskerk:2009pn,
Fitzpatrick:2010zm,
Albayrak:2019gnz,
Dodelson:2022eiz}.
These contributions contain short-distance information that is not fixed by
the universal multi-stress-tensor sector. A central aim of this work is to
isolate this short-distance sector of the HHLL correlator and organize it systematically in
terms of local worldline interactions. The corresponding Wilson coefficients
will provide a parametrization of the internal structure and response of the
heavy state, whose imprint can be identified directly in the CFT data.

In the long-wavelength regime,
Eq.~\eqref{eq:long-wavelength-regime}, this organization is achieved by
integrating out bulk physics at distances of order $R_H$. We introduce a
matching scale, $\Lambda_{\rm UV}\sim R_H^{-1}$, and replace the extended object
by a point particle carrying the most general local interactions along its
worldline that are compatible with the symmetries
\cite{Goldberger:2004jt,Porto:2016pyg}. Equivalently, the localized
worldline action determines the short-distance boundary conditions obeyed by
the light field near the position of the heavy object
\cite{Burgess:2016lal}. The resulting effective action
takes the schematic form
\begin{equation}
    S_{\rm EFT}
    =
    -M\int d\tau
    +
    \sum_i C_i
    \int d\tau\,
    \mathcal O_i
    \bigl[\phi,\nabla\phi,\ldots\bigr].
    \label{eq:SEFT}
\end{equation}
The Wilson coefficients $C_i$ encode the internal structure and response of
the heavy state to the probe light field $\phi$, and their effects are organized systematically in powers of
$\omega R_H$. They are the analogues of polarizabilities and multipole
response coefficients in ordinary effective field theory (EFT). When the probe is
gravitational, these coefficients include the tidal Love numbers, which
characterize the conservative response of the object to a slowly varying
external tidal field
\cite{Goldberger:2004jt,Kol:2011vg}.
More generally, the response is frequency dependent and can contain poles
associated with internal excitations of the compact object
\cite{Chakrabarti:2013xza,
Steinhoff:2016rfi}. Such dynamical effects can be
described by additional degrees of freedom localized on the worldline.
Dissipative effects are encoded in the absorptive part of the corresponding
response functions and can be included via an in-in, or Schwinger--Keldysh, formulation
\cite{Goldberger:2005cd,
Goldberger:2020fot,Galley:2012hx}.

It was recently shown that, within this effective theory, the light
probe obeys a classical equation of motion, which we refer to as the
effective wave equation
\cite{Correia:2024jgr,
Caron-Huot:2025tlq,Correia:2026utp},
\begin{equation}
\label{eq:ewe}
    \left(
        \Box
        +V_{\rm long}
        +V_{\rm short}
    \right)\phi=0.
\end{equation}
Here, $V_{\rm long}$ describes the universal long-distance gravitational
interaction, while $V_{\rm short}$ denotes a tower of contact interactions
localized on the worldline, of the schematic form 
$V_\text{short} \propto C_i\,\partial^p\delta^{(d)}(\mathbf x)\,\partial^q$, whose coefficients $C_i$ are the
Wilson coefficients appearing in the effective action
\eqref{eq:SEFT}. These coefficients can be determined by matching the
low-frequency solutions of the effective wave equation to a microscopic
calculation.\footnote{For black holes, the microscopic response is obtained from black-hole
perturbation theory: perturbations of Schwarzschild black holes are governed
by the Regge--Wheeler and Zerilli equations, while perturbations of Kerr black
holes are described by the Teukolsky formalism
\cite{Regge:1957td,Zerilli:1970se,
Teukolsky:1972my,Teukolsky:1973ha,
Pound:2021qin}.
Matching the near- and far-zone solutions, or equivalently the source,
response, and absorption coefficients, to the worldline EFT determines the
conservative and dissipative response coefficients
\cite{Ivanov:2022hlo,Ivanov:2024sds,
Saketh:2023bul,Caron-Huot:2025tlq,Chang:2026eti,Combaluzier--Szteinsznaider:2025eoc,Apostolidis:2026qsg, Correia:2026utp}.}

In this work, we develop the AdS counterpart of the worldline effective description and isolate precisely this short-range sector. We specialize to weakly self-gravitating compact objects whose physical size is parametrically larger than their Schwarzschild radius $R_s$,
\begin{equation}
R_s \ll R_H \ll L_{\rm AdS}.
\label{eq:hierarchy}
\end{equation}
In this limit, the universal gravitational potential may be neglected to leading order, $V_{\rm long}\to 0$, allowing us to focus directly on the short-range potential $V_\text{short}$ that encodes the internal response of the heavy state.

In particular, we show that the HHLL Witten-diagram expansion generated by these local heavy--light interactions reorganizes into the Born series of the effective wave equation Eq.~\eqref{eq:ewe}. This establishes explicitly, within the worldline effective theory, the interpretation of the heavy operator as generating an effective background for the light field in AdS.

From the boundary perspective, the $\mathcal O_H \times \mathcal O_L$ OPE decomposition (which we call the $s$--channel) of the HHLL correlator determines the dimensions and OPE coefficients of heavy--light composite operators. The local coefficients $C_i$ characterize the short-distance part of the interaction between $\mathcal O_H$ and $\mathcal O_L$. After universal long-range exchanges are separated, these data admit a natural interpretation as response coefficients of the heavy CFT state. In this sense, the worldline Wilson coefficients, and in particular the tidal Love numbers, possess a direct CFT characterization.

The paper is organized as follows. In section~\ref{The heavy--light correlator}, we show that Witten diagrams in the heavy--light limit reorganize into the Born series of the worldline effective theory. In section~\ref{Wave equation}, we formulate the corresponding effective wave equation in AdS and show how the tidal coefficients are encoded in its boundary conditions. In section~\ref{CFT} we identify the CFT data associated with the tidal response of the heavy state and compare the heavy limit with the lightcone-bootstrap and eikonal limits. 

The appendices are organized as follows. In Appendix~\ref{app:bornregime}, 
we provide a dictionary relating our holographic setup to standard flat-space scattering kinematics, 
reviewing the Born regime of \cite{Correia:2024jgr}. Appendix~\ref{app:heavyprop} derives the heavy limit of 
the bulk-to-bulk propagator from the DeWitt--Schwinger expansion, 
together with the composition rule for internal bulk integrations used in section~\ref{Loop diagrams}. 
In Appendix~\ref{app:yukawa}, we analyze a scalar Yukawa theory in AdS,
explicitly demonstrating how $t$-channel exchanges and one-loop diagrams factorize into a Born series iterating an effective potential. 
Appendix~\ref{app:anom2} contains the detailed evaluation of the second-order anomalous dimension in terms of the tidal coefficients.
Appendix~\ref{app:nrel} explores the non-relativistic two-body limit directly from the CFT perspective. 
Finally, Appendix~\ref{sec:graviton_heavy} discusses the heavy limit of a graviton exchange Witten diagram in Mellin space.
In the remainder of the paper, we set the AdS radius to unity, $L_{\text{AdS}}=1$.

\section{Heavy-light limit of Witten diagrams and the Born series}\label{The heavy--light correlator}

\begin{figure}
    \centering
    \includegraphics[width=0.4\linewidth]{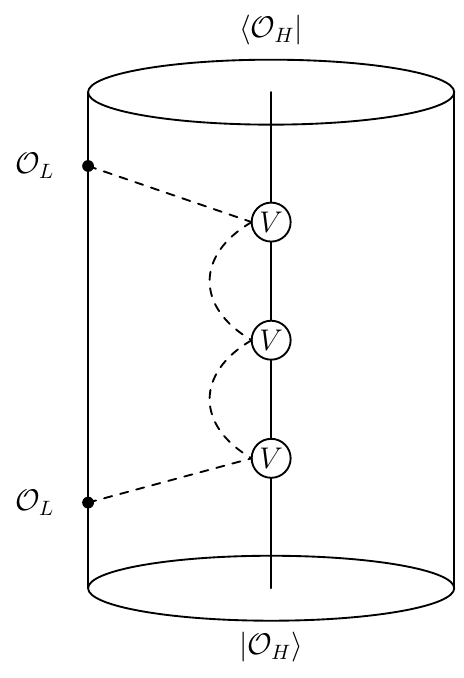}
    \caption{Schematic representation of the Born series for a light field interacting with a heavy particle propagating along a classical worldline, where $V$ denotes an insertion of the effective potential.}
    \label{fig:born}
\end{figure}

In this section, we study the HHLL correlator,
$G(P_i)$, from the bulk perspective and show how its Witten diagrams reorganize into a
Born series. We take the heavy and light scalar operators to be dual to bulk fields
$\chi$ and $\phi$, with masses $M$ and $m$, respectively, and consider the
bulk action in AdS
\begin{equation}
    S
    =
    \frac{1}{2}
    \int d^{d+1}x\,\sqrt{\abs{g}}\,
    \left[
        (\nabla\phi)^2+m^2\phi^2
        +
        (\nabla\chi)^2+M^2\chi^2
        +
        \frac{\lambda}{2}\phi^2\chi^2
    \right],
    \label{eq:bulk-action}
\end{equation}
where $g^{\mu \nu}$ denotes the AdS metric and $\nabla$ the covariant derivative in AdS. The masses  $m$ and $M$ relate to the dimensions of the dual CFT operators according to the usual relation \cite{Aharony:1999ti}
\begin{equation}
    m^2=\Delta_L(\Delta_L-d),
\qquad
M^2=\Delta_H(\Delta_H-d).
\end{equation}
We compute the HHLL correlator in the heavy--light limit where $\Delta_H\gg1$ and $\Delta_L \sim \mathcal{O}(1)$. In this limit,
the propagation of the heavy field becomes semiclassical and localizes on a
bulk geodesic. Repeated insertions of the interaction term in
\eqref{eq:bulk-action} then act, from the perspective of the light
field, as repeated interactions with an effective potential supported on the
heavy trajectory. We will show here explicitly that the corresponding sequence of Witten
diagrams takes the form of a Born series associated with an underlying wave equation, as depicted in Figure~\ref{fig:born}. The iterated effective potential $V$ in this case will be purely of short-range nature and, as we show later in section~\ref{Wave equation}, can be captured systematically within the worldline EFT formalism.

This section is organized as follows. In section~\ref{GESC}, we review the
embedding-space formalism and introduce the kinematic ingredients used
throughout. In section~\ref{Heavy limit of Witten diagrams}, we show that
non-derivative tree-level contact Witten diagrams simplify considerably in
the heavy limit, $\Delta_H\gg1$. In section~\ref{Mellin}, we reproduce the
same simplification using the Mellin-space representation. In
section~\ref{Loop diagrams} we compute the one-loop bubble Witten diagrams, using the heavy limit of the bulk-to-bulk propagator and a composition rule for internal bulk integrations whose derivation is given in Appendix~\ref{app:heavyprop}. Finally, in section~\ref{all-loop} we show how the all-loop result is obtained.

\subsection{Embedding space coordinates and geodesic motion}\label{GESC}
Consider Euclidean AdS, $\mathrm{EAdS}_{d+1}$\footnote{For simplicity of notation we will just write AdS.}, realized as the unit
hyperboloid in the flat embedding space \cite{Costa:2011mg,Costa:2014kfa},
\begin{equation}
   \mathrm{EAdS}_{d+1} = \{ X^A \in \mathbb{R}^{d+1,1}:X^A X_A=-1 \},
\end{equation}
where embedding-space indices are contracted with the flat space metric, $\eta_{AB}$, of
signature $(d+1,1)$. Let $X_1^A,Y_1^A\in\mathrm{EAdS}_{d+1}$ be two bulk
points, and denote by $\gamma$ the geodesic connecting them.\footnote{Unless
stated otherwise, we work in Euclidean signature. We will indicate explicitly
when the discussion is continued to Lorentzian AdS.} Parametrizing the
geodesic by $\tau\in[\tau_i,\tau_f]$, its endpoints satisfy
\begin{equation}
    X^A(\tau_i)=X_1^A,
    \qquad
    X^A(\tau_f)=Y_1^A.
\end{equation}
Then the geodesic distance between $X_1^A$ and $Y_1^A$ is
\begin{equation}
    L \equiv \int_{\tau_i}^{\tau_f} d\tau \sqrt{\abs{\eta_{AB}\derivative{X^A}{\tau}\derivative{X^B}{\tau}}}.
\end{equation}
This distance, $L$, must be a function of the only $SO(d+1,1)$ invariant one can construct with 
$X_1^A$ and $Y_1^A$, 
namely, $X_1 \cdot Y_1$. This inner product computes the hyperbolic angle between the two points.
Just as two points on the unit sphere separated by geodesic distance, $\theta$, satisfy
$\cos(\theta) = X\cdot Y$, two points on the unit hyperboloid satisfy
\begin{equation}
    \cosh(L) = -X_1\cdot Y_1.
\end{equation}
To compute correlators in the boundary theory, we must evaluate bulk-to-boundary propagators.
Since the boundary is a conformal boundary,
the geodesic distance between a boundary point and a bulk point will diverge. 
To evaluate bulk-to-boundary propagators, we can 
regularize the boundary point to extract a finite renormalized geodesic distance. 
Consider the bulk point
\begin{equation}
    X_\varepsilon^A = \frac{1}{\varepsilon} P^A + \varepsilon Q^A,
\end{equation}
where $P^A$ is a null boundary point, $P^2=0$, $\varepsilon$ is taken to be
infinitesimal, and $Q^A$ is chosen satisfying
\begin{equation}
    P\cdot Q=-\frac12 \qquad{\text{ and } \qquad Q^2 =0,}
\end{equation}
such that $
    X_\varepsilon\cdot X_\varepsilon
    =
    -1
    $,
implying that $X_\varepsilon$ stays on 
the AdS hyperboloid as $\varepsilon\to0$.
We can now compute the geodesic distance between $X$ and $X_\varepsilon$,
\begin{equation}
    L_\varepsilon = \mathrm{arccosh}\left(-\frac{1}{\varepsilon}X\cdot P - \varepsilon X\cdot Q\right).
\end{equation}
Expanding around small $\varepsilon$ gives
\begin{equation}
    L_\varepsilon \xrightarrow[]{\varepsilon \rightarrow 0} \log\left(-\frac{2 X \cdot P}{\varepsilon}\right)+\mathcal O(\varepsilon^2),
\end{equation}
which allows one to define the renormalized geodesic length by minimal subtraction as
\begin{equation}
    \mathcal{L} \equiv \lim_{\varepsilon\rightarrow 0} \left[
    L_\varepsilon - \log\left(\frac{1}{\varepsilon} \right)
    \right] = \log(-2 X\cdot P).
\end{equation}
This renormalized length allows the bulk-to-boundary propagator to be written as an exponential, 
which will be particularly useful for the heavy external legs, since the corresponding Witten 
integrals are dominated by a saddle in the large-$\Delta_H$ limit, as we now show.

\subsection{Heavy limit in a tree-level diagram}
\label{Heavy limit of Witten diagrams}

We consider the four-point function $G(P_i)$ in the limit
$\Delta_H\to\infty$ at fixed $\Delta_L$. We first show that the
tree-level contact diagram localizes onto the geodesic joining the
heavy insertions, and identify the resulting integral as the first
Born iteration for the light field. We then evaluate this integral
explicitly. Geodesic localization is familiar from semiclassical
worldline treatments of large-dimension AdS correlators and from
geodesic Witten diagrams
\cite{Minahan:2012fh,Hijano:2015zsa,Chen:2026ium,Chen:2025yxg}.
Here it provides the starting point for the reduction of loop
diagrams to higher terms in the Born series.

We begin by evaluating the connected tree-level contact Witten diagram
generated by the interaction $\lambda\phi^2\chi^2$ in
\eqref{eq:bulk-action}. At first order in $\lambda$, the tree-level
diagram is given by the standard $D$-function
\cite{DHoker:1999kzh},
\begin{equation}
    G^{\rm tree}(P_i)
    \equiv -\lambda\,D_{\Delta_H\Delta_L\Delta_L\Delta_H}(P_i),
    \label{eq:contact-D-function}
\end{equation}
where $G_{\rm conn}=G^{\rm tree}+\mathcal{O}(\lambda^2)$ and
\begin{equation}
    D_{\Delta_H\Delta_L\Delta_L\Delta_H}(P_i)
    =\int_{\mathrm{AdS}}dX\,
    \Pi_{\Delta_H}^{\partial}(P_1,X)
    \Pi_{\Delta_L}^{\partial}(P_2,X)
    \Pi_{\Delta_L}^{\partial}(P_3,X)
    \Pi_{\Delta_H}^{\partial}(P_4,X).
    \label{eq:D-function-definition}
\end{equation}
Here $dX$ is the invariant AdS measure. Using the renormalized
geodesic length introduced in the previous subsection, the heavy
bulk-to-boundary propagators take the form
\begin{equation}
    \Pi_{\Delta_H}^{\partial}(P,X)
    =\frac{\sqrt{\mathcal{C}_{\Delta_H}}}
          {(-2P\cdot X)^{\Delta_H}}
    =\sqrt{\mathcal{C}_{\Delta_H}}
      e^{-\Delta_H\mathcal{L}(P,X)},
    \label{eq:bulktoboundary}
\end{equation}
where unit normalization of the boundary two-point function fixes
\begin{equation}
\label{eq:CdH}
    \mathcal{C}_{\Delta_H}
    =\frac{\Gamma(\Delta_H)}
          {2\pi^{d/2}\Gamma(\Delta_H-\frac d2+1)}.
\end{equation}
Defining $S(X)\equiv\mathcal{L}(P_1,X)+\mathcal{L}(P_4,X)$,
we can therefore write
\begin{equation}
    D_{\Delta_H\Delta_L\Delta_L\Delta_H}(P_i)
    =\int_{\mathrm{AdS}}dX\,
    e^{-\Delta_H{S(X)}}\mathcal{K}_L(P_2,P_3;X),
    \label{eq:contact-saddle-integral}
\end{equation}
with the two light propagators collected in
\begin{equation}
    \mathcal{K}_L(P_2,P_3;X)
    =\mathcal{C}_{\Delta_H}
    \Pi_{\Delta_L}^{\partial}(P_2,X)
    \Pi_{\Delta_L}^{\partial}(P_3,X).
    \label{eq:light-kernel}
\end{equation}
At large $\Delta_H$, the integral is dominated by the minimum
of $S(X)$, which satisfies
\begin{equation}
    \nabla_X\left[\mathcal{L}(P_1,X)+\mathcal{L}(P_4,X)\right]=0.
    \label{boundary_saddle}
\end{equation}
This condition places $X$ on the geodesic $\gamma$ joining
$P_1$ and $P_4$: the gradients of the two lengths are equal and
opposite there \cite{Wald:1984rg}.
To compute the transverse fluctuations, we use an AdS isometry
to choose global coordinates in which $\gamma$ lies at $r=0$.
The metric is
\begin{equation}
    ds^2=(1+r^2)dt^2+\frac{dr^2}{1+r^2}+r^2d\Omega_{d-1}^2,
    \label{eq:globalAdS}
\end{equation}
and $t$ measures proper length along $\gamma$. The corresponding
bulk embedding and heavy endpoints can be chosen as
\begin{equation}
\begin{aligned}
X^A(r,t,\hat n)
&=\bigl(\sqrt{1+r^2}\cosh t,\,
        \sqrt{1+r^2}\sinh t,\,r\hat n\bigr),\\
P_{1,4}^A
&=\frac{\sqrt{P_{14}}}{2}(1,\pm1,\mathbf{0}),
\qquad P_{14}\equiv-2P_1\cdot P_4>0.
\end{aligned}
\label{eq:tree-global-embedding}
\end{equation}
It follows that
\begin{equation}
(-2P_1\cdot X)(-2P_4\cdot X)=P_{14}(1+r^2).
\end{equation}
Writing $r=\sinh\rho$, where $\rho$ is the proper distance from
$\gamma$, the exponent becomes
\begin{equation}
    {S(X)=\log P_{14}+2\log\cosh\rho
    =\log P_{14}+\rho^2+O(\rho^4).}
    \label{eq:heavy-transverse-exponent}
\end{equation}
Therefore, $S$ is minimized at $\rho=0$ and is constant along
$\gamma$. The heavy limit, $\Delta_H \to \infty$, then  localizes
the transverse integration in  Eq.~\eqref{eq:contact-saddle-integral} to $\rho=\mathcal{O}(\Delta_H^{-1/2})$. The invariant measure obtained from  Eq.~\eqref{eq:globalAdS} is
\begin{equation}
dX
=dt\,d\rho\,d\Omega_{d-1}\,
\cosh\rho\,\sinh^{d-1}\rho
=dt\,d\rho\,d\Omega_{d-1}\,
\rho^{d-1}\left[1+O(\rho^2)\right].
\end{equation}
At fixed $\Delta_L$ and separated light insertions, the light
propagators in  Eq.~\eqref{eq:light-kernel} are smooth across this
narrow region with $\rho = \mathcal{O}(\Delta_H^{-1/2})$. Their linear transverse variations vanish upon
angular integration, while quadratic terms give relative
corrections of order $1/\Delta_H$. At leading order, we may
therefore evaluate $\mathcal K_L$ on $\gamma$ and perform the
radial Gaussian integral:
\begin{equation}
{
\operatorname{vol}(S^{d-1})
\int_0^\infty d\rho\,\rho^{d-1}e^{-\Delta_H\rho^2}
=
\left(\frac{\pi}{\Delta_H}\right)^{d/2}.
}
\label{gaussian transverse integral}
\end{equation}
After substituting into  Eq.~\eqref{eq:contact-saddle-integral}, what remains is the integration over the geodesic of the heavy field at $r = 0$, which expressed in terms of the original
boundary points gives
\begin{equation}
    D_{\Delta_H\Delta_L\Delta_L\Delta_H}(P_i)
    {\approx
    \frac{\mathcal{C}_{\Delta_H}}{P_{14}^{\Delta_H}}
    \left(\frac{\pi}{\Delta_H}\right)^{d/2}
    \int_\gamma dt\,
    \Pi_{\Delta_L}^{\partial}(P_2,X(t))
    \Pi_{\Delta_L}^{\partial}(P_3,X(t)).}
    \label{contour_integral}
\end{equation}
The factor $P_{14}^{-\Delta_H}$ is the heavy two-point
function, while  Eq.~\eqref{eq:CdH} gives
\begin{equation}
\mathcal{C}_{\Delta_H}
\left(\frac{\pi}{\Delta_H}\right)^{d/2}
=\frac{1}{2\Delta_H}\left[1+O(\Delta_H^{-1})\right].
\end{equation}
Dividing out the heavy two-point function and using
 Eq.~\eqref{eq:contact-D-function}, the result takes the Born form
\begin{equation}
    {P_{14}^{\Delta_H}G^{\rm tree}(P_i)\approx}
    \int_{\mathrm{AdS}}dX\,
    \Pi_{\Delta_L}^{\partial}(P_2,X)
    V(X)\Pi_{\Delta_L}^{\partial}(X,P_3),
    \label{eq:Gtree}
\end{equation}
where the effective contact potential is identified as\footnote{Here the delta function is normalized with respect to the
invariant measure, so that
\[
\int_{\mathrm{AdS}}dX\,\delta^{(d+1)}(X-Y)f(X)=f(Y).
\]}
\begin{equation}
    V(X)=-\frac{\lambda}{2\Delta_H}
    {\int_\gamma dt\,\delta^{(d+1)}(X-X(t)).}
    \label{eq:contact potential with no derivatives}
\end{equation}
This is the first Born correction to the
light-field two-point function in a potential supported on the
heavy geodesic. The choice $r=0$ was made using an AdS isometry.
Since the worldline measure and delta function are invariantly
defined, the result applies to arbitrary boundary insertion points.

\paragraph{Explicit integration over the heavy geodesic.} To evaluate  Eq.~\eqref{contour_integral},
we make use of the standard cylinder configuration \cite{Rychkov:2016iqz} where the heavy
operators are placed at the origin and infinity, one light operator at
$x=1$, and the other at radius $e^{-\tau}$ and angular separation $\theta$, 
with  $z=e^{-\tau+i\theta}$ and $\bar z=e^{-\tau-i\theta}$. Under the map $x=e^{-t}\hat n$, the light insertions lie
at cylinder times $0$ and $\tau$. Including the Weyl factor,
the contact diagram  Eq.~\eqref{eq:contact-D-function} becomes
\begin{equation}
\label{eq:gcyltree}
G_{\rm cyl}^{\rm tree}(\tau,\theta)
\equiv
-\lambda e^{-\Delta_L\tau}
D_{\Delta_H\Delta_L\Delta_L\Delta_H}
\bigl(0,(z,\bar z),1,\infty\bigr).
\end{equation}
The insertion at infinity in the $D$-function is understood with
the standard normalization $\mathcal{O}_H(\infty)
    \equiv
    \lim_{x\to\infty}|x|^{2\Delta_H}\mathcal{O}_H(x)$. The Weyl factor
$e^{-\Delta_L\tau}$ comes from the light insertion at
$|z|=e^{-\tau}$, with the insertion at unit radius not contributing.

Boundary points in these cylinder coordinates have embedding
vectors $P^A(t_p,\hat n_p)
=
\bigl(\cosh t_p,\sinh t_p,\hat n_p\bigr)$. Together with Eq.~\eqref{eq:tree-global-embedding}, this gives
\begin{equation}
P\cdot X
=
r\,\hat n\cdot\hat n_p
-\sqrt{1+r^2}\cosh(t-t_p)
\xrightarrow{r\to0}
-\cosh(t-t_p).
\end{equation}
Thus Eq.~\eqref{eq:bulktoboundary}, with
$\Delta_H$ replaced by $\Delta_L$, gives the cylinder-normalized
light propagators on the heavy geodesic:
\begin{equation}
\label{eq:LLp}
\Pi_{\Delta_L}^{\partial}(P_i,X(t))
=
\frac{\sqrt{\mathcal C_{\Delta_L}}}
     {[2\cosh(t-t_{P_i})]^{\Delta_L}},
\qquad
t_{P_2}=0,\quad t_{P_3}=\tau.
\end{equation}
Substituting these propagators and the potential
\eqref{eq:contact potential with no derivatives} into
\eqref{eq:Gtree}, with $t_{P_2}=0$ and $t_{P_3}=\tau$, gives
\begin{equation}    
    G_{\mathrm{cyl}}^{\rm tree}(\tau,\theta)
    \approx
    -\frac{\lambda\mathcal C_{\Delta_L}}{2\Delta_H}
    \int_{-\infty}^{\infty}dt\,
    \frac{1}{(2\cosh t)^{\Delta_L}}
    \frac{1}{(2\cosh(t-\tau))^{\Delta_L}}.
\end{equation}
The leading
result is independent of $\theta$, so only the $\ell=0$
partial wave contributes. Taking the Fourier transform and
using the convolution theorem gives
\begin{equation}   
    \begin{aligned}
    \tilde G_{\mathrm{cyl}}^{\rm tree}(\omega,\theta)
    &\equiv
    \int_{-\infty}^{\infty}d\tau\,
    e^{-i\omega\tau}
    G_{\mathrm{cyl}}^{\rm tree}(\tau,\theta)
    \approx
    -\frac{\lambda\mathcal C_{\Delta_L}}{2\Delta_H}
    \left[
    \int_{-\infty}^{\infty}dt\,
    \frac{e^{-i\omega t}}{(2\cosh t)^{\Delta_L}}
    \right]^2.
    \end{aligned}
    \label{eq:convprop}
\end{equation}
The substitution $y=e^{2t}$ gives the Euler beta-function
identity\footnote{The general identity \cite{Gradshteyn:1943cpj} is
\begin{equation}
\begin{aligned}
B(x+iy,x-iy)
&=
2^{1-2x}\alpha e^{-2i\beta y}
\int_{\mathbb R}dt\,
\frac{e^{2i\alpha yt}}
     {\cosh^{2x}(\alpha t-\beta)}=\frac{\Gamma(x+iy)\Gamma(x-iy)}{\Gamma(2x)},\nonumber
\end{aligned}
\end{equation}
valid for $\Re x>0$, $\alpha>0$, and real $y,\beta$.
The identity used here follows by setting
$x=\Delta/2$, $y=-\omega/2$, $\alpha=1$, and $\beta=0$.} 
\begin{equation}
\int_{-\infty}^{\infty}
\frac{e^{-i\omega t}\,dt}{(2\cosh t)^\Delta}
=
\frac12 B\!\left(
\frac{\Delta+i\omega}{2},
\frac{\Delta-i\omega}{2}
\right),
\end{equation}
valid for $\operatorname{Re}\Delta>0$ and real $\omega$.
Therefore,
\begin{equation}  
    \tilde G_{\mathrm{cyl}}^{\rm tree}(\omega,\theta)
    \approx
    -\frac{\lambda\mathcal C_{\Delta_L}}{8\Delta_H}
    B^2\!\left(
    \frac{\Delta_L+i\omega}{2},
    \frac{\Delta_L-i\omega}{2}
    \right).
    \label{FT_contact}
\end{equation}
What we have obtained is a partial-wave decomposition:
the absence of angular dependence reflects the fact that,
at leading order in the heavy limit, the non-derivative contact
diagram contributes only to the $\ell=0$ partial wave.
This spectral representation also makes the connection to the
heavy--light $s$--channel OPE explicit. For $\tau>0$, the inverse
Fourier transform can be evaluated by closing the contour in
the upper half-plane, picking up double poles at
$\omega=i(\Delta_L+2n)$, with $n=0,1,\ldots$.
Their residues differentiate the factor $e^{i\omega\tau}$,
producing terms proportional to
$\tau e^{-(\Delta_L+2n)\tau}$.
Since $\tau=-\frac12\log(z\bar z)$, these generate the logarithms
associated with anomalous dimensions.
We analyze the extraction of OPE data in section~\ref{CFT}.

\subsection{Heavy limit in Mellin space}\label{Mellin}
Mellin amplitudes are well known to provide an excellent framework for analyzing holographic correlators. In the previous section we obtained a compact formula for the contact Witten diagram in the limit where one of the operators is heavy. The goal of this section is to highlight that the same result can be obtained from Mellin amplitudes \cite{Mack:2009mi,Penedones:2010ue}. The correlator can be written as
\begin{equation}
        \langle 
        \mathcal O_H(P_1)
        \mathcal O_L(P_2)
        \mathcal O_L(P_3)
        \mathcal O_H(P_4)
        \rangle
        =\int \left[d\delta_{ij}\right] M(\delta_{ij} )
        \prod_{i<j}\frac{\Gamma(\delta_{ij})}{(P_{ij})^{\delta_{ij}}},
\end{equation}
with the constraint,
\begin{equation}
    \sum_{j\neq i}\delta_{ij} = \Delta_i =
    \left\{ 
        \begin{array}{ll}
            \Delta_H, \quad i = 1,4 \\
            \Delta_L, \quad i = 2,3 \\
        \end{array}
    \right..
\end{equation}
We choose the independent Mellin variables to be
\begin{equation}
    \begin{aligned}
&\delta_{12}=\delta_{34}
=
\frac{\Delta_H+\Delta_L-s}{2},
\qquad
\delta_{14}
=
\Delta_H-\frac{t}{2}, \\
&\delta_{13}=\delta_{24}
=
\frac{s+t-\Delta_H-\Delta_L}{2},
\qquad
\delta_{23}
=
\Delta_L-\frac{t}{2},
\end{aligned} 
\end{equation}
and the prefactor for the reduced correlator, 
\begin{equation}
        \langle 
        \mathcal O_H(P_1)
        \mathcal O_L(P_2)
        \mathcal O_L(P_3)
        \mathcal O_H(P_4)
        \rangle = 
        \frac{
        \left(\frac{P_{24}}{P_{14}}\right)^{\frac{\Delta_H-\Delta_L}{2}}
        \left(\frac{P_{14}}{P_{13}}\right)^{\frac{\Delta_L-\Delta_H}{2}}
    }{
                \left(P_{12} P_{34}\right)^{\frac{\Delta_H + \Delta_L}{2}}
        } \mathcal G (u,v),
        \label{eq:redGuv} 
\end{equation}
where we defined the standard cross ratios by
\begin{equation}
u=\frac{P_{12}P_{34}}{P_{13}P_{24}},
\qquad
v=\frac{P_{14}P_{23}}{P_{13}P_{24}}.
\end{equation}
The reduced correlator then admits the Mellin representation
\begin{equation}\label{eq:DefinitionMellin}
    \begin{aligned}
        \mathcal G(u,v)=\int\frac{ds}{4\pi i}
        \int\frac{dt}{4\pi i}\,
&u^{s/2} 
v^{\frac{t}{2} - \Delta_L} M(s,t)
\Gamma \left(\Delta_H-\frac{t}{2}\right) 
\Gamma \left(\Delta_L-\frac{t}{2}\right) \\
              \times&\Gamma \left(\frac{\Delta_L + \Delta_H - s}{2}\right)^2 
\Gamma \left(\frac{s + t - \Delta_L - \Delta_H }{2}\right)^2.
\end{aligned}
\end{equation}
For the contact contribution generated by the interaction
in Eq.~\eqref{eq:bulk-action}, the Mellin amplitude is independent
of $s$ and $t$. Using the Mellin representation of the contact Witten diagram
derived in \cite{Penedones:2010ue}, with the propagator
normalization Eq.~\eqref{eq:bulktoboundary}, we have
\begin{equation}
M(s,t)=M_0
\equiv
-\frac{\lambda\pi^{d/2}}{2}
\frac{
\mathcal C_{\Delta_H}\mathcal C_{\Delta_L}
\Gamma(\Delta_H+\Delta_L-\frac d2)
}{
\Gamma(\Delta_H)^2\Gamma(\Delta_L)^2
},
\label{eq:contact-Mellin-amplitude}
\end{equation}
with $\mathcal{C}_{\Delta_H}$ and $\mathcal{C}_{\Delta_L}$ given by Eq.~\eqref{eq:CdH}. The contours in Eq.~\eqref{eq:DefinitionMellin} are chosen such that the real parts of the arguments
of all Gamma functions are positive.
This condition implies the inequalities,
\begin{equation}
    \Re(s)<\Delta_H + \Delta_L, \quad 
    \Re(t)<2\Delta_L, \quad
    \Re(s+t) > \Delta_H + \Delta_L.
\end{equation}
To isolate the dependence on $\Delta_H$ in Eq.~\eqref{eq:DefinitionMellin}, it is convenient to use
$\sigma \equiv s-\Delta_H$. We find
\begin{equation}
    \begin{aligned}
        \mathcal G(u,v)={M_0}\, u^{\frac{\Delta_H}{2}}
        \int\frac{d{\sigma}}{4\pi i}
        \int\frac{dt}{4\pi i}\,
&u^{\frac{\sigma}{2}} 
v^{\frac{t}{2} - \Delta_L}
\Gamma \left(\Delta_H-\frac{t}{2}\right) 
\Gamma \left(\Delta_L-\frac{t}{2}\right) \\
        \times&\Gamma \left(\frac{\Delta_L  - {\sigma}}{2}\right)^2 
\Gamma \left(\frac{{\sigma} + t - \Delta_L  }{2}\right)^2 .
\end{aligned} 
\end{equation}
Taking the heavy limit $\Delta_H \to \infty$ via Stirling's approximation, $\Gamma(\Delta_H - \frac{t}{2}) \approx \Gamma(\Delta_H) \Delta_H^{-t/2}$, the reduced correlator simplifies to:
\begin{equation}
    \begin{aligned}
        \mathcal G(u,v)&\xrightarrow{\Delta_H\rightarrow\infty} {M_0}\,
\Gamma \left(\Delta_H\right) 
        u^{\frac{\Delta_H}{2}}
v^{ - \Delta_L}
        \int\frac{d{\sigma}}{4\pi i}\int\frac{dt}{4\pi i}\,
u^{{\sigma}/2} 
\left(\frac{v}{\Delta_H}\right)^{\frac{t}{2}}
\Gamma \left(\Delta_L-\frac{t}{2}\right) \\
              &\times\Gamma \left(\frac{\Delta_L  - {\sigma}}{2}\right)^2 
\Gamma \left(\frac{{\sigma} + t - \Delta_L  }{2}\right)^2 .
\end{aligned}
\end{equation}
The integral in $t$ can now be computed by residues by deforming the contour to the right. The weight factor
$\left(\frac{v}{\Delta_H}\right)$ naturally organizes the contribution of the residues in the heavy 
limit. The leading-order contribution in $1/\Delta_H$ is completely dominated by the pole at  $t = 2 \Delta_L $.  The final result is given by
\begin{equation}
\begin{aligned}
        \mathcal G(u,v)&\xrightarrow{\Delta_H\rightarrow\infty}
        {M_0}\,\frac{\Gamma(\Delta_H)}{(\Delta_H)^{\Delta_L}}
        u^{\frac{\Delta_H}{2}}
        \int\frac{d{\sigma}}{4\pi i}u^{{\sigma}/2} 
\Gamma \left(\frac{\Delta_L-{\sigma}  }{2}\right)^2 
\Gamma \left(\frac{  \Delta_L +{\sigma} }{2}\right)^2 \label{eq:ContactWittenHyper}\\
&={M_0}\,\frac{\Gamma(\Delta_H)\Gamma^4(\Delta_L)}{(\Delta_H)^{\Delta_L} \Gamma(2\Delta_L)} u^{\frac{\Delta_H+\Delta_L}{2}}{}_2F_1\bigg(\Delta_L,\Delta_L;2\Delta_L;1-u\bigg).
\end{aligned}
\end{equation}
Evaluating the prefactor in Eq.~\eqref{eq:redGuv} in the frame
$(0,z,1,\infty)$ and multiplying by the factor
$e^{-\Delta_L\tau}=u^{\Delta_L/2}$ from the map to the cylinder,
we obtain
\begin{equation}
G_{\rm cyl}^{\rm tree}(\tau,\theta)
=u^{-\Delta_H/2}\mathcal G(u,v),
\qquad u=e^{-2\tau}.
\end{equation}
From Eq.~\eqref{eq:contact-Mellin-amplitude} in the heavy limit $\Delta_H \to \infty$ we have
\begin{equation}
M_0\frac{\Gamma(\Delta_H)}{\Delta_H^{\Delta_L}}
\approx
-\frac{\lambda\mathcal C_{\Delta_L}}
{4\Delta_H\Gamma(\Delta_L)^2},
\end{equation}
and setting $\sigma=-i\omega$ in the Mellin integral in Eq.~\eqref{eq:ContactWittenHyper},
we obtain
\begin{equation}
G_{\rm cyl}^{\rm tree}(\tau,\theta)
\approx
-\frac{\lambda\mathcal C_{\Delta_L}}{8\Delta_H}
\int_{-\infty}^{\infty}\frac{d\omega}{2\pi}\,
e^{i\omega\tau}
B\!\left(
\frac{\Delta_L+i\omega}{2},
\frac{\Delta_L-i\omega}{2}
\right)^2.
\end{equation}
This is precisely the inverse Fourier transform of
\eqref{FT_contact}.

A feature of this result is its manifest independence of the cross-ratio $v$. This will be important when we do the OPE analysis in the crossed channel. Furthermore, this Mellin representation provides a simple framework for extracting the $u\rightarrow 0$ limit.

\subsection{Loop diagrams}\label{Loop diagrams}
Having shown the localization of the tree-level contact diagram and its simplicity in the heavy limit, we now extend this analysis to the infinite series of higher-loop bubble diagrams. Computing any diagram within this infinite class involves two new elements: bulk-to-bulk propagators and internal bulk vertices that do not anchor directly to the asymptotic boundary.

The exact bulk-to-bulk propagator for a scalar field of dimension $\Delta$ is
\begin{equation}
    \Pi^{\text{B}}_{\Delta}(X,Y)=
    \frac{\mathcal{C}_\Delta }{\left(-2X\cdot Y \right)^\Delta}
    \ _2 F_1\left( \frac{\Delta}{2}, \frac{\Delta+1}{2} ; \Delta-\frac{d}{2} + 1; \frac{1}{(-X\cdot Y)^2}\right).
    \label{bulk-to-bulk}
\end{equation}
For $m_H\approx\Delta_H-d/2\to\infty$ the propagator has the
leading asymptotic form \cite{Maxfield:2017rkn}
\begin{equation}
\Pi_{\Delta_{H}}^{B}(X,Y)\xrightarrow{m_{H}\rightarrow\infty}\frac{e^{-m_{H}L(X,Y)}}{2m_{H}}\left(\frac{m_{H}}{2\pi \sinh L(X,Y)}\right)^{d/2},
\label{eq:heavy-bulk-propagator}
\end{equation}
where $L(X,Y) =\operatorname{arccosh}(-X\cdot Y)$ is the geodesic distance between two bulk points introduced in section~\ref{GESC}.
The exponential is the contribution of the classical
geodesic joining $X$ and $Y$, while the prefactor accounts
for fluctuations around it through the AdS Van Vleck
determinant. 
We derive this result in Appendix~\ref{app:heavyprop} from the
DeWitt--Schwinger representation of the propagator.

A typical intermediate bulk integration contains a structure of the form
\begin{equation}
   I[\mathcal R] \equiv  \int_{\mathrm{AdS}} dX\,
    \Pi_{\Delta_H}^{B}(X_1,X)\,
    \Pi_{\Delta_H}^{B}(X,X_2)\,
    \mathcal R(X),
    \label{eq:bulkintegration}
\end{equation}
where $\mathcal R(X)$ denotes the remaining light-field propagators and interactions. In the heavy limit, the exponential dependence of the heavy propagators forces the intermediate point $X$ onto the geodesic joining $X_1$ and $X_2$. For the higher-loop construction, however, the location of the saddle is not sufficient. The semiclassical prefactors of the
heavy propagators must combine consistently with the Gaussian fluctuations generated by
the integration over $X$. In Appendix~\ref{app:heavyprop} we show, using the worldline representation of the heavy propagator, that the Van Vleck prefactors combine with the Gaussian fluctuations of the intermediate vertex in such a way that the heavy propagation composes along a single classical geodesic. The result is the composition rule
\begin{equation}
    I[\mathcal R]
\approx
\frac{1}{2m_H}\,
\Pi^{\rm B}_{\Delta_H}(X_1,X_2)
\int_{\gamma_{12}} d\tau\, \mathcal R\bigl(X_*(\tau)\bigr),
\label{eq: rule for bulk integrals}
\end{equation}
where $\gamma_{12}$ is the geodesic connecting $X_1$ and $X_2$, $\tau$ is its length parameter and $X_*(\tau)$ the corresponding point on it. The prefactor $1/2m_H$ is the same $1/2 \Delta_H$ that appears in Eq.~\eqref{eq:contact potential with no derivatives}, up to quadratic corrections,
\begin{equation}
    m_H = \Delta_H + \mathcal O(1)
\Rightarrow
\frac{1}{2m_H}
=
\frac{1}{2\Delta_H}
+
\mathcal O\!\left(\Delta_H^{-2}\right).
\end{equation}

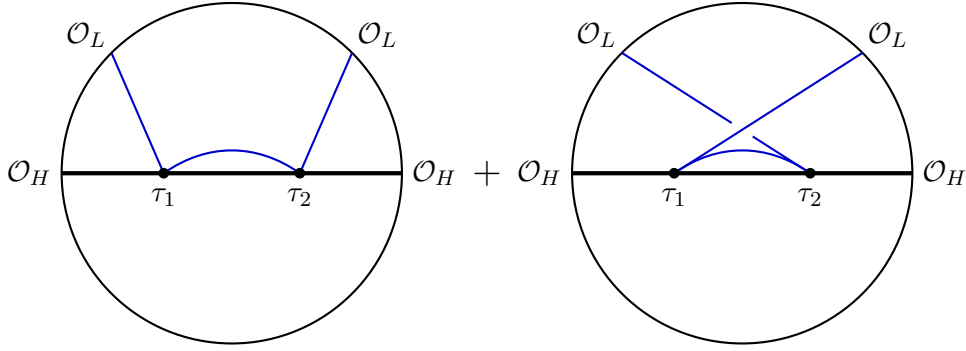
\begin{figure}[htbp]
    \centering
    \begin{tikzpicture}[scale=1.5]
        
        \begin{scope}[shift={(0,0)}]
            \draw[thick] (0,0) circle (1.5);
            
            \draw[ultra thick, black] (-1.5,0) -- (1.5,0);
            
            \node[left, black] at (-1.5,0) {$\mathcal{O}_H$};
            \node[right, black] at (1.5,0) {$\mathcal{O}_H$};

            \coordinate (x1) at (-0.6,0);
            \coordinate (x2) at (0.6,0);

            \draw[thick, blue!80!black] (135:1.5) node[above left=-2pt, black] {$\mathcal{O}_L$} -- (x1);
            \draw[thick, blue!80!black] (45:1.5) node[above right=-2pt, black] {$\mathcal{O}_L$} -- (x2);

            \draw[thick, blue!80!black] (x1) to[out=35, in=145] (x2);

            \filldraw[black] (x1) circle (1.2pt) node[below=2pt] {$\tau_1$};
            \filldraw[black] (x2) circle (1.2pt) node[below=2pt] {$\tau_2$};
        \end{scope}
        \node at (2.25,0) {\Large $+$};

        \begin{scope}[shift={(4.5,0)}]
            \draw[thick] (0,0) circle (1.5);
            
            \draw[ultra thick, black] (-1.5,0) -- (1.5,0);
            
            \node[left, black] at (-1.5,0) {$\mathcal{O}_H$};
            \node[right, black] at (1.5,0) {$\mathcal{O}_H$};

            \coordinate (x1) at (-0.6,0);
            \coordinate (x2) at (0.6,0);

            \draw[thick, blue!80!black] (x1) to[out=35, in=145] (x2);

            \draw[thick, blue!80!black] (135:1.5) node[above left=-2pt, black] {$\mathcal{O}_L$} -- (x2);
            
            \filldraw[white] (0, 0.383) circle (3pt);

            \draw[thick, blue!80!black] (45:1.5) node[above right=-2pt, black] {$\mathcal{O}_L$} -- (x1);

            \filldraw[black] (x1) circle (1.2pt) node[below=2pt] {$\tau_1$};
            \filldraw[black] (x2) circle (1.2pt) node[below=2pt] {$\tau_2$};
        \end{scope}

    \end{tikzpicture}
    \caption{Witten diagrams contributing at one loop to the correlator for the quartic vertex $\chi^2\phi^2$.
    The heavy geodesic is the thick black line and light-field propagators are drawn in blue. The two orderings of the vertices along the worldline contribute at the same order in the Born series.}
    \label{fig:loop_diagrams}
\end{figure}

With Eq.~\eqref{eq: rule for bulk integrals} in hand, the remaining subtlety arises when several internal vertices are integrated. Let us begin by considering the following integral,
\begin{equation}
    I_{\Delta_H}(X_1,X_4) \equiv \int_{\text{AdS}}dX_2 dX_3 \ 
        \Pi_{\Delta_H}^B(X_1,X_2)
        \Pi_{\Delta_H}^B(X_2,X_3)
        \Pi_{\Delta_H}^B(X_3,X_4) \mathcal R(X_2,X_3),
\end{equation}
for some kernel $\mathcal R(X_2,X_3)$. When we study the integral $dX_3$, we arrive at the integral
over the geodesic connecting $X_2$ with $X_4$,
\begin{equation} 
    I_{\Delta_H}(X_1,X_4)  \approx 
\frac{1}{2\Delta_H}
\int_{\text{AdS}}dX_2 \ 
        \Pi_{\Delta_H}^B(X_1,X_2)
        \Pi_{\Delta_H}^B(X_2,X_4)
\int_{\gamma_{24}}d\tau_1 \ 
         \mathcal R(X_2,X_*(\tau_1)).
\end{equation}
Repeating the same process, we arrive at the geodesic connecting $X_1$ and $X_4$, but notice that the bound on the $\tau_1$ integral depends on the location of $X_2$ (the variable that is now being integrated). This induces a time ordering of the parameters: the $\tau_1$ integral extends at most up to $\tau_2$, which is itself integrated along the worldline,
\begin{equation}
    I_{\Delta_H} (X_1,X_4) = 
    \left(\frac{1}{2\Delta_H}\right)^2
    \Pi_{\Delta_H}^B(X_1,X_4)
\int_{\tau_2>\tau_1}d\tau_1 d\tau_2 \ 
\mathcal{R}(X_*(\tau_2),X_*(\tau_1)).
\end{equation}
This ordering is essential for the emergence of the Born series.
To see this explicitly, we can compute higher-order Witten diagrams in the Born regime,
starting with the one-loop diagrams shown
in Figure \ref{fig:loop_diagrams}. This will
serve as a consistency check of the calculation of the next section.

Let $W \equiv W_{\text{bubble}}+W_{\text{cross}}$ denote the 1-loop contribution from the two diagrams,
\begin{equation}
    \begin{aligned}
        W_{\text{bubble}} \equiv\lambda^2\int_{\text{AdS}} dX_1 dX_2 \ \Big[
    &\Pi_{\Delta_H}^{\partial}(P_1,X_1)
    \Pi_{\Delta_H}^{\text{B}}(X_1,X_2)
    \Pi_{\Delta_H}^{\partial}(P_4,X_2) \\ 
        \times&\Pi_{\Delta_L}^{\partial}(P_2,X_1)
        \Pi_{\Delta_L}^{\text{B}}(X_1,X_2)
        \Pi_{\Delta_L}^{\partial}(P_3,X_2)
    \Big].
    \end{aligned}
\end{equation}
\begin{equation}
    \begin{aligned}
        W_{\text{cross}} \equiv \lambda^2\int_{\text{AdS}} dX_1 dX_2 \ \Big[
    &\Pi_{\Delta_H}^{\partial}(P_1,X_1)
    \Pi_{\Delta_H}^{\text{B}}(X_1,X_2)
    \Pi_{\Delta_H}^{\partial}(P_4,X_2) \\ 
        \times&\Pi_{\Delta_L}^{\partial}(P_2,X_2)
        \Pi_{\Delta_L}^{\text{B}}(X_1,X_2)
        \Pi_{\Delta_L}^{\partial}(P_3,X_1)
    \Big].
    \end{aligned}
\end{equation}
We now place the heavy insertions at $0$ and $\infty$,
with a normalization such that
their two-point function is unity. For separated interaction vertices, the composition
rule,  Eq.~\eqref{eq: rule for bulk integrals}, localizes the heavy propagators onto the
geodesic joining the heavy insertions.
For the two diagrams, the localized contributions are
\begin{equation}
    \begin{aligned}
        \left(\frac{2 \Delta_H}{\lambda}\right)^2 W \approx &\int^{\tau_1>\tau_2}_{\tau_1,\tau_2 \in \mathbb R} d\tau_1 d\tau_2 \ 
        \Pi_{\Delta_L}^\partial(P_2,X(\tau_1)) 
        \Pi_{\Delta_L}^{\text{B}}(X(\tau_1),X(\tau_2)) 
        \Pi_{\Delta_L}^\partial(P_3,X(\tau_2))\\
        +&
        \int^{\tau_2>\tau_1}_{\tau_1,\tau_2 \in \mathbb R} d\tau_1 d\tau_2 \ 
        \Pi_{\Delta_L}^\partial(P_2,X(\tau_1)) 
        \Pi_{\Delta_L}^{\text{B}}(X(\tau_1),X(\tau_2)) 
        \Pi_{\Delta_L}^\partial(P_3,X(\tau_2))
    \end{aligned}
\end{equation}
After relabeling the integration variables, the two
diagrams have the same integrand and complementary
domains, $\tau_1>\tau_2$ and $\tau_2>\tau_1$.
Their sum therefore gives the integral over
$\mathbb{R}^2$. The
same mechanism appears in the defect description of HHLL correlators of giant
gravitons \cite{Chen:2025yxg,Chen:2026ium}. In the heavy limit, the $t$- and
$u$-channel exchange Witten diagrams reduce to the two time orderings of the
vertices along the defect, and their sum recombines into a single defect-channel
exchange diagram.

We can now reuse the potential $V(X)$ in  Eq.~\eqref{eq:contact potential with no derivatives},
giving precisely the second Born correction,
\begin{equation}
\label{eq:Born2}
    \begin{aligned}
        W \approx     \int_{\text{AdS}} dX_1 dX_2
            &\Pi^\partial_{\Delta_L}(P_2,X_1)V(X_1)\Pi^B_{\Delta_L}(X_1,X_2)V(X_2)
            \Pi^\partial_{\Delta_L}(X_2,P_3).
    \end{aligned}
\end{equation}
This is the defining feature of the Born regime \cite{Correia:2024jgr}. The $2\to2$ process becomes a one-body problem, the light field scattering off a potential that corrects its propagator.
From a perturbative point of view, we are computing corrections to the two point function
by iterating the free propagator with the potential,
\begin{align}
    &G(0,P_2,P_3,\infty)\approx 
    \langle \mathcal{O}_L(P_2) \mathcal{O}_L(P_3)\rangle 
    + \int_{\text{AdS}} dX \ \Pi^\partial_{\Delta_L}(P_2,X) V(X)\Pi^\partial_{\Delta_L}(X,P_3) \nonumber \\
    &
    + \int_{\text{AdS}} dX_1 dX_2 \ \Pi^\partial_{\Delta_L}(P_2,X_1)  V(X_1)
    \Pi^{\text{B}}_{\Delta_L}(X_1,X_2)
    V(X_2)\Pi^\partial_{\Delta_L}(X_2,P_3) +\cdots
    \label{Born series def}
    \end{align}
This is the Born series. It can be formally resummed into the Lippmann--Schwinger equation, 
whose solution solves the equation of motion of the light field. This has a natural implication: 
instead of computing these integrals, we can 
just solve the equation of motion, analyze the near boundary behavior, 
and compute directly the Green's function, from the wave function\footnote{
We will show this in section~\ref{Wave equation}.}.

\paragraph{Explicit integration over the heavy geodesic.} We now evaluate the second Born contribution in the
cylinder frame used in section~\ref{Heavy limit of Witten diagrams}, with the light
insertions at times $\tau$ and $0$. We denote this
contribution by $W_{\mathrm{cyl}}(\tau,\theta)$.
The remaining embedding-space invariant we need is
\begin{equation}
\label{eq:X1X2}
    X_1\cdot X_2 =  r_1 r_2\, \hat n_1\cdot \hat n_2 -
    \sqrt{(1+r_1^2)(1+r_2^2)} \cosh(t_1 - t_2)    \xrightarrow{\,r\to 0\,} 
    -\cosh(t_1-t_2).
\end{equation}
Substituting the potential
\eqref{eq:contact potential with no derivatives} into
\eqref{eq:Born2} localizes both bulk integrations onto the heavy
geodesic. Using Eq.~\eqref{eq:X1X2} and the light bulk-to-boundary
propagators in Eq.~\eqref{eq:LLp}, we obtain
\begin{equation}
W_{\mathrm{cyl}}(\tau,\theta)
\approx
\frac{\lambda^2\mathcal C_{\Delta_L}}{4\Delta_H^2}
\int_{\mathbb R^2}dt_1\,dt_2\,
\frac{
\Pi_{\Delta_L}^{B}\bigl(X(t_1),X(t_2)\bigr)
}{
[2\cosh(t_1-\tau)]^{\Delta_L}
[2\cosh t_2]^{\Delta_L}
}.
\end{equation}
By Eq.~\eqref{eq:X1X2}, the bulk-to-bulk propagator $\Pi_{\Delta_L}^{B}\bigl(X(t_1),X(t_2)\bigr)$ depends only
on $t_1-t_2$, so this integral is a convolution.
Its Fourier transform is therefore the product of the
three propagator transforms. The bulk-to-boundary
transforms were computed in Eq.~\eqref{eq:convprop}. It remains
to evaluate the bulk-to-bulk transform along the geodesic:
\begin{equation}
    \frac{\tilde{\Pi}_{\Delta}^{\text{B}}(\omega)}{\mathcal C_\Delta} =
    \int_{-\infty}^{\infty}dt \ \frac{e^{-i \omega t}}{(2 \cosh{t})^\Delta} 
    \ _2 F_1 \left(
        \frac{\Delta}{2}, 
        \frac{\Delta+1}{2};
        1+\Delta-\frac{d}{2};
        \frac{1}{(\cosh t)^2}
    \right).
\end{equation}
We first evaluate this integral for $\operatorname{Re}d<2$,
where the singularity at $t=0$ is integrable,
and subsequently analytically continue in $d$.
Expanding the hypergeometric function,
\begin{equation}
    \ _2 F_1 \left( a,b;c;z\right) = 
        \sum_{n=0}^{\infty} \frac{
        (a)_n 
        (b)_n
    }{(c)_n}\frac{z^n}{n!},
\end{equation}
and integrate each term in the series
\begin{equation}
\int_{-\infty}^{\infty} dt\, 
\frac{e^{-i\omega t}}{(\cosh t)^{\Delta + 2 n}}
=
\frac{
\Gamma\!\left(\frac{\Delta+i\omega }{2} + n\right)
\Gamma\!\left(\frac{\Delta-i\omega}{2} + n\right)
}{
\Gamma(\Delta+2n)
}
\frac{1}{2^{1-\Delta-2 n } }.
\end{equation}
We can now separate the $n$-dependent terms using two properties of the Gamma function, namely
the definition of the Pochhammer symbol
\begin{equation}
\Gamma\!\left(\frac{\Delta\pm i\omega}{2}+n\right)
=
\Gamma\!
\left(\frac{\Delta\pm i\omega}{2}\right)
\left(\frac{\Delta\pm i\omega}{2}\right)_n,
\end{equation}
and the Legendre duplication formula
\begin{equation}
\Gamma(\Delta+2n) =
\frac{2^{\Delta+2n-1}}{\sqrt{\pi}}
\Gamma\!\left(\frac{\Delta}{2}\right)
\Gamma\!\left(\frac{\Delta+1}{2}\right)
\left(\frac{\Delta}{2}\right)_n
\left(\frac{\Delta+1}{2}\right)_n,
\end{equation}
to write
\begin{equation}
    \frac{\tilde{\Pi}_{\Delta}^{\text{B}}(\omega)}{\mathcal C_\Delta} =
    \frac{\pi^{1/2}}{2^\Delta}
        \frac{
    \Gamma\left(\frac{\Delta+i\omega}{2}\right)
    \Gamma \left(\frac{\Delta-i\omega}{2} \right)
}{
\Gamma\!\left(\frac{\Delta}{2}\right)
\Gamma\!\left(\frac{\Delta+1}{2}\right)
}
\sum_{n=0}^{\infty} \frac{
\left(\frac{\Delta+ i\omega}{2}\right)_n
\left(\frac{\Delta - i\omega}{2}\right)_n
}{
    \left(1+ \Delta- \frac{d}{2}\right)_n
}
\frac{1}{n!}.
\end{equation}
The series sums to the hypergeometric
${}_2F_1(a,b;c;1)$, with
$a=(\Delta+i\omega)/2$, $b=(\Delta-i\omega)/2$ and
$c=1+\Delta-d/2$.
Gauss's summation formula applies for
$\operatorname{Re}(c-a-b)=1-\operatorname{Re}d/2>0$
and gives
\begin{equation}
\sum_{n=0}^{\infty} \frac{
\left(\frac{\Delta+ i\omega}{2}\right)_n
\left(\frac{\Delta - i\omega}{2}\right)_n
}{
\left(1+ \Delta- \frac{d}{2}\right)_n
}
\frac{1}{n!}= 
\frac{
    \Gamma\left( \Delta+1-\frac{d}{2}\right)
    \Gamma\left(1-\frac{d}{2} \right)
}{
    \Gamma\left(1-\frac{d-\Delta+i\omega}{2}\right)
    \Gamma\left(1-\frac{d-\Delta-i\omega}{2}\right)
},
\label{eq:269}
\end{equation}
leading to the final result
\begin{align}
\tilde{\Pi}_{\Delta}^{\text{B}}(\omega) &= 
\frac{\mathcal C_\Delta }{2^\Delta}
\frac{\pi^{1/2} 
    \Gamma\left( \Delta+1-\frac{d}{2}\right)
    \Gamma\left(1-\frac{d}{2} \right)
}{
\Gamma\!\left(\frac{\Delta}{2}\right)
\Gamma\!\left(\frac{\Delta+1}{2}\right)
}
\frac{
    \Gamma\left(\frac{\Delta+i\omega}{2}\right)
    \Gamma \left(\frac{\Delta-i\omega}{2} \right)
}{
    \Gamma\left(1-\frac{d-\Delta+i\omega}{2}\right)
    \Gamma\left(1-\frac{d-\Delta-i\omega}{2}\right)
} \nonumber\\
&=\frac{\Gamma\!\left(1-\frac{d}{2}\right)}{4\pi^{d/2}}
\frac{
\Gamma\!\left(\frac{\Delta+i\omega}{2}\right)
\Gamma\!\left(\frac{\Delta-i\omega}{2}\right)
}{
\Gamma\!\left(1+\frac{\Delta-d+i\omega}{2}\right)
\Gamma\!\left(1+\frac{\Delta-d-i\omega}{2}\right)
}.
\end{align}
Combining the above with the bulk-to-boundary Fourier
transform used in  Eq.~\eqref{FT_contact}, the second Born
contribution becomes
\begin{equation}
\widetilde W_{\mathrm{cyl}}(\omega)
\approx
\frac{\lambda^2\mathcal C_{\Delta_L}}{16\Delta_H^2}
B\!\left(
\frac{\Delta_L+i\omega}{2},
\frac{\Delta_L-i\omega}{2}
\right)^2
\widetilde\Pi_{\Delta_L}^{B}(\omega).
\label{eq:one-loop-fourier}
\end{equation}
After subtracting local UV divergences where necessary,
the bulk-to-bulk transform has simple poles at
$\omega=\pm i(\Delta_L+2n)$, $n\in\mathbb Z_{\geq0}$.
Together with the double poles of the squared beta
function, these produce triple poles in
 Eq.~\eqref{eq:one-loop-fourier}. For $\tau>0$, their inverse
Fourier transform contains terms proportional to
$\tau^2e^{-(\Delta_L+2n)\tau}$, as expected from expanding
the shifted heavy--light energies to second order.
The required worldline counterterms and the running
associated with logarithmic divergences are discussed
in section~\ref{sec:renorm}.

\subsection{All-loop result and resummation}
\label{all-loop}
The one-loop analysis of the previous subsection already contains the essential
ingredient required for the all-order result. Each contraction fixes a particular
ordering of the interaction vertices along the heavy worldline, while the sum over
the possible orderings reconstructs the unrestricted integration domain appearing
in the Born series. This structure persists at arbitrary loop order.

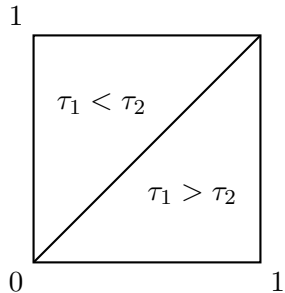
\begin{figure}
    \centering
    \begin{tikzpicture}[scale=3]
    \draw[thick] (0,0) rectangle (1,1);

    \draw[thick] (0,0) -- (1,1);

    \node[below left]  at (0,0) {$0$};
    \node[below right] at (1,0) {$1$};
    \node[above left]  at (0,1) {$1$};
    
    \node at (0.7,0.3) {$\tau_1>\tau_2$};
    \node at (0.3,0.7) {$\tau_1<\tau_2$};
\end{tikzpicture}
\caption{Triangulation of a square. This reproduces the region of integration of each Witten diagram
    at one loop.}
    \label{triangulation of a square}
\end{figure}

To see this, consider a rescaling of the global time to the unit 
interval\footnote{This is not strictly mandatory but it does help with visualization.}, 
and let $Q^n$ denote the $n$-dimensional cube.
The domain of integration of each 1-loop Witten diagram is then mapped to a region in $Q^2$
given by the triangulation of the square in Figure~\ref{triangulation of a square}. At $p$ loops, the class of diagrams considered here
contains $p+1$ interaction vertices connected along
a heavy line and a light line. Fixing their order along
the heavy line leaves $(p+1)!$ possible orders along
the light line.
This will impose an ordering of positions, $X_i$,
(for which there are $(p+1)!$) that
will be mapped to an ordering of parameters, $\tau_i$, in the contour integral, after taking the heavy limit.
This will again define a region, $\Delta_{\sigma}$, in $Q^{p+1}$, for a given ordering
$\sigma \in S_{p+1}$, called an \emph{ordered simplex},
\begin{equation}
    \Delta_{\sigma} =  \{
        (\tau_1,\cdots,\tau_{p+1})\in  Q^{p+1}\ | \  \tau_{\sigma(1)}>
       \tau_{\sigma(2)}>
       \cdots>\tau_{\sigma(p+1)}
    \}.
\end{equation}
The collection $\{\Delta_\sigma, \sigma \in S_{p+1}\}$ gives a triangulation of the unit cube $Q^{p+1}$.
After relabeling the vertices in their order along
the light line, the diagrams have the same integrand
and complementary ordered integration domains.
Summing over these domains gives one unrestricted
worldline integral, with no additional factorial.
The $p$-loop contribution therefore equals the Born
term with $p+1$ potential insertions and $p$ internal
light propagators. Since the worldline integrals are
convolutions, their Fourier transform is
\begin{equation}
\widetilde G_{\mathrm{cyl,conn}}^{(p\text{-loop})}(\omega,\theta)
\simeq
\left(-\frac{\lambda}{2\Delta_H}\right)^{p+1}
\left[\widetilde\Pi_{\Delta_L}^{\partial}(\omega)\right]^2
\left[\widetilde\Pi_{\Delta_L}^{B}(\omega)\right]^p,
\qquad p\geq0.
\label{eq:p-loop-born}
\end{equation}
Here $p=0$ denotes the tree contribution.
The cases $p=0$ and $p=1$ reproduce
 Eq.~\eqref{FT_contact} and  Eq.~\eqref{eq:one-loop-fourier},
respectively. Summing over $p$ gives the formal
geometric series
\begin{equation}
    \tilde G_\text{cyl,conn}(\omega,\theta) = -\frac{
        \lambda \frac{
              \tilde{\Pi}^\partial_{\Delta_L} (\omega) 
              \tilde{\Pi}^\partial_{\Delta_L} (\omega) 
        }{
            2\Delta_H
    }
    }{
        1+ \lambda \frac{\tilde{\Pi}^{\text{B}}_{\Delta_L} (\omega)}{2\Delta_H}
    }.
    \label{no derivative geometric series}
\end{equation}
This expression sums the connected contact
contributions and is independent of $\theta$. The full cylinder correlator
also includes the free light two-point function,
as in  Eq.~\eqref{Born series def}. 

This diagrammatic resummation yields the exact same non-perturbative structure as solving an effective AdS wave
equation with localized contact potentials, establishing,
for the contact interaction, Eq.~\eqref{eq:bulk-action}, 
the equivalence between the Witten-diagram expansion in the heavy limit and semiclassical bulk wave scattering, as we show in the following section.

\section{Worldline effective field theory and effective wave equation}\label{Wave equation}

In this section we consider the worldline effective field theory (WEFT) for a heavy particle localized in the bulk and probed by the light field \cite{Goldberger:2004jt,Porto:2016pyg}. We first introduce the effective action and derive the associated wave equation in AdS where the Wilson coefficients of the EFT enter manifestly as boundary conditions, following the formalism of \cite{Caron-Huot:2025tlq}. We reproduce the Witten-diagram resummation of section~\ref{The heavy--light correlator} and generalize it to arbitrary contact interactions with derivatives and discuss renormalization. In this section we work in Lorentzian signature.

\subsection{Worldline effective action and effective wave equation in AdS}

In the regime where long-distance gravity can be neglected we simply have the short-distance contribution to the EFT. The schematic relation in  Eq.~\eqref{eq:SEFT} takes the specific form, in Lorentzian signature, 
\begin{equation}
    S_\text{WEFT} = -M \int d\tau + \sum_{\ell,n} \frac{\bar \mu^{3-d}}{\ell !} \int d \tau \big[ C_{\ell,n} (D_L \phi_+) \partial_\tau^n (D^L \phi_-) \big] ,
\end{equation}
where we employ the Schwinger--Keldysh, or in-in, formalism in the
Keldysh basis
\cite{Schwinger:1960qe,Keldysh:1964ud,Galley:2012hx}.
Here, $\phi_+ = \frac{1}{2}(\phi_1 + \phi_2 )$ is the average, or classical, field, while $\phi_- = \phi_1 - \phi_2$ is the
difference, or response, field. The physical equation of motion for
$\phi_+$ is obtained by varying the in-in effective action with respect to
$\phi_-$ and subsequently taking the physical limit $\phi_-\to0$. The
doubling of the fields allows causal and non-conservative effects to be
incorporated at the level of the effective action. In worldline EFT, this
formalism provides a systematic description of absorptive tidal response,
including tidal heating and torquing of compact objects
\cite{Goldberger:2005cd,
Goldberger:2020fot,Ivanov:2024sds}.

The mass $M$ represents the mass of the heavy object and $C_{\ell,n}$ the Wilson coefficients of the effective action. Here $\bar\mu$ is the scale introduced by dimensional regularization, needed to keep the dimensions of the Wilson coefficients fixed: $[C_{\ell,n}] = -2\ell - n - 1$.\footnote{The fact that the dimension depends on $\ell$ leads to exponential behavior in spin of these short-distance effects and associated CFT data, as we show in section~\ref{CFT}.} Moreover, the heavy object is placed in the center of AdS with pure timelike velocity $u^\mu = (1,\vec{0})$ where the spatial derivative is given by $D^\mu = (g^{\mu \nu} + u^\mu u^\nu) \nabla_\nu$. The notation $D_L$ denotes the symmetrized trace-free (STF) product of $\ell$ spatial derivatives with the multi-index notation $L = \mu_1, \mu_2, \dots,\mu_\ell$.\footnote{The first cases are given by
\begin{equation}
 \ell = 1:\;    D_{L} = D_{i}\,,  \qquad \ell=2: \;D_{L} = D_{(i} D_{j)} - \frac{1}{d } g_{i j} D_a D^a.
\end{equation}}

Away from the worldline of the heavy field the light probe $\phi$ propagates inside AdS with bulk action
\begin{equation}
    S_\text{AdS} =  - \int d^{d+1}x \sqrt{-g}\,\,\big( g_{\mu \nu} \,\partial^\mu \phi_+ \partial^\nu \phi_- +m^2 \phi_+ \phi_-\big),
\end{equation}
where $m$ is the mass of the light field. Including the gravitational interaction between the heavy object and the probe would amount to replacing the AdS metric by Schwarzschild--AdS (see e.g. \cite{Dodelson:2022yvn}), understood in the long-wavelength EFT as an expansion in powers of Newton's constant $G_N$ in dimensional regularization \cite{Ivanov:2024sds, Caron-Huot:2025tlq}.

The dynamics of the light probe are described by the equations of motion
\begin{equation}
\label{eq:dS}
    \frac{ \delta}{\delta \phi_-}(S_\text{AdS} + S_\text{WEFT}) = 0 \;\; \implies   \;\;  (\Box_{\text{AdS}} - m^2 + V_\text{WEFT})\, \phi_+ = 0,
\end{equation}
where $\Box_{\text{AdS}} \phi \equiv \frac{1}{\sqrt{-g}} \partial_\mu(\sqrt{-g} \,g^{\mu \nu} \partial_\nu \,\phi )$ is the d'Alembert operator for propagation in AdS and the potential coming from the WEFT action reads
\begin{equation}
\label{eq:VWEFT}
    V_\text{WEFT}(x) = - \bar \mu^{3-d} \sum_{\ell,n} C_{\ell,n} \frac{(-1)^{\ell +n}}{\ell!} D^L \delta^{(d)}(\mathbf x) D_L\, \partial^n_\tau,
\end{equation}
where the delta-function is localized on the worldline of the heavy object placed at the center of global coordinates, $r= 0$. 
As we will see shortly the short-range potential $V_\text{WEFT}$ amounts to a boundary condition at $r = 0$ for a free field propagating in AdS. We start by reviewing the free propagating field in AdS.

\subsection{Free solutions in AdS}
 Away from the worldline $r \neq 0$ the light field $\phi$ propagates freely in AdS obeying the wave equation
\begin{equation}
\label{eq:homog}
    (\Box_{\text{AdS}} - m^2) \phi = 0,
\end{equation}
which can be written in global AdS coordinates. As reviewed e.g. in \cite{Nastase:2007kj}, the scalar field admits the partial wave decomposition
\begin{equation}
\label{eq:phix}
    \phi(x) = \int_{-\infty}^{\infty}\frac{d\omega}{2\pi} \sum_{\ell,\vec{m}}
    e^{-i \omega t}
    \psi_{\omega,\ell}(r)
    Y_{\ell,\vec{m}}(\Omega) ,
\end{equation}
with angular eigenvalues $-\ell(\ell+d-2)$. The wave equation reduces to the radial form,
\begin{equation}
    \left[\frac{1}{r^{d-1}}\derivative{}{r}\left(r^{d-1}(1+r^2)\derivative{}{r}\right) + \frac{\omega^2}{1+r^2} - 
    \frac{\ell(\ell+d-2)}{r^2} - \Delta_L(\Delta_L-d)\right]\psi_{\omega,\ell}(r) = 0,
    \label{radial kg equation}
\end{equation}
where we replaced $m^2 = \Delta_L(\Delta_L - d)$. The radial solution is a linear combination of hypergeometric functions,
\begin{equation}
    \psi_{\omega,\ell} (r) = A_{\text{rsp}}(\omega,\ell) \psi_+ (r) + A_{\text{src}}(\omega,\ell) \psi_- (r),
    \label{eq:bdybasis}
\end{equation}
where,
\begin{equation}
    \psi_\pm(r) =
    \frac{ r^\ell}{ \left(1+r^2\right)^{\frac{ (\Delta_\pm +\ell)}{2}}} \, _2F_1\left(\frac{ (\ell+\Delta_\pm -\omega )}{2},
    \frac{ (\ell+\Delta_\pm +\omega)}{2};1-\frac{d}{2}+\Delta_\pm ;\frac{1}{1+r^2}\right),
\end{equation}
and $\Delta_+ = \Delta_L$ and $\Delta_- = d-\Delta_L$.

The asymptotic behavior near the boundary, $r\rightarrow \infty$, is
\begin{equation}
    \psi_{\omega,\ell}(r) \xrightarrow[r\rightarrow\infty]{}
    A_{\text{src}}(\omega,\ell )r^{\Delta_L-d}
    + A_{\text{rsp}}(\omega,\ell) r^{-\Delta_L}.
\end{equation}
Similarly, near the worldline the radial solution has the leading power behavior
\begin{equation}
\label{eq:rto0}
    \psi_{\omega,\ell}(r)
    \underset{r\to 0}{\longrightarrow}
    B_{\mathrm{reg}}(\omega,\ell)\,r^\ell
    +
    B_{\mathrm{irr}}(\omega,\ell)\,r^{2-d-\ell},
\end{equation}
which follows from analyzing the wave equation,  Eq.~\eqref{radial kg equation}, at small $r$, where the AdS metric is locally flat. An exact basis of solutions adapted to this behavior is
\begin{equation}
\label{eq:psiregirr}
    \psi_{\omega,\ell}(r)
    =
    B_{\mathrm{reg}}(\omega,\ell)\,\Psi_{\mathrm{reg}}(r)
    +
    B_{\mathrm{irr}}(\omega,\ell)\,\Psi_{\mathrm{irr}}(r),
\end{equation}
where the regular solution reads
\begin{equation}
\label{eq:reg0}
      \Psi_{\mathrm{reg}}(r)
    =
    \frac{r^\ell}{(1+r^2)^{\frac{\Delta_L+\ell}{2}}}
    {}_2F_1\left(
        \frac{\ell+\Delta_L-\omega}{2},
        \frac{\ell+\Delta_L+\omega}{2};
        \frac{d}{2}+\ell;
        \frac{r^2}{1+r^2}
    \right),
\end{equation}
and the irregular solution is given by\footnote{For even integer $d$, the expression for
$\Psi_{\mathrm{irr}}$ is understood by analytic continuation in $d$ and may
contain a subleading term proportional to $r^\ell\log r$. In the degenerate
case $d=2$, $\ell=0$, the irregular solution is logarithmic at the
origin.}
\begin{equation}
    \Psi_{\mathrm{irr}}(r) = \Psi_{\mathrm{reg}}(r)\big|_{\ell \,\to\, 2-d-\ell}.
\end{equation}
These solutions are normalized such that
\begin{equation}
\label{eq:regirr0}
    \Psi_{\mathrm{reg}}(r)
    \underset{r\to0}{\longrightarrow}r^\ell,
    \qquad
    \Psi_{\mathrm{irr}}(r)
    \underset{r\to0}{\longrightarrow}r^{2-d-\ell}.
\end{equation}
The two bases $\{\psi_+,\psi_-\}$ and $\{\Psi_{\rm reg},\Psi_{\rm irr}\}$ span the same two-dimensional space of solutions of  Eq.~\eqref{radial kg equation}: the first is adapted to the conformal boundary, where it separates source from response, the second to the worldline, where it separates regular from singular behavior. In terms of the coefficients introduced in  Eq.~\eqref{eq:bdybasis} and  Eq.~\eqref{eq:psiregirr}, the boundary data are linear in the worldline data,
\begin{equation}
    \begin{pmatrix}
        A_{\mathrm{rsp}}(\omega,\ell)
        \\
        A_{\mathrm{src}}(\omega,\ell)
    \end{pmatrix}
    =
    \begin{pmatrix}
        \mathcal C_{\mathrm{rsp}}^{\mathrm{reg}}
        &
        \mathcal C_{\mathrm{rsp}}^{\mathrm{irr}}
        \\
        \mathcal C_{\mathrm{src}}^{\mathrm{reg}}
        &
        \mathcal C_{\mathrm{src}}^{\mathrm{irr}}
    \end{pmatrix}
    \begin{pmatrix}
        B_{\mathrm{reg}}(\omega,\ell)
        \\
        B_{\mathrm{irr}}(\omega,\ell)
    \end{pmatrix},
\end{equation}
where
\begin{equation}
\begin{aligned}
\mathcal C_{\mathrm{rsp}}^{\mathrm{reg}}
&=
\frac{
    \Gamma\left(\ell+\frac{d}{2}\right)
    \Gamma\left(\frac{d}{2}-\Delta_L\right)
}{
    \Gamma\left(\frac{\ell+d-\Delta_L+\omega}{2}\right)
    \Gamma\left(\frac{\ell+d-\Delta_L-\omega}{2}\right)
},
\qquad
&
\mathcal C_{\mathrm{rsp}}^{\mathrm{irr}}
&=
\left.
\mathcal C_{\mathrm{rsp}}^{\mathrm{reg}}
\right|_{\ell\to 2-d-\ell},
\\[2mm]
\mathcal C_{\mathrm{src}}^{\mathrm{reg}}
&=
\frac{
    \Gamma\left(\ell+\frac{d}{2}\right)
    \Gamma\left(\Delta_L-\frac{d}{2}\right)
}{
    \Gamma\left(\frac{\ell+\Delta_L-\omega}{2}\right)
    \Gamma\left(\frac{\ell+\Delta_L+\omega}{2}\right)
},
\qquad
&
\mathcal C_{\mathrm{src}}^{\mathrm{irr}}
&=
\left.
\mathcal C_{\mathrm{src}}^{\mathrm{reg}}
\right|_{\ell\to 2-d-\ell}.
\end{aligned}
\label{eq:connection}
\end{equation}
\paragraph{Regular normalizable solution.} 
For a solution regular at the worldline, $B_{\mathrm{irr}}(\omega,\ell)=0$,
so that
\begin{equation}
    \psi_{\omega,\ell}(r)
    =
    B_{\mathrm{reg}}(\omega,\ell)\,
    \Psi_{\mathrm{reg}}(r).
\end{equation}
Its non-normalizable coefficient at the conformal boundary is therefore
\begin{equation}
    A_{\mathrm{src}}(\omega,\ell)
    =
    B_{\mathrm{reg}}(\omega,\ell)\,
    \mathcal C_{\mathrm{src}}^{\mathrm{reg}}(\omega,\ell),
\end{equation}
with $\mathcal C_{\mathrm{src}}^{\mathrm{reg}}$ given in  Eq.~\eqref{eq:connection}.
A non-trivial solution is normalizable only when
$\mathcal C_{\mathrm{src}}^{\mathrm{reg}}(\omega,\ell)=0$.
This quantizes the frequency according to
\begin{equation}
    \omega_{n,\ell}
    =
    \pm\left(\Delta_L+\ell+2n\right),
    \qquad
    n\in\mathbb N^0.
    \label{eq:AdSfreq}
\end{equation}
For positive-frequency modes, we take
$\omega_{n,\ell}=\Delta_L+\ell+2n$.

The overall coefficient $B_{\mathrm{reg}}$ is not fixed by regularity or
normalizability and is determined by imposing unit normalization under the
Klein--Gordon inner product,
\begin{equation}
    (\psi_1,\psi_2)
    \equiv
    i\int_{\Sigma} d^d x\,\sqrt{-g}\,g^{tt}
    \left[
        (\partial_t\psi_1^*)\psi_2
        -
        \psi_1^*\partial_t\psi_2
    \right],
\end{equation}
where $\Sigma$ is a hypersurface of fixed global time. The normalized radial
wavefunction for which $(\psi_{n, \ell} , \psi_{n,\ell}) = 1$ is then
\begin{equation}
    \psi_{n,\ell}(r)
    =
    \left.
    \frac{1}{N_{n,\ell}}\,
    \Psi_{\mathrm{reg}}(r)
    \right|_{\omega=\omega_{n,\ell}},
\end{equation}
corresponding to
\begin{equation}
    B_{\mathrm{reg}}(\omega_{n,\ell},\ell)
    =
    \frac{1}{N_{n,\ell}},
    \qquad
    B_{\mathrm{irr}}(\omega_{n,\ell},\ell)=0,
\end{equation}
with
\begin{equation}
    N_{n,\ell}
    \equiv
    (-1)^n
    \sqrt{
        \frac{
            n!\,
            \Gamma^2\left(\ell+\frac{d}{2}\right)
            \Gamma\left(\Delta_L+n-\frac{d-2}{2}\right)
        }{
            \Gamma\left(n+\ell+\frac{d}{2}\right)
            \Gamma\left(\Delta_L+n+\ell\right)
        }
    }.
    \label{eq:Nnl}
\end{equation}

\subsection{Worldline interaction and irregular solution}

We now determine how the worldline interaction fixes the relative coefficient
of the regular and irregular solutions at the origin. Suppressing the $+$
label in  Eq.~\eqref{eq:dS} for notational simplicity, we begin
with the full wave equation
\begin{equation}
    \left(
        \Box_{\mathrm{AdS}}
        -m^2
        +V_{\mathrm{WEFT}}
    \right)\phi
    =
    0,
    \label{eq:worldline-full-wave-equation}
\end{equation}
where $V_{\mathrm{WEFT}}$ is given in  Eq.~\eqref{eq:VWEFT}. Consider a field of definite frequency and angular momentum,
\begin{equation}
    \phi_\omega(t,r,\Omega)
    =
    e^{-i\omega t}
    \sum_{\ell,\vec m}
        \psi_{\omega,\ell}(r)
    Y_{\ell,\vec m}(\Omega),
    \label{eq:frequency-space-field}
\end{equation}
which is related to the full field in  Eq.~\eqref{eq:phix} by
\begin{equation}
    \phi(x)
    =
    \int_{-\infty}^{\infty}
    \frac{d\omega}{2\pi}\,
    \phi_\omega(t,r,\Omega).
\end{equation}
Since the heavy object is static at the center of AdS, its proper time agrees
with global time, $\tau=t$. Therefore,
\begin{equation}
    (-1)^n\partial_\tau^n e^{-i\omega t}
    =
    (i\omega)^n e^{-i\omega t}.
\end{equation}
It is convenient to introduce the frequency-dependent combination of
worldline Wilson coefficients
\begin{equation}
    F_\ell(\omega)
    \equiv
    \sum_{n=0}^{\infty}
    (i\omega)^n C_{\ell,n}.
    \label{eq:worldline-response-function}
\end{equation}
Writing the STF derivative introduced above with the multi-index
$L=i_1\cdots i_\ell$, the worldline potential acting on a
frequency eigenmode becomes
\begin{equation}
    V_{\mathrm{WEFT}}\,\phi_\omega
    =
    -\bar \mu^{3-d}
    \sum_{\ell=0}^{\infty}
    F_\ell(\omega)
    \frac{(-1)^\ell}{\ell!}
 D_L\delta^{(d)}(\mathbf x)
    \left[
      D^L\phi_\omega
    \right]_{\mathbf x=0},
    \label{eq:frequency-space-worldline-potential}
\end{equation}
which amounts to a localized source at $\mathbf{x} = 0$. Following  Eq.~\eqref{eq:psiregirr} we can split $\phi_\omega$ according to regular and irregular solutions
\begin{equation}
    \phi_\omega(t,\mathbf{x}) 
    =
 \phi^\text{reg}_\omega(t,\mathbf{x})  +  \phi^\text{irr}_\omega(t,\mathbf{x}),
 \label{eq:split}
\end{equation}
with
\begin{equation}
    \begin{aligned}
     &\phi^\text{reg}_\omega(t,\mathbf{x}) \equiv
     e^{- i \omega t} \sum_{\ell,\vec{m}} B_\text{reg}(\omega,\ell) \Psi_\text{reg}(r)
     Y_{\ell,m}(\Omega) ,
     \\
     &\phi^\text{irr}_\omega(t,\mathbf{x}) \equiv
     e^{- i \omega t} \sum_{\ell,\vec{m}} B_\text{irr}(\omega,\ell) \Psi_\text{irr}(r)
     Y_{\ell,m}(\Omega) .
    \end{aligned}
\end{equation}

\paragraph{Contribution of the regular solution to the worldline interaction.}  We first evaluate the worldline potential on the regular part of the field.
Near the origin, the regular component can be expanded as
\begin{equation}
    \phi_\omega^{\mathrm{reg}}(t,\mathbf x)
    =
    e^{-i\omega t}
    \sum_{\ell',\vec m}
    B_{\mathrm{reg}}(\omega,\ell')
    \,r^{\ell'}\,Y_{\ell',\vec m}(\Omega)
    \left[1+\mathcal O(r^2)\right].
    \label{eq:regular-field-expansion}
\end{equation}
Introducing the STF tensors $\mathcal Y_{\ell,\vec m}^{L}$ through
\begin{equation}
    r^\ell\, Y_{\ell,\vec m}(\Omega)
    =
    \mathcal Y_{\ell,\vec m}^{L}\,x_L,
    \qquad
    x_L=x_{i_1}\cdots x_{i_\ell}\,,
    \label{eq:STF-solid-harmonic}
\end{equation}
the rank-$\ell$ STF derivative at the origin obeys
\begin{equation}
    \left.
 D^L
    \left[
        r^{\ell'}Y_{\ell',\vec m}(\Omega)
    \right]
    \right|_{\mathbf x=0}
    =
    \ell!\,
    \delta_{\ell\ell'}\,
    \mathcal Y_{\ell,\vec m}^L.
    \label{eq:STF-angular-momentum-projection}
\end{equation}
Note that the higher-order terms in the regular expansion do not modify this result.
Terms of degree greater than $\ell$ vanish after evaluation at the origin,
while terms with the same total degree but lower angular momentum contain
traces and are annihilated by the STF projection. Consequently,
\begin{equation}
    \left.
    D^L
    \phi_\omega^{\mathrm{reg}}
    \right|_{\mathbf x=0}
    =
    e^{-i\omega t}\,
    \ell!
    \sum_{\vec m}
    B_{\mathrm{reg}}(\omega,\ell)
\,    \mathcal Y_{\ell,\vec m}^L.
    \label{eq:regular-worldline-multipole}
\end{equation}
Thus, the term proportional to $F_\ell(\omega)$ in the potential,  Eq.~\eqref{eq:frequency-space-worldline-potential}, projects onto the partial
wave with angular momentum $\ell$. As we show below, this source generates an irregular term with the same angular momentum. Rotational invariance therefore makes the worldline interaction diagonal in both $\ell$ and $\vec m$. The contribution of the regular solution to
the potential is
\begin{equation}
    V_{\mathrm{WEFT}}
    \phi_\omega^{\mathrm{reg}}
    =
    -\bar \mu^{3-d} \sum_{\ell,\vec{m}}
    (-1)^\ell
    F_\ell(\omega)
    B_{\mathrm{reg}}(\omega,\ell)
    \mathcal Y_{\ell,\vec m}^{L}
D_L\delta^{(d)}(\mathbf x).
    \label{eq:regular-worldline-source}
\end{equation}

\paragraph{Contribution of the irregular solution to the worldline interaction.}

The potential formally also acts on the irregular part of the field given in  Eq.~\eqref{eq:split}. The irregular solution has the expansion
\begin{equation}
    \phi_\omega^{\mathrm{irr}}(t,\mathbf{x})
    =
    e^{-i\omega t}
    \sum_{\ell, \vec{m}}B_{\mathrm{irr}}(\omega,\ell)
    Y_{\ell,\vec m}(\Omega)
    \,r^{2-d-\ell}
    \sum_{k=0}^{\infty}
    a_k(\omega,\ell)\,r^{2k},
    \qquad
    a_0=1.
    \label{eq:irregular-Frobenius-series}
\end{equation}
After applying $\ell$ spatial derivatives, its terms scale as
\begin{equation}
   D_L\phi_\omega^{\mathrm{irr}}
    \sim
    r^{2-d-2\ell+2k},
    \qquad
    k\in\mathbb N_0.
    \label{eq:irregular-derivative-scaling}
\end{equation}
For generic complex $d$, none of these powers corresponds to a constant
Taylor coefficient at the origin. We define the worldline value by
analytic continuation in $d$ for which these terms vanish as
$r\to0$. In dimensional regularization, the purely local irregular
self-field therefore gives
\begin{equation}
    \left[
        D_L
        \phi_\omega^{\mathrm{irr}}
    \right]_{\mathbf x=0}^{\mathrm{DR}}
    =
    0,
    \label{eq:irregular-field-vanishes-in-DR}
\end{equation}
and thus only the regular solution contributes to the worldline interaction. 

However, at special values of $d$ where the regular and irregular expansions 
may overlap, singularities in $d$ can appear. Their local contributions are
removed by renormalizing the corresponding worldline coefficients (counterterms) $C_{\ell,n}$, as we show explicitly later.

\paragraph{AdS wave equation near the worldline.}

To isolate the terms responsible for the localized source, consider a single
partial wave,
\begin{equation}
    \phi_{\omega,\ell,\vec m}(t,r,\Omega)
    =
    e^{-i\omega t}
    \psi_{\omega,\ell}(r)
    Y_{\ell,\vec m}(\Omega).
\end{equation}
In the global coordinates, the free wave operator in  Eq.~\eqref{radial kg equation} takes the form
\begin{align}
    \left(
        \Box_{\mathrm{AdS}}-m^2
    \right)\phi_{\omega,\ell,\vec m}
    =
    e^{-i\omega t}Y_{\ell,\vec m}(\Omega)
    \Bigg[
        &
        \partial_r^2
        +\frac{d-1}{r}\partial_r
        -\frac{\ell(\ell+d-2)}{r^2}
        \nonumber\\
        &+
        r^2\partial_r^2
        +(d+1)r\partial_r
        +\frac{\omega^2}{1+r^2}
        -m^2
    \Bigg]
    \psi_{\omega,\ell}(r).
    \label{eq:local-AdS-operator}
\end{align}
The terms in the first line constitute the flat-space Laplacian in $d$
spatial dimensions projected onto angular momentum $\ell$. The coefficients
of the terms in the second line are regular at $r=0$.

The regular and irregular solutions both obey the homogeneous
free-field equation for $r>0$. The regular solution is smooth at the origin
and thus does not generate any source term. On the other hand, the irregular solution $\phi_\omega^\text{irr}$ also obeys the homogeneous equation away from the
worldline, but its leading term,
$r^{2-d-\ell}Y_{\ell,\vec m}(\Omega)$, produces an $\ell$-th
derivative of a delta function when acted on by the flat-space Laplacian.
The remaining terms in  Eq.~\eqref{eq:local-AdS-operator} determine the subleading terms in the
irregular power series but, for generic $d$, do not produce distributions
supported at the origin. The distributional part of the bulk wave operator is
therefore entirely determined by the flat-space Laplacian acting on the
irregular solution.

\paragraph{Relation between irregular solution and Love numbers.} In summary, we have concluded that the wave equation,  Eq.~\eqref{eq:worldline-full-wave-equation}, in the limit $r \to 0$, upon the split,  Eq.~\eqref{eq:split}, reduces to
\begin{equation}
- \nabla^2 \phi_\omega^\text{irr} = V_\text{WEFT} \,\phi_\omega^\text{reg},
    \label{eq:worldline-waveeq}
\end{equation}
where $\nabla^2$ denotes the flat-space Laplacian in $d$ spatial dimensions. Equating both sides allows us to fix the coefficient of the irregular solution $B_\text{irr}$ in terms of the Love numbers in $F_\ell(\omega)$. From Eq.~\eqref{eq:regular-worldline-source} we have
\begin{equation}
    -\nabla^2\phi_\omega^{\mathrm{irr}}
    =
    -\bar \mu^{3-d}
    e^{-i\omega t}
       \sum_{\ell,\vec m}(-1)^\ell
   \, F_\ell(\omega)
   \, B_{\mathrm{reg}}(\omega,\ell)
  \,  \mathcal Y_{\ell,\vec m}^{L}
\, D_L\delta^{(d)}(\mathbf x).
    \label{eq:distributional-multipole-equation}
\end{equation}
Introducing the local Green's function satisfying
\begin{equation}
    -\nabla^2G_0(\mathbf x)
    =
    \delta^{(d)}(\mathbf x),
\end{equation}
which is given by
\begin{equation}
    G_0(\mathbf x)
    =
    \frac{1}{(d-2)\Omega_{d-1}}\,
    r^{2-d},
    \qquad
    \Omega_{d-1}
    =
    \frac{2\pi^{d/2}}{\Gamma(d/2)},
    \label{eq:local-flat-Green-function}
\end{equation}
we have that the irregular solution is proportional to $G_0$,
\begin{equation}
    \phi_\omega^{\mathrm{irr}}
    =
    -\bar \mu^{3-d} \sum_{\ell,\vec m}
    (-1)^\ell
    F_\ell(\omega)
    B_{\mathrm{reg}}(\omega,\ell)
    \mathcal Y_{\ell,\vec m}^{L}
   \, D_LG_0(\mathbf x).
    \label{eq:irregular-field-from-Green-function}
\end{equation}
Using the definition of the spherical harmonics in
 Eq.~\eqref{eq:STF-solid-harmonic}, one finds
\begin{equation}
    \mathcal Y_{\ell,\vec m}^{L}
    D_L r^{2-d}
    =
    (-1)^\ell 2^\ell
    \left(\frac d2-1\right)_\ell
    r^{2-d-\ell}Y_{\ell,\vec m}(\Omega),
    \label{eq:STF-derivative-power-law}
\end{equation}
and we get
\begin{equation}
    \phi_\omega^{\mathrm{irr}}
    =
    -\bar \mu^{3-d} \, e^{-i\omega t} \sum_{\ell,\vec m}
  F_\ell(\omega)
    B_{\mathrm{reg}}(\omega,\ell)
    \frac{
        2^\ell
        \left(\frac d2-1\right)_\ell
    }{
        (d-2)\Omega_{d-1}
    }\,
    r^{2-d-\ell}
    Y_{\ell,\vec m}(\Omega).
    \label{eq:irregular-field-final}
\end{equation}
Comparing with the expression for $\phi_\omega^\text{irr}$ in  Eq.~\eqref{eq:irregular-Frobenius-series} we obtain the desired relation
\begin{equation}
   \frac{ B_{\mathrm{irr}}(\omega,\ell)}{  B_{\mathrm{reg}}(\omega,\ell)}
    =
    -\bar \mu^{3-d}
   \, \mathcal{N}_{d,\ell} \, 
    F_\ell(\omega), \qquad \mathcal{N}_{d,\ell} =  \frac{
        2^\ell
        \left(\frac d2-1\right)_\ell
    }{
        (d-2)\Omega_{d-1}
    }.
    \label{eq:Birr-in-terms-of-Fell}
\end{equation}
From Eq.~\eqref{eq:psiregirr} we see that the worldline interaction selects the radial solution
\begin{equation}
    \psi_{\omega,\ell}(r)
    =
    B_{\mathrm{reg}}(\omega,\ell)
    \left[
        \Psi_{\mathrm{reg}}(r)
        -
        \bar \mu^{3-d}
       \mathcal{N}_{d,\ell}\,
        F_\ell(\omega)
        \Psi_{\mathrm{irr}}(r)
    \right],
    \label{eq:worldline-selected-radial-solution}
\end{equation}
where the tidal Love numbers, or  WEFT Wilson coefficients $C_{\ell,n}$, contained in the Taylor expansion of $F_\ell(\omega)$ according to Eq.~\eqref{eq:worldline-response-function} directly source an irregular solution $\Psi_\text{irr}$ on the worldline.

\subsection{Retarded Green's function for weakly gravitating compact objects in AdS}

We now proceed to calculate the retarded Green's function at the boundary $r \to \infty$ by solving the connection problem. 
Conveniently, as we showed in the previous section, the worldline interaction encoded in the Love numbers $C_{\ell,n}$ amounts to 
a simple boundary condition at $r = 0$, and thus we can write the connection problem as
\begin{equation}
\label{eq:ACB}
    \begin{pmatrix}
        A_{\mathrm{rsp}}(\omega,\ell)
        \\
        A_{\mathrm{src}}(\omega,\ell)
    \end{pmatrix}
    =
    \begin{pmatrix}
        \mathcal C_{\mathrm{rsp}}^{\mathrm{reg}}
        &
        \mathcal C_{\mathrm{rsp}}^{\mathrm{irr}}
        \\
        \mathcal C_{\mathrm{src}}^{\mathrm{reg}}
        &
        \mathcal C_{\mathrm{src}}^{\mathrm{irr}}
    \end{pmatrix}
    \begin{pmatrix}
     1
        \\
    -\bar \mu^{3-d}\mathcal{N}_{d,\ell} \,   F_\ell(\omega)
    \end{pmatrix},
\end{equation}
where the AdS connection coefficients are given in Eq.~\eqref{eq:connection} and $F_\ell(\omega)$ is given in terms of $C_{\ell,n}$ in Eq.~\eqref{eq:worldline-response-function}.

The retarded Green's function at the boundary is the ratio of source and response coefficients:
\begin{equation}
    G_R(\omega,\ell) =
\frac{ \pi ^{d/2} \Gamma \left(\Delta_L -\frac{d}{2}\right)}{\Gamma (\Delta_L )}
    \frac{A_{\text{rsp}}(\omega,\ell)}{A_{\text{src}}(\omega,\ell)},
\end{equation}
and where the normalization was chosen such that the two point function has unit norm (see section~\ref{CFT}). From Eq.~\eqref{eq:ACB} we can write explicitly in terms of the Love numbers, 
\begin{equation}
\label{eq:GRCF}
    G_R(\omega,\ell) = \frac{ \pi ^{d/2} \Gamma \left(\Delta_L -\frac{d}{2}\right)}{\Gamma (\Delta_L )}
    \frac{  
        \mathcal C_{\mathrm{rsp}}^{\mathrm{reg}} -
        \bar \mu^{3-d}
        \mathcal{N}_{d,\ell} 
        F_\ell(\omega)
        \mathcal C_{\mathrm{rsp}}^{\mathrm{irr}}   
        }{
        \mathcal C_{\mathrm{src}}^{\mathrm{reg}}
        -
        \bar \mu^{3-d}
        \mathcal{N}_{d,\ell}
        F_\ell(\omega)
        \mathcal C_{\mathrm{src}}^{\mathrm{irr}}   
        }.
\end{equation}
This structure is in fact universal and generalizes beyond the AdS background. If the compact object is strongly gravitating and has a long-range gravitational field, as in the case of a black hole, only the connection coefficients $\mathcal{C}$ change, and they can be computed perturbatively in Newton's constant.

\paragraph{Matching to the Witten diagram resummation.} Another way to write the retarded Green's function, which connects more closely to the bubble-diagram resummation of section~\ref{The heavy--light correlator}, is the following. We separate the free part $G_R^{(0)}(\omega,\ell)$, given by
\begin{equation}
    G^{(0)}_R(\omega,\ell)=
    \frac{
        \pi^{\frac{d}{2}}
        \Gamma(\frac{d}{2}-\Delta_L)
}{
\Gamma(\Delta_L)
}
    \frac{
        \Gamma\left(
            \frac{\Delta_L+\ell +\omega}{2}
        \right)
        \Gamma\left(
            \frac{\Delta_L + \ell-\omega}{2}
        \right)
    }
        {
        \Gamma\left(
            \frac{d+\ell-\Delta_L+\omega}{2}
        \right)
        \Gamma\left(
            \frac{d+\ell-\Delta_L-\omega}{2}
        \right)
    },
\end{equation}
and the interacting Green's function as
\begin{equation}
    G_R(\omega,\ell) = G_R^{(0)}(\omega,\ell) + \frac{V(\omega,\ell)}
    {1-  V(\omega,\ell)(G^{(0)}_{R}(\omega,\ell))^{-1}\xi_{\Delta_L,\ell}(\omega)}.
    \label{Full correlator}
\end{equation}
The Love numbers in $F_\ell(\omega)$ are encoded in the potential term
\begin{equation}
\label{eq:Vwl}
    V(\omega,\ell)
    =
    -\bar \mu^{3-d}
   \, 2^{\ell-2}\,F_\ell(\omega)
    \frac{
        \Gamma^2\left(\frac{\ell+\Delta_L-\omega}{2}\right)
        \Gamma^2\left(\frac{\ell+\Delta_L+\omega}{2}\right)
    }{
        \Gamma(\Delta_L)
        \Gamma\left(\Delta_L-\frac d2+1\right)
        \Gamma\left(\frac d2+\ell\right)
    },
\end{equation}
and the kinematic factor reads
\begin{equation}
    \xi_{\Delta_L,\ell}(\omega)
=
\frac{
\cos(\pi\omega)-\cos\!\left[\pi(d-\Delta_L+\ell)\right]
}{
\cos\!\left[\pi(\Delta_L+\ell)\right]
-
\cos\!\left[\pi(d-\Delta_L+\ell)\right]
}.
\end{equation}

The form Eq.~\eqref{Full correlator} highlights several features of the result. Namely, the free particle Green's function, $G_R^{(0)}(\omega,\ell)$, is written
in such a way that its residues on the physical poles give the mean-field-theory (MFT) OPE coefficients for heavy--light correlators,
after multiplying by the universal $SO(d)$ factor from the addition theorem,
\begin{equation}
    \sum_{\vec{m}} 
    Y_{\ell,\vec{m}}(\Omega)
    Y^*_{\ell,\vec{m}}(\Omega') = 
\frac{\ell + \nu}{\nu}\frac{1}{\Omega_{d-1}}
C_\ell^{(\nu)}(\cos\theta), \quad \cos \theta \equiv \Omega\cdot \Omega',
\label{addition theorem}
\end{equation}
where $\nu\equiv\frac{d-2}{2}$, $C^{(\nu)}_\ell$ is a Gegenbauer polynomial and $\Omega_{d-1}$ is the volume of the unit sphere, Eq.~\eqref{eq:local-flat-Green-function}.
The factor, $\xi_{\Delta_L,\ell}(\omega)$, should be viewed as the partial-wave representation of the monodromy projection \cite{Simmons-Duffin:2012juh} selecting the physical poles. It equals one on the physical global-AdS/MFT tower $\omega = \pm (\Delta_L+2n +\ell)$, and vanishes on the shadow tower $\omega = \pm (d-\Delta_L+2n +\ell)$. Hence it removes the shadow singularities, while leaving the physical poles unchanged. 

Finally, $V(\omega,\ell)$ is recognizable: at $\ell=0$ it coincides with the contact Witten diagram Eq.~\eqref{FT_contact}. The retarded correlator Eq.~\eqref{Full correlator} therefore reproduces the bubble-diagram resummation of section~\ref{The heavy--light correlator} in the heavy limit, $\Delta_H \to \infty$.  In the next section we proceed to the renormalization of the divergences that occur at even dimension $d$.

\subsection{Renormalization}
\label{sec:renorm}
For generic $d$, all four connection coefficients are finite. However, the
coefficients associated with the irregular solution contain the common factor
\begin{equation}
    \mathcal C_{\rm rsp}^{\rm irr},\, \mathcal C_{\rm src}^{\rm irr}
    \propto
    \Gamma\left(2-\frac{d}{2}-\ell\right).
\end{equation}
For integer $\ell$, this factor develops poles at even values of $d$. More
precisely, denoting by $d_\star$ the physical dimension, the singularities
occur whenever
\begin{equation}
    2-\frac{d_\star}{2}-\ell=-n_\ell,
    \qquad
    n_\ell \equiv \ell+\frac{d_\star}{2}-2
    \in\mathbb Z_{\geq0}.
\end{equation}
Thus, for even $d_\star\geq4$, the irregular connection coefficients are
singular for all integer $\ell$, whereas for odd $d_\star$ the argument of
the Gamma function is a half-integer and the coefficients remain finite.
As we will see below, these poles are associated with the logarithmic mixing
of the regular and irregular solutions at the origin.

To isolate the divergence, we analytically continue the spatial dimension as $d=d_\star-2\epsilon$,
and we expand the remaining Gamma functions around $d=d_\star$.
One then finds that
the divergent part of each irregular connection coefficient is proportional
to the corresponding regular coefficient,
\begin{equation}
    \mathcal C_a^{\rm irr}(d)
    =
    \frac{\kappa_\ell(\omega)}{\epsilon}
    \mathcal C_a^{\rm reg}(d)
    +
    \mathcal C_a^{\rm irr,fin}(\omega)
    +
    O(\epsilon),
    \qquad
    a={\rm rsp},{\rm src},
\end{equation}
where
\begin{equation}
    \kappa_\ell(\omega)
    =
    \frac{(-1)^{n_\ell}}
    {n_\ell!(n_\ell+1)!}
    \left(
        \frac{2-\ell-\Delta_L+\omega}{2}
    \right)_{n_\ell+1}
    \left(
        \frac{2-\ell-\Delta_L-\omega}{2}
    \right)_{n_\ell+1}.
\end{equation}
Notice that $\kappa_\ell(\omega)$ is an even polynomial in $\omega$. The
divergent contribution is therefore analytic in frequency and can be
absorbed into local worldline operators.

We define the finite part of the irregular connection coefficients by
minimal subtraction,
\begin{equation}
    \mathcal C_a^{\rm irr,fin}(\omega)
    \equiv
    \lim_{\epsilon\rightarrow0}
    \left[
        \mathcal C_a^{\rm irr}(d)
        -
        \frac{\kappa_\ell(\omega)}{\epsilon}
        \mathcal C_a^{\rm reg}(d)
    \right].
\end{equation}
Inserting this decomposition into the connection problem gives
\begin{equation}
    A_a
    =
    \left(
        B_{\rm reg}
        +
        \frac{\kappa_\ell(\omega)}{\epsilon}B_{\rm irr}
    \right)
    \mathcal C_a^{\rm reg}
    +
    B_{\rm irr}\mathcal C_a^{\rm irr,fin}.
\end{equation}
The first term shows that the pole amounts to a mixing of the irregular
solution into the regular one. Since the Green's function depends only on
the ratio $A_{\rm rsp}/A_{\rm src}$, the overall normalization of the
solution is irrelevant. It is therefore natural to define the renormalized ratio
\begin{equation}
    \left(
        \frac{B_{\rm irr}}{B_{\rm reg}}
    \right)_{\!R}
    \equiv
    \frac{B_{\rm irr}}
    {B_{\rm reg}
    +\dfrac{\kappa_\ell(\omega)}{\epsilon}B_{\rm irr}} .
\end{equation}
The same subtraction can then be expressed directly as a renormalization of the
worldline response function $F_\ell(\omega)$ through
\eqref{eq:Birr-in-terms-of-Fell},
\begin{equation}
\frac{1}{F_\ell^{R}(\omega)}
\equiv
\frac{1}{F_\ell^{B}(\omega)}
-
\frac{\bar \mu^{3-d}\mathcal{N}_{d,\ell}\kappa_\ell(\omega)}{\epsilon}.
\end{equation}

The bare short-distance boundary condition is independent of the arbitrary
scale $\bar \mu$.
The renormalization-group equation can be written directly as
\begin{equation}
\bar \mu\frac{d}{d\bar \mu}
F_\ell^{R}(\omega,\bar \mu)
=
(d_\star-3)F_\ell^{R}(\omega,\bar \mu)
-
2\bar \mu^{3-d_\star}
\mathcal{N}_{d_\star,\ell}
\kappa_\ell(\omega)
\left[F_\ell^{R}(\omega,\bar \mu)\right]^2.
\label{eq: full RG equation}
\end{equation}

The first term reflects the canonical scaling of $F_\ell$, inherited from
the factor $\bar \mu^{3-d}$. The second
term is the anomalous contribution generated by the logarithmic mixing of
the regular and irregular solutions. It is present only at the
even values of $d_\star$ for which the irregular connection coefficients
develop a pole.
Thus, although the bulk equation has been solved non-perturbatively, at even
$d_\star\geq 4$, the existence of this logarithmic mixing 
introduces a dependence on the sliding scale $\bar \mu$.
Indeed one can write Eq.~\eqref{eq: full RG equation} in a more compact form
\begin{equation}
    \bar \mu\frac{d}{d\bar \mu}
\left[
\frac{\bar \mu^{d_\star-3}}
{F_\ell^{R}(\omega,\bar \mu)}
\right]
=
2\mathcal{N}_{d_\star,\ell}\kappa_\ell(\omega),
\end{equation}
fully exhibiting the logarithmic behavior. For a given initial data $F_\ell^{R}(\omega,\bar \mu_0)$, fixed by the UV theory (which describes
the physics of the compact object), the scale dependence of the worldline response function
is
\begin{equation}
\frac{\bar \mu^{d_\star-3}}
{F_\ell^{R}(\omega,\bar \mu)}
=
\frac{\bar \mu_0^{d_\star-3}}
{F_\ell^{R}(\omega,\bar \mu_0)}
+
2\mathcal{N}_{d_\star,\ell}\kappa_\ell(\omega)
\log\frac{\bar \mu}{\bar \mu_0},
\end{equation}
which can be solved directly for $F_\ell^{R}(\omega,\bar \mu)$:
\begin{equation}
F_\ell^{R}(\omega,\bar \mu)
=
\left(
\frac{\bar \mu}{\bar \mu_0}
\right)^{d_\star-3}
\frac{
F_\ell^{R}(\omega,\bar \mu_0)
}{
1
+
2\bar \mu_0^{3-d_\star}
\mathcal{N}_{d_\star,\ell}
\kappa_\ell(\omega)
F_\ell^{R}(\omega,\bar \mu_0)
\log\frac{\bar \mu}{\bar \mu_0}
}.
\end{equation}
This form allows for a decomposition in terms of a geometric series,
\begin{equation}
F_\ell^{R}(\omega,\bar \mu)
=
\left(
\frac{\bar \mu}{\bar \mu_0}
\right)^{d_\star-3}
F_\ell^{R}(\omega,\bar \mu_0)
\sum_{n=0}^{\infty}
\left[
-2\bar \mu_0^{3-d_\star}
\mathcal{N}_{d_\star,\ell}
\kappa_\ell(\omega)
F_\ell^{R}(\omega,\bar \mu_0)
\log\frac{\bar \mu}{\bar \mu_0}
\right]^n ,
\end{equation}
that admits a simple diagrammatic interpretation. The first UV divergence appears at the first iteration of the contact potential,
a consequence of the coincident-point singularity of the bulk-to-bulk propagator.
Correspondingly, the first UV renormalization requires two insertions of the worldline response function.

\section{Perturbative CFT data}
\label{CFT}
In this section we study the CFT interpretation of the tidal response encoded by the bulk response function $F_\ell(\omega)$. We first perform the conformal block decomposition of the heavy--light four-point function in the $s$ and $t$--channels and extract the corresponding CFT data. We then discuss the holographic interpretation of the heavy--light double-twist operators and their relation to the normalizable modes of the light field in AdS. Finally, we derive the perturbative corrections to their anomalous dimensions and OPE coefficients for a general bulk potential, establishing their direct relation with standard quantum-mechanical perturbation theory. We conclude with a recap of the results.
\subsection{Conformal block decomposition}\label{sec:CB decomp}
To extract the CFT data we begin by first performing a
conformal block decomposition of the reduced correlator, $\mathcal G(u,v)$, defined in Eq.~\eqref{eq:redGuv}.

\paragraph{S-channel.} 
In the $s$--channel $(12)(34)$ 
this corresponds to the OPE $\mathcal{O}_H(P_1)\times \mathcal{O}_L(P_2)$ and $\mathcal{O}_L(P_3)\times \mathcal{O}_H(P_4)$,
\begin{equation}
    \mathcal G(u,v)
    = \sum_{\mathcal O_{\Delta,\ell} \in \mathcal{O}_H \times \mathcal{O}_L}
    \lambda_{\mathcal{O}_H \mathcal{O}_L \mathcal O_{\Delta,\ell}}^{\,2}\;
    g_{\Delta,\ell}^{(\alpha,\beta)}(u,v),
\end{equation}
where the external dimension differences are
$\alpha=\Delta_H-\Delta_L$, $\beta=\Delta_L-\Delta_H=-\alpha$.
In the heavy limit, $\Delta_H \to \infty$, the conformal block simplifies \cite{Jafferis:2017zna, Chen:2026ium},
\begin{equation}
\label{eq:confgegen}
     g_{\Delta,\ell}^{(\alpha,-\alpha)}(u,v)\approx 
    u^{\frac{\Delta}{2}} 
    \frac{\ell!}{(\nu)_\ell}\,
      C_\ell^{(\nu)}\left(
      \frac{1+u-v}{2\sqrt{u}}
      \right) 
    +\mathcal O\left(\Delta_H^{-1}\right), \quad \nu = \frac{d-2}{2},
\end{equation}
so that the four-point function can be written as
\begin{equation}
    \langle 
    \mathcal O_H(0)
    \mathcal O_L(z)
    \mathcal O_L(1)
    \mathcal O_H(\infty) \rangle \approx  
   \sum_{\mathcal O_{\Delta,\ell}} 
    \lambda_{\mathcal{O}_H \mathcal{O}_L \mathcal O_{\Delta,\ell}}^{\,2}\;
    u^{\frac{\Delta- \Delta_H - \Delta_L}{2}} 
    \frac{\ell!}{(\nu)_\ell}\,
      C_\ell^{(\nu)}\left(
    \frac{1+u-v}{2\sqrt{u}}
      \right),
\end{equation}
upon choosing standard kinematics. 

For the contact interactions and their Born iterations
considered in this work, the $s$--channel contribution is exhausted
by the conformal families continuously connected to the
double-trace operators
$[\mathcal O_H\mathcal O_L]_{n,\ell}$.
The worldline interactions correct their dimensions and OPE coefficients,
with renormalized scaling dimension
$\Delta_{n,\ell} = \Delta_H + \Delta_L + 2 n + \ell + \gamma_{n,\ell}$.
We organize the anomalous dimensions and the OPE coefficients as series in the number of insertions of the potential,
\begin{equation}
    \gamma_{n,\ell} = \sum_{i=1}^{\infty} \gamma_{n,\ell}^{(i)},
    \quad
    \lambda^2_{\mathcal O_H \mathcal O_L \left[\mathcal O_H \mathcal O_L\right]_{n,\ell}}
    \equiv \lambda^2_{n,\ell} = \sum_{i=0}^{\infty} a_{n,\ell}^{(i)},
\end{equation}
where $\gamma^{(i)}_{n,\ell}$ and $a^{(i)}_{n,\ell}$ are homogeneous of degree $i$ in the worldline coupling $\bar\mu^{3-d}F_\ell$. This produces a power-log expansion that can be reorganized order by order,
\begin{equation}
\lambda_{n,\ell}^{2}\,
u^{n+\frac{\ell}{2}+\frac{\gamma_{n,\ell}}{2}}\,
=
u^{n+\frac{\ell}{2}}\,
\left[
\mathcal T_{n,\ell}^{(0)}
+
\mathcal T_{n,\ell}^{(1)}
+
\mathcal T_{n,\ell}^{(2)}
+\cdots
\right].
\end{equation}
The zeroth order result is given by mean field theory \cite{Li:2020dqm},
\begin{equation}
\mathcal T_{n,\ell}^{(0)}
=
a_{n,\ell}^{(0)}=   
    \frac{
    (\Delta_L-\nu)_n
    (\Delta_L)_{n+\ell}
    }{
    n!\,\ell!\,(\nu+\ell+1)_{n}
    }.
    \label{eq:MFT0}
\end{equation}
At first order,
\begin{equation}
\mathcal T_{n,\ell}^{(1)}
=
a_{n,\ell}^{(1)}
+
\frac{1}{2}
a_{n,\ell}^{(0)}
\gamma_{n,\ell}^{(1)}
\log u,
\end{equation}
and so on. Using the cylinder map $u=e^{-2\tau}$, this becomes
\begin{equation}
    \mathcal T_{n,\ell}^{(1)}
    =
    a_{n,\ell}^{(1)}
    -
    a_{n,\ell}^{(0)}
    \gamma_{n,\ell}^{(1)}
    \tau .
    \label{CFT compare}
\end{equation}
We extract the anomalous dimensions from the poles
of the retarded Green's function computed in worldline EFT given in Eq.~\eqref{eq:GRCF}.
In particular, we can expand the
first-order correction, Eq.~\eqref{eq:Vwl}, around the free
frequencies $\omega_{n,\ell}=\Delta_L+\ell+2n$.
Matching to Eq.~\eqref{CFT compare}, with the angular normalization
\eqref{addition theorem}, gives
\begin{equation}
\gamma_{n,\ell}^{(1)}
=
\frac{2^{\ell-1}\bar\mu^{1-2\nu}}{\pi^{\nu+1}}
F_\ell(\omega_{n,\ell})
\frac{
\Gamma(\Delta_L+\ell+n)
\Gamma(\nu+\ell+n+1)
}{
n!\,\Gamma(\nu+\ell+1)
\Gamma(\Delta_L+n-\nu)
}. \label{eq:gamma1}
\end{equation}
The corresponding correction
to the squared OPE coefficient is
\begin{equation}
a_{n,\ell}^{(1)}
=
a_{n,\ell}^{(0)}\gamma_{n,\ell}^{(1)}
\left[
\psi(\Delta_L+\ell+n)-\psi(n+1)
+
\frac{F_\ell'(\omega_{n,\ell})}
     {F_\ell(\omega_{n,\ell})}
\right].
\end{equation}
Here $\omega_{n,\ell}=\Delta_L+\ell+2n$ are the free AdS frequencies Eq.~\eqref{eq:AdSfreq}, and $\nu=\frac{d-2}{2}$. The computation of the anomalous dimensions is particularly simple at any order, since they are determined by the shifts of the AdS frequencies read off from the poles of $G_R(\omega,\ell)$ in Eq.~\eqref{Full correlator}, i.e. by the transcendental equation\footnote{Equivalently, they follow from requiring the bulk solution to be normalizable at the
AdS boundary. From Eq.~\eqref{eq:ACB}, the relation Eq.~\eqref{eq:transcendental} follows from
\begin{equation*}
\mathcal{C}_{\mathrm{src}}^{\mathrm{reg}}(\omega,\ell)
-
\bar\mu^{3-d}\mathcal N_{d,\ell}F_\ell(\omega)
\mathcal{C}_{\mathrm{src}}^{\mathrm{irr}}(\omega,\ell)
= A_\text{src}(\omega,\ell) = 0.
\end{equation*}}
\begin{equation}
G^{(0)}(\omega,\ell) = \xi_{\Delta_L,\ell}(\omega) V(\omega,\ell).
\label{eq:transcendental}
\end{equation}
At second order we have:
\begin{equation}
\begin{aligned}
\gamma_{n,\ell}^{(2)}
&=
\frac{\left(\gamma_{n,\ell}^{(1)}\right)^2}{2}
\Bigg[
H_{\Delta_L+\ell+n-1}
+H_{\frac d2+\ell+n-1}
-H_n
-H_{\Delta_L-\frac d2+n}
\\
&\hspace{3.6cm}
+\pi\cot\!\left(\frac{\pi d}{2}\right)
+2\frac{F_\ell'(\omega_{n,\ell})}{F_\ell(\omega_{n,\ell})}
\Bigg],
\label{second energy correction}
\end{aligned}
\end{equation}
where $H_x\equiv\psi(x+1)+\gamma_E$ denotes the harmonic number analytically continued to non-integer argument. 
The last term accounts for evaluating the response at
the shifted frequency and vanishes for a
frequency-independent response.
At even $d$, the expression is understood by dimensional
continuation and must be combined with the worldline
counterterms of section~\ref{sec:renorm} to obtain finite anomalous
dimensions. We give the quantum-mechanical derivation
for a frequency-independent interaction in
section~\ref{sec:genformulas} and Appendix~\ref{app:anom2}.

\paragraph{T-channel.}

In the $t$-channel $(14)(23)$, we pair
$\mathcal O_H(P_1)\times\mathcal O_H(P_4)$ and
$\mathcal O_L(P_2)\times\mathcal O_L(P_3)$.
With the reduced-correlator definition in Eq.~\eqref{eq:redGuv},
\begin{equation}
\mathcal G(u,v)
=
\frac{u^{(\Delta_H+\Delta_L)/2}}{v^{\Delta_L}}
\sum_{\Delta,J}
b_{J,\Delta}\,
g_{\Delta,J}^{(0,0)}(v,u),
\qquad
b_{J,\Delta}
\equiv
\lambda_{HH\mathcal O_{\Delta,J}}
\lambda_{LL\mathcal O_{\Delta,J}}.
\label{eq:tchannel-blocks}
\end{equation}
The external dimension differences vanish, $\alpha = \beta = 0$, because each OPE
pairs identical operators. To determine these coefficient products, we compare
\eqref{eq:tchannel-blocks} with the Fourier representation
of the cylinder correlator. Using
$\mathcal G=u^{\Delta_H/2}G_{\mathrm{cyl}}$,
as established below Eq.~\eqref{eq:ContactWittenHyper},
and the addition theorem Eq.~\eqref{addition theorem}, we obtain
\begin{equation}
\mathcal G(u,v)
=
u^{\Delta_H/2}
\sum_\ell
\frac{\ell+\nu}{\nu\Omega_{d-1}}
C_\ell^{(\nu)}
\!\left(\frac{1-v+u}{2\sqrt u}\right)
\int_{-\infty}^{\infty}\frac{d\omega_E}{2\pi}\,
u^{-i\omega_E/2}
\widetilde G_{\mathrm{cyl}}(\omega_E,\ell).
\label{eq:tchannel-fourier}
\end{equation}
For the conservative analytic response considered below,
the Euclidean partial waves follow
from Eq.~\eqref{eq:GRCF},
$\widetilde G_{\mathrm{cyl}}(\omega_E,\ell)
\approx G_R(i\omega_E,\ell)$ in the heavy limit.

At first Born order,
\eqref{Full correlator} gives
$G_R=G_R^{(0)}+V+O(F_\ell^2)$.
Substituting Eq.~\eqref{eq:Vwl} into
\eqref{eq:tchannel-fourier}, the static frequency integral
is evaluated using the Barnes identity underlying
\eqref{eq:ContactWittenHyper}, with
$\Delta_L$ replaced by $\Delta_L+\ell$.
The conservative frequency dependence is included through
$F_\ell(i\omega_E)=\sum_k C_{\ell,2k}\omega_E^{2k}$:
each power $\omega_E^{2k}$ acts on the Fourier transform as
$(-1)^k(2u\partial_u)^{2k}$. The result for the first Born correction to the reduced correlator, $\mathcal G_{\mathrm{cons}}^{(1)}(u,v)$, in the conservative Wilson coefficients is
\begin{equation}
\begin{aligned}
\mathcal G_{\mathrm{cons}}^{(1)}(u,v)
\approx{}&
-\frac{\bar\mu^{3-d}u^{\Delta_H/2}}
{4\pi^{d/2}\Gamma(\Delta_L)
\Gamma(\Delta_L-\frac d2+1)}
\\
&\times
\sum_\ell
\frac{2^\ell}{(\nu)_\ell}
C_\ell^{(\nu)}
\!\left(\frac{1+u-v}{2\sqrt u}\right)
F_\ell^{\mathrm{cons}}(2u\partial_u)
\mathcal H_{\Delta_L+\ell}(u),
\end{aligned}
\label{eq:tchannel-conservative}
\end{equation}
where
\begin{equation}
\mathcal H_a(u)
\equiv
\frac{\Gamma^4(a)}{\Gamma(2a)}
u^{a/2}{}_2F_1(a,a;2a;1-u),
\qquad
F_\ell^{\mathrm{cons}}(2u\partial_u)
=
\sum_k(-1)^k C_{\ell,2k}(2u\partial_u)^{2k}.
\end{equation}
The differential operator acts only on
$\mathcal H_{\Delta_L+\ell}(u)$, and both
sums are truncated at the chosen order in the derivative
expansion.

To extract the lowest-twist data, we take $v\to0$ at fixed
$u$. The result Eq.~\eqref{eq:tchannel-conservative} only depends on $v$ via the Gegenbauer polynomial $C_\ell^{(\nu)}\big(\frac{1 - v + u  }{2 \sqrt{u}} \big) \sim v^0$, which is
finite and nonzero in this limit. The small-$v$ limit of the $t$-channel conformal blocks reads
\begin{equation}
\label{eq:gtchan}
g^{(0,0)}_{\Delta,J}(v,u) = v^{\frac{\Delta-J}{2}}\left[ (1-u)^J {}_2F_1\left(\frac{\Delta+J}{2},\frac{\Delta+J}{2};\Delta+J;1-u\right) + \mathcal{O}(v) \right] ,
\end{equation}
where we used the same normalization as the $s$--channel block in Eq.~\eqref{eq:confgegen}. 

Fixing the conformal data in t-channel at leading order thus amounts to use the small $v$ expansion of the conformal block Eq.~\eqref{eq:gtchan} in Eq.~\eqref{eq:tchannel-blocks} and compare with the expression Eq.~\eqref{eq:tchannel-conservative} linear in the worldline coefficients $C_{\ell,k}$. Given the behavior $g_{\Delta,J}^{(0,0)}(v,u)\sim v^{(\Delta-J)/2}$, we
identify the lowest exchanged twist as
$\Delta-J=2\Delta_L$ at this order. We identify them with the leading-twist light--light double-trace
operators $[\mathcal O_L\mathcal O_L]_{0,J}$,
with $J=0,2,4,\ldots$. 

Expanding around $u=1$ determines the coefficients
recursively: the constant term fixes the scalar
contribution, and subtracting its complete collinear
block allows the spin-two coefficient to be read from
the quadratic term, and so on.
Only even spins occur because the light operators are
identical. Higher-twist data require the subleading
powers of $v$, after subtracting descendants of the
lower-twist families.

For a static response at first Born order, the cases
examined explicitly contain lowest-twist exchanges with
$J=0,2,\ldots,2\ell$.
This restriction need not persist for frequency-dependent
response or higher Born iterations. 

A simple example illustrates both the matching and its
extension beyond first order. Take $d=3$, $\Delta_L=2$,
and retain only a static monopole response
$F_0(\omega)=C_{0,0}$. Equation
\eqref{eq:tchannel-conservative} reduces to
\begin{equation}
u^{-(\Delta_H+2)/2}
\mathcal G_{\mathrm{cons}}^{(1)}(u,0)
=
-\frac{C_{0,0}}{12\pi^2}
{}_2F_1(2,2;4;1-u).
\end{equation}
Comparison with
\eqref{eq:tchannel-blocks} using Eq.~\eqref{eq:gtchan}
gives $b^{(1)}_{0,4}=-C_{0,0}/(12\pi^2)$,
with all higher-spin coefficients in the lowest-twist
family vanishing. At second Born order, expanding Eq.~\eqref{Full correlator}
gives
\begin{equation}
G_R^{(2)}(\omega,\ell)
=
\frac{V(\omega,\ell)^2}
     {G_R^{(0)}(\omega,\ell)}
\,\xi_{\Delta_L,\ell}(\omega).
\label{eq:tchannel-second-born}
\end{equation}
Repeating the Fourier transform and matching for the same
monopole example yields\footnote{Including the angular
normalization $1/\Omega_2=1/(4\pi)$, the Fourier transform
of Eq.~\eqref{eq:tchannel-second-born} gives
\[
u^{-(\Delta_H+2)/2}
\mathcal G_{\mathrm{cons}}^{(2)}(u,0)
=
-\frac{C_{0,0}^{\,2}}{48\pi^4}
\left[
3+3(1-u)+\frac52(1-u)^2
+O\bigl((1-u)^3\bigr)
\right].
\]
The scalar collinear block starts as
$1+(1-u)+\frac9{10}(1-u)^2+\cdots$,
while the spin-two block starts as
$(1-u)^2+\cdots$.
Matching these and higher powers gives
Eq.~\eqref{eq:monopole-second-order-ope}.}
\begin{equation}
b^{(2)}_{0,4}
=-\frac{C_{0,0}^{\,2}}{16\pi^4},
\qquad
b^{(2)}_{2,6}
=\frac{C_{0,0}^{\,2}}{240\pi^4},
\qquad
b^{(2)}_{4,8}
=\frac{C_{0,0}^{\,2}}{12096\pi^4}, \qquad \cdots
\label{eq:monopole-second-order-ope}
\end{equation}
Thus a static bulk monopole generates a spin-two lowest-twist
exchange at second order, demonstrating that the
first-order static restriction $J\leq2\ell$ does not
persist under iteration.

At a finite truncation in $\ell$, these Born contributions
remain polynomial in $v$, which enters only through the
Gegenbauer polynomials in Eq.~\eqref{eq:tchannel-fourier}.
They therefore generate no $\log v$ terms, indicating that the light--light double-trace $[\mathcal{O}_L \mathcal{O}_L]_{n,J}$ dimensions
are not corrected.

\subsection{Perturbative corrections for a general potential}\label{sec:generic_potential}

We now extend the $s$--channel analysis to a static, rotationally
invariant potential $V(r)$, which need not be localized on the
worldline. In the heavy limit, perturbation theory for the
normalizable light-field modes gives the corresponding shifts of
heavy--light dimensions and OPE coefficients. We first explain the
state--mode correspondence and then derive the first-order shifts
from matrix elements of $V(r)$.
 We note that, as we showed in Eq.~(\ref{eq:transcendental}), the
computation of the anomalous dimension in the s-channel is directly connected with the dynamics of the light particle, namely, there is a connection between their Hilbert spaces. We explore this in section \ref{sec:twoparticle}.

After the map is fully understood we use it to compute generic double twist data as a function of matrix elements of the potential in section \ref{sec:genformulas}.
\subsubsection{Two-particle Hilbert space in AdS/CFT}
\label{sec:twoparticle}
In the heavy limit, the dynamics are governed by a wave equation, which implies the interaction action can be written as
\begin{equation}
    S_{\text{int}} = \frac{1}{2} \int_{\text{AdS}} dX \ \phi(X) V(X) \phi(X).
    \label{int_action}
\end{equation}
The GKPW prescription \cite{Gubser:1998bc,Witten:1998qj} computes the heavy--light four-point function from the on-shell action with non-normalizable boundary conditions. If instead one imposes normalizable boundary conditions in Eq.~\eqref{int_action}, one is solving the ``Schr\"odinger''
equation for the AdS frequency modes that are now shifted by the presence of the potential. These corrections are computed by the familiar time independent perturbation
theory formulas from QM.
The Born regime implies that these corrections are the only data required to solve the theory. One must then identify both the OPE coefficients and the anomalous dimensions of double-twist operators, at each order in the potential, in terms of simple QM perturbation theory \cite{Kulaxizi:2018dxo,Maxfield:2022nat,Maxfield:2022hkd}. The key aspect that connects the double-twist data with the QM of the light field is the simplification of the double-twist operator in the heavy limit. This can be easily seen by computing the first double-twist operator, $\left[\mathcal O_H \mathcal O_L\right]_{0,1}$, using the operator-state correspondence, and constructing directly the two-particle Hilbert
space as the tensor product of the Verma modules \cite{OsbornCFTNotes} of the two primaries,
\begin{equation}
    \mathcal H_{HL} \equiv \mathcal V_{\Delta_H,0} \otimes \mathcal V_{\Delta_L,0} = 
    \bigoplus_{n,\ell\geq 0}^{\infty} \mathcal V_{\Delta_H + \Delta_L + 2n + \ell,\ell}
\end{equation}
The $n=0$, $\ell=1$ state is then
\begin{equation}
\ket{\left[\mathcal O_H \mathcal O_L\right]_{0,1}} =
c_1 \left(P_\mu \ket{\mathcal O_H}\right) \otimes \ket{\mathcal O_L}+
c_2\ket{\mathcal O_H}\otimes \left(P_\mu \ket{\mathcal O_L}\right).
\end{equation}
The requirement that this state is a primary is just that it has to be annihilated by the special 
conformal generator, $K_\mu$,
\begin{equation}
K_\mu\ket{\left[\mathcal O_H \mathcal O_L\right]_{0,1}} = 0 \Leftrightarrow
 c_1 \Delta_H + c_2 \Delta_L =0,
 \end{equation}
 where we used the conformal algebra, $\comm{K_\mu}{P_\nu} = 2\delta_{\mu\nu} D - 2 M_{\mu\nu}$. 
 Writing the state in terms of the normalization constant $c_2$ alone, we get
\begin{equation} 
\ket{\left[\mathcal O_H \mathcal O_L\right]_{0,1}} =c_2\left(
\ket{\mathcal O_H}\otimes \left(P_\mu \ket{\mathcal O_L}\right)-
 \frac{\Delta_L}{\Delta_H}\left(P_\mu \ket{\mathcal O_H}\right) \otimes \ket{\mathcal O_L}
 \right).
    \end{equation}
This means that
all the contributions from the descendants of the heavy states are suppressed by powers of $\Delta_H$, and the heavy--light double-twist operator is the heavy primary tensored with descendants of the light operator, 
 \begin{equation}
 \ket{\left[\mathcal O_H \mathcal O_L\right]_{n,\ell}} \approx
 \ket{\mathcal O_H} \otimes 
 \left( P_{\langle\mu_1} \cdots P_{\mu_\ell\rangle} (P^2)^n\ket{\mathcal O_L}\right).
\end{equation}
Equivalently, this identifies the heavy--light space of primaries with the Verma module of the light state. Since the descendants of the light operator are dual to AdS excitations
of the light field, we can identify the resolution of the identity in AdS and in the CFT.
The interpretation is that the conformal block decomposition (resolution of the identity in the CFT) is related to the normalizable mode expansion of the dual light field in AdS. We note that this behavior is also manifested in the non-relativistic limit as we show in Appendix \ref{app:nrel}.

\subsubsection{Generic formulas for CFT data}
\label{sec:genformulas}
 In the heavy limit, the four-point function
is given by the propagator of the heavy state (which we will not write for simplicity), multiplied by the propagator of the light state, which  will receive corrections from the potential. The propagator of the light particle will solve a Klein--Gordon equation in AdS, and hence, its propagator can be written
as a sum of normalizable modes
\begin{equation}
\Pi^{\text{B}}(X,X')  =
\int_{-\infty}^{\infty} \frac{d\omega}{2\pi} 
\sum_{n,\ell,\vec{m}}
e^{i\omega (t-t')}
\frac{
u_{n,\ell}(r)
u_{n,\ell}(r')
}{
 \omega_{n,\ell}^2-\omega^2+i\epsilon
}Y_{\ell,\vec{m}}(\Omega)
Y^*_{\ell,\vec{m}}(\Omega'),
\end{equation}
where $u_{n,\ell}(r) \equiv \sqrt{2\omega_{n,\ell}}\psi_{n,\ell}(r)$ are the Sturm--Liouville (SL)
normalized wavefunctions\footnote{We used these solutions since the Green's function is simpler for the derivation. 
We return to the usual normalization in the end.}, such that the completeness relation is
\begin{equation}
   \sum_{n\geq0}w(r) u_{n,\ell}(r) u_{n,\ell}(r')  = \delta(r-r'),\quad w(r) = \frac{r^{d-1}}{1+r^2}.
\end{equation}
Taking one point to the conformal boundary allows us to write the bulk-to-boundary propagator as a sum of normalizable modes,
\begin{equation}
    \Pi^{\partial}(P,X)  = \frac{1}{\sqrt{\mathcal C_{\Delta_L}}}
\int_{-\infty}^{\infty} \frac{d\omega}{2\pi} 
\sum_{n,\ell,\vec{m}}
e^{i\omega (t-t_p)}
\frac{
u_{n,\ell}(r)
\tilde k_{n,\ell}
}{
 \omega_{n,\ell}^2-\omega^2+i\epsilon
}Y_{\ell,\vec{m}}(\Omega)
Y^*_{\ell,\vec{m}}(\Omega_p),
\end{equation}
where $P = (t_p,\Omega_p)$. The factor $\mathcal C_{\Delta_L}^{-1/2}$ converts the boundary
value of a KG-normalized bulk mode into the normalization of the
light operator used in Eq.~\eqref{eq:CdH}. Note that $\tilde k_{n,\ell} \equiv \sqrt{2\omega_{n,\ell}} k_{n,\ell}$ is the SL boundary value\footnote{This means $k_{n,\ell}$ is defined as $\psi_{n,\ell} \xrightarrow[]{r\rightarrow\infty} k_{n,\ell}
\;r^{-\Delta_L}$.}.
Now we consider the first correction in the Born series given by Eq.~\eqref{Born series def}, and expand in normalizable modes. For simplicity, consider the correction to the propagator coming from a static, isotropic potential $V(r)$. The propagator then becomes
\begin{equation}
\begin{aligned}
 G(0,P_2,P_3,\infty)&= \frac{1}{\mathcal C_{\Delta_L}} 
   \int_{-\infty}^{\infty} \frac{d\omega}{2\pi} e^{i \omega t_2}
   \int_{-\infty}^{\infty} \frac{d\omega'}{2\pi} e^{-i \omega'  t_3}
\sum_{n,\ell,\vec{m}}
\sum_{n',\ell',\vec{m}'}
   V^{\ell,\ell'}_{n,n'} (\omega,\omega')\\
   & 
   \times\frac{\tilde k_{n,\ell}}{\omega^2_{n,\ell} - \omega^2 + i\epsilon}
   \frac{\tilde k_{n',\ell'}}{\omega^2_{n',\ell'} - \omega'^2 + i\epsilon}
   Y_{\ell,\vec{m}}(\Omega_2)
   Y^*_{\ell',\vec{m}'}(\Omega_3),
\end{aligned}
\end{equation}
where we defined the matrix element
\begin{equation}
   V^{\ell,\ell'}_{n,n'}(\omega,\omega')
   \equiv \int_{\text{AdS}} dX \ 
   e^{i (\omega - \omega') t}
   u_{n,\ell}(r)
   V(r)
   u_{n',\ell'}(r)
   Y^*_{\ell,\vec{m}}(\Omega)
   Y_{\ell',\vec{m}'}(\Omega).
\end{equation}
The time and angular integral just give orthogonality relations,
\begin{equation}
   V^{\ell,\ell'}_{n,n'}(\omega,\omega') =2\pi \delta(\omega-\omega')\delta_{\ell,\ell'} \delta_{\vec{m},\vec{m}'} \mathcal V^{\ell}_{n,n'},
\end{equation}
where $\mathcal V^\ell_{n,n'}$ is the radial integral,
\begin{equation}
\mathcal V_{n,n'}^\ell = \int_{0}^{\infty} dr \ w(r) u_{n,\ell}(r) V(r) u_{n',\ell}(r).
\end{equation}
We can then compute
one of the frequency integrals and one of the angular momentum sums. These fix a frequency, $\omega$, and angular quantum numbers, $(\ell,\vec{m})$, and the sum over $\vec{m}$ is again given by the addition theorem Eq.~\eqref{addition theorem}.
Thus the spectral representation of the full propagator can then be written as 
\begin{equation}
 G(0,P_2,P_3,\infty)= 
\frac{1}{\mathcal C_{\Delta_L}}  \sum_{n,n',\ell}  \frac{ (\ell + \nu)}{\nu \Omega_{d-1}}
  \tilde k_{n,\ell}
  \tilde k_{n',\ell}
   \mathcal V_{n,n'}^\ell
   \mathcal I_{n,n'}^\ell 
   C_\ell^{(\nu)}(\cos \theta),
\end{equation}
where all that is left is the integral,
\begin{equation}
      \mathcal I_{n,n'}^\ell \equiv 
  \int_{-\infty}^{\infty} \frac{d\omega}{2\pi}
      \frac{e^{i\omega T} }
   {(\omega_{n,\ell}^2 - \omega^2+i\epsilon)
   (\omega_{n',\ell}^2 - \omega^2+i\epsilon)} ,
\end{equation}
where $T = t_2-t_3$, is the time difference which we will assume to be positive.
When $n\neq n'$, we pick two positive-frequency poles in $\omega_{n,\ell}$ and $\omega_{n',\ell}$,
\begin{equation}
      \mathcal I_{n\neq n'}^\ell
 = 
   \frac{i}
   {\omega^2_{n,\ell} - \omega^2_{n',\ell} }
   \left[
   \frac{e^{i\omega_{n,\ell}T}}{2\omega_{n,\ell} } - 
   \frac{e^{i\omega_{n',\ell}T}}{2\omega_{n',\ell} } 
   \right].
\end{equation}
Note that the full sum is symmetric under $n\leftrightarrow n'$. This allows us to reorganize the sum,
\begin{equation}
\sum_{\substack{n, n'\\ n\neq n'}} A_{n, n'}^\ell \mathcal I_{n ,n'}^\ell
=
i \sum_n 
\frac{e^{i\omega_{n,\ell} T}}{\omega_{n,\ell}}
\sum_{n'\neq n}
\frac{A_{n ,n'}^\ell}
{\left(\omega_{n,\ell}^2-\omega_{n',\ell}^2\right)} ,
\end{equation}
where $A^\ell_{n,n'}$ is any matrix symmetric in $n\leftrightarrow n'$.
When $n=n'$, we pick a double positive frequency pole,
\begin{equation}
      \mathcal I_{n=n'}^\ell
 = 
 -e^{i\omega_{n,\ell} T} \left[
 \frac{T}{( 2\omega_{n,\ell})^2 } 
 +\frac{2i}{(2\omega_{n,\ell})^3}
 \right].
\end{equation}
We can now collect the powers of $T$, Wick rotate ($T\rightarrow i\tau$), and compare with Eq.~\eqref{CFT compare}.
This gives us an equation for the anomalous dimension, $\gamma_{n,\ell}^{(1)}$ and correction of the squared OPE coefficient, $a^{(1)}_{n,\ell}$ in terms of AdS quantities,
\begin{equation}
\gamma_{n,\ell }^{(1)} = -\frac{\mathcal V^\ell_{n,n}}{2\omega_{n,\ell}} .
\end{equation}
In particular, the free-mode residue obeys
\begin{equation}
a^{(0)}_{n,\ell}
=
\frac{\ell+\nu}
{\nu\Omega_{d-1}\mathcal C_{\Delta_L}}\,
k_{n,\ell}^{\,2},
\end{equation}
which reproduces Eq.~\eqref{eq:MFT0} for the mean-field coefficient, and
\begin{equation}
a^{(1)}_{n,\ell}
=
\frac{\ell+\nu}{\nu\Omega_{d-1}\mathcal C_{\Delta_L}}
\left[
\frac{\widetilde k_{n,\ell}^{\,2}\mathcal V^\ell_{n,n}}
     {4\omega_{n,\ell}^{3}}
-
\frac{\widetilde k_{n,\ell}}{\omega_{n,\ell}}
\sum_{n'\ne n}
\frac{\widetilde k_{n',\ell}\mathcal V^\ell_{n,n'}}
     {\omega_{n,\ell}^{2}-\omega_{n',\ell}^{2}}
\right].
\end{equation}
If we instead use the KG normalized wavefunctions, then these formulas are written with respect to 
$\psi_{n,\ell}(r) \equiv \braket{r}{n,\ell} $,
\begin{equation}
\gamma_{n,\ell }^{(1)} = -\bra{n,\ell}V\ket{n,\ell},
\label{eq:gammaV}
\end{equation}
\begin{equation}
\frac{a^{(1)}_{n,\ell}}{a^{(0)}_{n,\ell}}
=
\frac{\langle n,\ell|V|n,\ell\rangle}{\omega_{n,\ell}}
-
2\sum_{n'\ne n}
\frac{k_{n',\ell}}{k_{n,\ell}}\,
\frac{2\omega_{n',\ell}
\langle n',\ell|V|n,\ell\rangle}
{\omega_{n,\ell}^{2}-\omega_{n',\ell}^{2}} .
\label{first order OPE formula}
\end{equation}
These are the formulas of first-order perturbation theory: the energy shift gives the anomalous dimension and the wavefunction correction gives the OPE coefficient.
As a consistency check, we compute the anomalous dimension up to second order and OPE coefficient up to first order in Appendix \ref{app:anom2} using this formalism.

\subsection{Recap of different regimes: eikonal, lightcone and Born}
Having extracted the CFT data governing the heavy--light double-twist operators, it is instructive to place our effective field theory results within the landscape of CFT limits. The heavy--light correlator encodes a rich variety of bulk physics, which can be  separated by studying the asymptotic limits of the boundary data. Table \ref{tabela:cft_regimes} summarizes the three primary regimes of interest.

\begin{table}[htpb]
    \centering
    \renewcommand{\arraystretch}{1.4}
    \begin{tabularx}{\linewidth}{@{}
        >{\bfseries\raggedright\arraybackslash\hyphenpenalty=10000\hsize=0.76\hsize}X
        >{\hsize=1.08\hsize}Y
        >{\hsize=0.98\hsize}Y
        >{\hsize=1.18\hsize}Y @{}}
    \toprule
    Regime & \textbf{CFT limit} & \textbf{CFT data} & \textbf{Bulk correspondence} \\
    \midrule
    Lightcone bootstrap
        & $\ell \to \infty$, fixed $n$ \mbox{(fixed twist $\tau$)}
        & Power-law decay, $\gamma \sim 1/\ell^{\tau_m}$
        & Long-range single exchange (e.g.\ massless graviton) \\
    Eikonal
        & $\ell, \Delta_H \to \infty$, fixed ratio $\ell/\Delta_H \sim b$
        & Eikonal phase shift $\delta(s,b)$
        & High-energy multi-graviton resummation \\
    Born / EFT (this work)
        & $\Delta_H \to \infty$, fixed $n$ and~$\ell$
        & Quantum mechanics, $\gamma \sim \bra{n,\ell}V\ket{n,\ell}$
        & Short-distance finite-size effects (Love numbers / contact terms) \\
    \bottomrule
    \end{tabularx}
    \caption{The three regimes of the heavy--light correlator discussed in the text: the CFT limit defining each, the form taken by the CFT data, and the bulk physics it captures.}
    \label{tabela:cft_regimes}
\end{table}

The lightcone bootstrap and the eikonal regimes are 
particularly useful for studying long-range gravitational interactions in the bulk.
The former 
isolates the leading low-twist exchanges by focusing on operators approaching the boundary lightcone, giving rise to a characteristic power-law decay in the anomalous dimensions, $\gamma_{n,\ell} \sim 1/\ell^{\tau_m}$, with $\tau_m$ the twist of the lightest exchanged operator \cite{Komargodski:2012ek,Fitzpatrick:2014vua}. The latter describes high-energy, small-angle scattering at a fixed impact parameter, resulting  in the exponentiation of large anomalous dimensions into an eikonal phase shift 
\cite{Cornalba:2006xk,Cornalba:2007zb}.

In this paper, we have focused on short-range interactions localized
on the heavy worldline. The Born-series method also applies to more
general bulk potentials, as discussed in section~\ref{sec:generic_potential}. For local
worldline interactions truncated at a fixed order in spatial
derivatives, the corrections have finite support in the bulk angular
momentum $\ell$. Our heavy-limit formulas determine the CFT data at
each fixed $\ell$, but do not by themselves establish a uniform
large-$\ell$ limit. It is therefore useful to compare them with
large-spin analyses of HHLL correlators.

In the lightcone-bootstrap approach, one typically derives the double-twist operator data by requiring consistency between the direct and crossed-channel decompositions of the conformal blocks. The anomalous dimensions of these operators admit an expansion of the form
\begin{equation}
\gamma_{n,\ell} = \sum_{i=1}^{\infty} \mu^i \hat\gamma_{n,\ell}^{(i)}, \qquad \mu =\frac{\Delta_H}{c} ,
\end{equation}
and similarly for the OPE coefficients (the coefficients $\hat\gamma^{(i)}_{n,\ell}$ of this gravitational expansion should not be confused with the $\gamma^{(i)}_{n,\ell}$ of section~\ref{sec:CB decomp}, which count insertions of the worldline potential). The analytic form of $\hat\gamma_{n,\ell}^{(i)}$ is constrained by the direct-channel contribution of $i$-th order multi-stress tensor exchanges. Typically, these gravitational formulas for the anomalous dimensions 
are only strictly valid at large spin. Their analytic continuation to arbitrary spin often exhibits spurious singularities at low spin. For instance, at second order in the expansion, $\mathcal{O}(\mu^2)$, unphysical divergences appear at $\ell = 0$ and $\ell=1$ (see Appendix D of \cite{Dodelson:2022yvn}).

Our framework naturally accommodates and resolves these divergences through dimensional regularization. In our approach, we also encounter singularities at second order in the Born series: Eq.~\eqref{second energy correction} diverges at $d=4$. However, these divergences arise precisely from the short-distance contact terms. This provides a clean physical picture consistent with standard effective-field-theory matching: the short-range local operators (the Love numbers) are the natural candidates for the counterterms that cure the spurious low-spin divergences generated by the long-range gravitational exchanges in the conformal bootstrap. By renormalizing these short-distance contributions, divergences are absorbed leading to well-defined, renormalized CFT data valid for all spins.

Finally, one might consider taking the large-spin asymptotic limit of our explicit formulas for the anomalous dimensions and OPE coefficients, such as the second-order correction in Eq.~\eqref{second energy correction}. However, it is important to note that these expressions are governed by the boundary-condition parameter $F_\ell$, which itself may carry a non-trivial spin dependence dictated by the specific ultraviolet completion of the heavy object.

\section{Discussion and outlook}

In this work, we have shown that HHLL correlators in holographic CFTs admit
a natural reorganization in terms of a Born series. The perturbative series of Witten diagrams generated by repeated interactions
between the heavy and light fields becomes the Born series associated with an
effective wave equation in AdS. The heavy state determines an effective
potential supported on its semiclassical trajectory, while the light field
propagates through this potential by successive scatterings. 

We verified this relation explicitly for a localized interaction. The direct
solution of the effective wave equation resums the contact Witten diagrams and
agrees with their diagrammatic expansion. The construction extends naturally
to interactions containing derivatives, which generate the tower of
multipolar and frequency-dependent operators appearing in the worldline
effective theory. In this way, the effective wave equation provides a common
description of both the minimal point-particle interaction and its finite-size
corrections.

The central result is a direct map between the Wilson coefficients of the
worldline EFT and heavy--light CFT data. After decomposing the light field into
frequency and angular-momentum modes, the worldline EFT coefficients $C_{\ell,n}$ determine a
frequency-dependent response function of the compact object $F_\ell(\omega)$. This
interaction fixes the ratio between the coefficients of the irregular and regular solutions, respectively $B_\text{irr}(\omega,\ell)$ and $B_\text{reg}(\omega,\ell)$, near
the worldline dual to the heavy state. The AdS connection problem then determines the boundary
response function,
\begin{equation}
    \text{WEFT coefficients}
    \;\longrightarrow\;
    F_\ell(\omega)
    \;\longrightarrow\;
    \frac{B_{\rm irr}(\omega,\ell)}
         {B_{\rm reg}(\omega,\ell)}
    \;\longrightarrow\;
    G_R(\omega,\ell)     \;\longrightarrow\; \text{CFT}.
    \label{eq:EFT-CFT-map}
\end{equation}
The four arrows are, respectively, Eq.~\eqref{eq:worldline-response-function}, Eq.~\eqref{eq:Birr-in-terms-of-Fell}, Eq.~\eqref{eq:GRCF}, and Eq.~\eqref{eq:gamma1}--\eqref{second energy correction} together with Eq.~\eqref{eq:gammaV}--\eqref{first order OPE formula}: the shifts of the poles and residues of $G_R(\omega,\ell)$ give, respectively,
the corrections to the dimensions and to the OPE coefficients of the heavy--light
double-twist operators. 

 On the CFT side the content of this map is simple. At first order in the worldline interaction, the anomalous dimension of $[\mathcal O_H\mathcal O_L]_{n,\ell}$ is the expectation value of the worldline potential in the corresponding normalizable mode of the light field, $\gamma^{(1)}_{n,\ell}=-\langle n,\ell|V|n,\ell\rangle$, and the correction to the squared OPE coefficient is the first-order correction to the wavefunction, Eq.~\eqref{eq:gammaV}--\eqref{first order OPE formula}. For a worldline operator of multipole order $\ell$ the matrix element is proportional to the square of the wavefunction at the origin, $N_{n,\ell}^{-2}$, and vanishes in every other partial wave: a given Wilson coefficient $C_{\ell,n}$ shifts only the heavy--light double-twist operators of spin $\ell$, for all $n$, and leaves the large-spin asymptotics untouched. The first-order data obey the derivative relation Eq.~\eqref{eq:derivrel}, which in the bulk arises from the resummation of the off-diagonal matrix elements. At second order, Eq.~\eqref{second energy correction}, the poles at even $d$ they are absorbed into the running of the renormalized response function $F^R_\ell(\omega,\bar\mu)$, and the renormalized CFT data depend on $\bar\mu$ only through it. In the crossed channel, the same interaction is reproduced by the light--light double-twist operators, $[\mathcal O_L\mathcal O_L]_{n,J}$, and receive no anomalous dimensions at any order. The crossed-channel OPE coefficients are also constrained in terms of the Wilson coefficients as we show explicitly in Eq.~\eqref{eq:monopole-second-order-ope}.

This dictionary gives a precise holographic meaning to the effective
properties of a compact object. Its universal long-distance field is
described by the propagation of light bulk degrees of freedom, whereas its
unresolved internal structure is parametrized by local operators on the
worldline
\cite{Goldberger:2004jt,Porto:2016pyg}. In gravitational applications, the
corresponding Wilson coefficients include static and dynamical tidal Love
numbers. The matching between the near-worldline solution and the boundary
correlator consequently gives a CFT definition of these response
coefficients, once an operator normalization and renormalization scheme have
been specified. Dimensional regularization is particularly convenient for
this purpose because it separates local counterterms, logarithmic running,
and finite response coefficients in the same scheme used in worldline EFT
and black-hole perturbation theory
\cite{Ivanov:2022hlo,Saketh:2023bul}. 

An important application concerns the small-$\mu$ expansion, with
$\mu\sim\Delta_H/c$, of correlators in semiclassical black-hole states. The
universal part of this expansion is organized by stress-tensor and
multi-stress-tensor exchange and captures the long-range gravitational field
of the heavy state
\cite{Karlsson:2019dbd,
Parnachev:2020fna,Huang:2024wbq}.
The worldline EFT identifies the complementary, non-universal part: the
dependence on the size, composition, and internal response of the object.
At any fixed order in the local derivative expansion, a worldline operator
contributes only to a bounded set of partial waves. Such contributions are
therefore absent from the asymptotic inverse-spin expansion and are
nonperturbative in spin in this precise sense
\cite{Heemskerk:2009pn,
Caron-Huot:2017vep,Albayrak:2019gnz,
Dodelson:2022eiz}.
The worldline EFT thus complements the lightcone bootstrap:
the latter determines the universal large-spin sector, while the worldline
coefficients parametrize the additional short-distance data needed to
complete the correlator.

This observation suggests that the worldline EFT should play a direct role
in the study of holographic thermal correlators. Their near-boundary and
near-lightcone expansions efficiently determine the multi-stress-tensor
sector, while global consistency conditions such as KMS periodicity constrain
the remaining double-trace data
\cite{Buric:2025anb}.
The present framework predicts a further decomposition of this data into a
universal gravitational contribution and object-dependent response terms.
It would be interesting to identify these terms directly in the
small-$\mu$ expansion of thermal correlators and compare them with
Wilson coefficients obtained independently from black-hole perturbation
theory.

A first extension is to include graviton exchange. The
effective potential would then contain both a long-range component,
$V_{\rm long}$, generated by gravitational propagation, and a short-range
component, $V_{\rm short}$, determined by worldline operators. Applying the
construction to a light graviton, or equivalently to HHLL correlators in
which the light operator is the stress tensor, would give direct access to
the gravitational Love numbers and their contribution to thermal
stress-tensor correlators. It would also provide a concrete setting in which
to combine the multi-stress-tensor expansion with finite-size response
coefficients. Love numbers of AdS$_5$ black holes and their CFT interpretation will be discussed in \cite{IossaKarlssonZhiboedov:toappear}.

A second direction concerns dynamical response and dissipation. Static local
operators describe only the conservative low-frequency expansion. Internal
excitations, horizon absorption, and quasinormal modes require
frequency-dependent response functions, which may be represented by
additional worldline degrees of freedom or by kernels that are nonlocal in
time. Their causal and dissipative components are most naturally formulated
using a Schwinger--Keldysh effective action
\cite{Goldberger:2005cd,
Goldberger:2020fot,Galley:2012hx,Ivanov:2022hlo} as implemented here.
On the CFT side, the corresponding information should be visible in the
analytic structure and spectral density of real-time HHLL correlators. The
map Eq.~\eqref{eq:EFT-CFT-map} should then relate absorptive worldline response
directly to widths and spectral weights in the heavy--light channel.

A further question is whether the formalism presented here generalizes to correlators with several heavy insertions. In flat-space EFT, coefficients
fixed from the linear response of a compact object also enter nonlinear
scattering and binary dynamics \cite{Caron-Huot:2025tlq,Ivanov:2024sds}. The holographic counterpart would be a
consistency relation between the response data extracted from HHLL
correlators and the local counterterms required in HHHH or more general
heavy--light correlators. Establishing such relations would test whether the
worldline coefficients define intrinsic properties of the heavy state,
independent of the particular probe used to measure them. This extension
would also require nonlinear worldline operators describing higher-order
tidal response and radiation.

Finally, the framework is not restricted to black holes. It applies whenever
the heavy state is localized on scales much smaller than $L_{\rm AdS}$ and
admits a controlled worldline description. Different compact objects are
then distinguished by their Wilson coefficients and by the analytic
structure of their response functions. Extended branes, bubbling geometries,
and other configurations that remain resolved on AdS scales instead require
a defect or extended-object EFT. Understanding the relation between these
descriptions may lead to a CFT-intrinsic characterization of when a heavy
operator admits a point-particle bulk dual.

The main conclusion is that HHLL correlators furnish a quantitative
holographic dictionary for compact-object effective field theory. The Born
expansion reorganizes Witten diagrams into the perturbative solution of a
bulk wave equation, while the near-worldline boundary condition translates
worldline Wilson coefficients into shifts of CFT dimensions and OPE
coefficients. In black-hole states, this is expected to separate the universal
multi-stress-tensor sector from additional response data that are
nonperturbative in spin. This provides a systematic route to extracting
conservative and dissipative properties of heavy states directly from
holographic correlation functions.

\acknowledgments
The authors are grateful to António Antunes, Bruno Fernandes, Francesco Aprile,
Joydeep Chakravarty, Miguel Costa, Giulia Isabella, Shota Komatsu,
Jos\'e Matos, Daniel Neves, Jo\~ao Penedones, Pedro Vieira and
Sasha Zhiboedov for helpful discussions.
M.C. is supported by
the National Science and Engineering Research Council of Canada
(NSERC) and the Canada Research Chair program, reference number CRC-2022-00421. 
V.G. is supported by Fundacao para a Ciencia e Tecnologia (FCT)
under the grant CEECIND/\allowbreak 03356/\allowbreak 2022,
by FCT grants 2024.00230.\allowbreak CERN, 2023.17511.\allowbreak ICDT
and HORIZON-MSCA-2023-SE-01-\allowbreak 101182937-HeI.
F.S. is supported by FCT -- Fundação para a Ciência e a Tecnologia, I.P.,
through doctoral grant 2024.06037.BD
(DOI: \href{https://doi.org/10.54499/2024.06037.BD}
{10.54499/2024.06037.BD}).
Centro de F\'isica do Porto
is partially funded by Fundacao para a Ciencia e a Tecnologia (FCT) under the grant UID04650-FCUP.

\appendix

\section{Dictionary with flat space}
\label{app:bornregime}
Consider $1+2\rightarrow 3+4$  elastic scattering of two masses, 
$m_1$ and $m_2$. The observable of interest
is the scattering amplitude, 
$\bra{p_3, p_4}\hat{S}\ket{p_2, p_1} $,
which is
a function of the usual Mandelstam invariants
\begin{equation}
    \begin{aligned}
        s &= -(p_1 +p_2)^2,  \\
        t &= -(p_1 -p_3)^2, \\
        u &= -(p_1 - p_4)^2.
    \end{aligned}
\end{equation}
In the center-of-mass (COM) frame, one can also define the COM energy, $ E_1 +E_2$, and
COM spatial ingoing and outgoing momentum, $\mathbf{p}$ and $\mathbf{p}'$, respectively.
In these variables, the momenta are
\begin{equation}
    \begin{aligned}
    &p_1^\mu = (E_1, \mathbf{p}), \quad
    p_2^\mu = (E_2, -\mathbf{p}), \\
    &p_3^\mu = (E_1, \mathbf{p}' =\mathbf{q} + \mathbf{p}), \quad
    p_4^\mu = (E_2, -\mathbf{p}'),
    \end{aligned}
\end{equation}
where $\mathbf{q}$ is the exchanged spatial momentum. The Mandelstam invariants can also
be written as
\begin{equation}
    |\mathbf{p}|^{2} = 
    \frac{\big[s - (m_{1} + m_{2})^{2}\big]\big[s - (m_{1} - m_{2})^{2}\big]}{4s},
    \qquad
    t = -|\mathbf{p}' - \mathbf{p}|^{2} = -|\mathbf{q}|^{2}.
\end{equation}

The Born regime is defined as a hierarchy of length scales
\cite{Correia:2024jgr},
where the Compton wavelength, $\lambda_{\text{C}}$, is parametrically smaller than 
the rest of the scales, namely the impact parameter, $b$, and the de Broglie wavelength, $\lambda_{\rm dB}$,
of the external particles, which are taken to be of the same order,
\begin{equation}
    b\sim \lambda_{\rm dB} \gg \lambda_{\text{C}}.
\end{equation}
In this regime, the scattering amplitude
can be reorganized into a geometric series of some potential, $V$, and some Green's function, $G$.
Let $T$ denote the connected part of the amplitude, then the structure of the Born amplitude is
\begin{equation}
    T \sim \frac{V}{1-\int GV},
\end{equation}
where the integral is to  be interpreted as a convolution.
The key point is that the dynamics of two body scattering can be seen as an effective particle\footnote{
    The properties of the effective particle are dictated by the form of the Green's function
appearing in the Born series.}
interacting with some potential. There are many ways to achieve this regime, the most common
one being by taking the non-relativistic limit in the COM variables. In this scenario,
the variable, $s$, is evaluated very close to the two-particle threshold, $(m_1+m_2)^2$,
and the effective particle has the reduced mass,
\begin{equation}
    \abs{\mathbf{p}}^2 = 2 m_{\rm red} E, \quad m_{\rm red} = \frac{m_1 m_2}{m_1 + m_2},
\end{equation}
where the COM scattering energy is $E\equiv E_1 +E_2 - m_1-m_2$, which is very
small compared to the mass of the external particles.
The Mandelstam variable, $t = -\abs{\mathbf{q}}^2 =-2\abs{\mathbf{p}}^2(1-\cos\theta)$,
also goes to zero, but the scattering angle, $\theta$, remains fixed. 
An equivalent way to ensure a Born-type regime is instead by
taking a spectral limit, namely the large mass limit, $m_1,m_2 
\rightarrow \infty$, with fixed ratio, and fixed momentum, $\abs{\mathbf{p}}$. 
The interpretation is simple: as long as $\abs{\mathbf{p}}/m$
is small, the amplitude is well approximated by the non-relativistic limit, which implies
a Schr\"odinger-like equation. Depending on the ratios of the masses, the
effective particle can have the reduced mass, or the mass of the light particle.

The case of interest in this work is when only one of the masses is very large, $m_2\rightarrow\infty$. In this
limit, the scattering amplitude behaves as the scattering of the light particle
with a potential generated by the heavy particle. This is a simple consequence of
the fact that the COM frame becomes the rest frame of the heavy particle.
In this case, the energy of the light
particle, $E_1$, can
be either smaller than the rest mass of the light particle or of the same order.
In the former, we again reach the non-relativistic limit, but in the latter we 
arrive at relativistic quantum mechanics,
\begin{equation}
    \abs{\mathbf{p}}^2 = E_1^2 - m_1^2.
\end{equation}
This means we can define a potential
for a Klein--Gordon equation and solve iteratively for the amplitude.
The validity of this expansion relies on the fact that the scattering must remain elastic,
i.e., no particle creation can occur. For this to happen we impose a gap in the
spectrum of the theory, such that, no inelastic excitations contribute 
for the order of energies that we consider.

\section{Heavy limit of internal bulk integrations}
\label{app:heavyprop}
In this appendix we derive the large-mass asymptotics Eq.~\eqref{eq:heavy-bulk-propagator} of the AdS bulk-to-bulk propagator Eq.~\eqref{bulk-to-bulk}, and the composition rule Eq.~\eqref{eq: rule for bulk integrals} for the internal bulk integrations Eq.~\eqref{eq:bulkintegration} used in section~\ref{Loop diagrams}. We proceed in two steps. In section~\ref{sec:adiabatic} we obtain the asymptotics from the DeWitt--Schwinger expansion. Besides the expected geodesic exponential, this derivation isolates the Van Vleck determinant controlling the fluctuations around the classical trajectory. In section~\ref{sec:worldline} we use the worldline representation of the propagator to study its composition under an intermediate bulk integration, and show that the Van Vleck prefactors combine with the Gaussian fluctuations of the intermediate vertex so that the heavy propagation composes along a single classical geodesic. In section~\ref{sec:congruence} we recover the same geometric data from the geodesic deviation equation.

\subsection{The adiabatic expansion}\label{sec:adiabatic}
The asymptotic form of the propagator Eq.~\eqref{bulk-to-bulk} in the heavy limit ($m_{H}\approx\Delta_{H}-\frac{d}{2} \to \infty$) has been studied before in the literature, see for instance \cite{Maxfield:2017rkn}. Instead of just quoting the final result, we derive it, since the derivation is also what simplifies the bulk integrations.

We start by reviewing the DeWitt--Schwinger representation of the Green's function \cite{DeWitt:1964mxt},  $G(x,x')$, 
for a generic spacetime\footnote{For generic spacetimes, we use lower case coordinates, $x^\mu$, so as todistinguish from
embedding space.}, $(M,g_{ab})$,
\begin{equation}
    G(x,x') = \mathcal D^{1/2}(x,x') \int_0^\infty dT \ (4\pi T)^{-\frac{d+1}{2}}
    e^{-m^2 T - \frac{\sigma(x,x')}{2T}} \mathcal F(x,x';T),
\end{equation}
where $\sigma(x,x') = \frac{1}{2}L^2(x,x')$ is Synge's world function, and
$\mathcal D$ is the Van Vleck determinant,
\begin{equation}
    \mathcal D(x,x') = \det\left[
       - \frac{ \partial}{\partial x^a}
        \frac{ \partial}{\partial (x')^b}
    \sigma(x,x') \right]\left[g(x)g(x')\right]^{-1/2},
\end{equation}
with $g(x) \equiv |\det g_{ab}(x)|$ denoting the determinant of the metric tensor. 
The function $\mathcal F(x,x';T)$ admits an adiabatic expansion 
\begin{equation}
    \mathcal F(x,x';T) \approx 
    a_0(x,x')+ 
    a_1(x,x')T +
    a_2(x,x')T^2  + \cdots,
\end{equation}
where $a_0(x,x') \equiv 1$, and the coefficients $a_n$  are strictly suppressed by increasing inverse powers of the AdS radius.\footnote{The  DeWitt coefficients $a_n$ are local geometric invariants constructed from the spacetime curvature. In the maximally symmetric geometry of $AdS_{d+1}$, the only available dimensionful geometric scale is the AdS radius. Consequently, dimensional analysis restricts the curvature corrections to scale as $a_n \propto L_{\rm AdS}^{-2n}$.}

For large $m$ we can now consider the saddle point approximation on the $T$ (Schwinger time) 
integral. The saddle point equation gives
\begin{equation}
    \left.\derivative{}{T}\left(-m^2 T -\frac{\sigma(x,x')}{2T}\right)\right|_{T=T_*} = 0 
    \Leftrightarrow T_* = \frac{L(x,x')}{2m},
\end{equation}
which means the leading behavior of the Green's function in the large mass\footnote{What controls this expansion is the product $mL \rightarrow \infty$.} limit is
\begin{equation}
    G(x,x') \xrightarrow{m\rightarrow \infty}
    \mathcal D^{1/2}(x,x')\frac{e^{-m L(x,x')}}{2 m} \left(\frac{m}{2\pi L(x,x')}\right)^{d/2}. 
\end{equation}
To obtain the explicit AdS propagator, we must use the Van Vleck determinant
\begin{align}
\mathcal D(X,Y)=\left(\frac{L(X,Y)}{\sinh L(X,Y)}\right)^{d}.
\end{align}
Substituting this geometric measure into the general saddle-point result gives the heavy bulk-to-bulk propagator in Eq.~\eqref{eq:heavy-bulk-propagator}. While the classical localization is a well-known feature of the heavy limit, the geometric prefactor is crucial at higher orders to cancel the Gaussian fluctuations of the loop integrals, as we show next.

\subsection{Internal bulk integrations}\label{sec:worldline}
To evaluate the nested bulk integrals over intermediate vertices, we turn to the worldline representation of the Green's function,
\begin{equation}
    G(x,x') = \int_0^\infty dT \ e^{-m^2T}\int_x^{x'} \mathscr{D} x(t) 
    \exp{-\frac{1}{4}\int_0^T dt \  g_{ab}\derivative{x^a}{t}\derivative{x^b}{t}}.
\end{equation}
Although the particle propagates in $d+1$ dimensions, the inner path integral can be interpreted as the Euclidean quantum-mechanical propagator, $K(X_1,X_2;T)$, of a non-relativistic particle in the $(d+2)$-dimensional embedding space constrained to the hyperboloid, where the Schwinger parameter $T$ plays the role of an auxiliary Euclidean time. By comparing this representation with the saddle-point integral, it is easy to see the heavy limit coincides with the short-time form of the propagator, since the saddle is at $T\sim 1/m$. We can therefore identify the short-time limit as
\begin{equation}
    K(x,x',T) = \mathcal D^{1/2}(x,x')(4 \pi T)^{-\frac{d+1}{2}}\exp{-\frac{\sigma(x,x')}{2T}}.
\end{equation}
Now consider the gluing identity of the worldline path integral, which expresses the semigroup (convolution) property of the heat kernel:
\begin{equation}
    K(x,x',T_1+T_2) = \int_M d^{d+1}y \ \sqrt{\abs{g(y)}} K(x,y,T_1) K(x',y,T_2).
\end{equation}
To evaluate this gluing identity in the heavy limit, we use again the saddle-point approximation, which gives the following condition:
\begin{equation}
    \nabla_y \left(
        \frac{\sigma(x,y)}{2 T_1} +
        \frac{\sigma(x',y)}{2 T_2}
    \right) = 0.
\end{equation}

This argument closely mirrors the tree-level analysis of section \ref{Heavy limit of Witten diagrams}. The
gradient of Synge's world function, $\nabla_y \sigma(x,y)$, is a vector of length,
$L(x,y)$, collinear to the geodesic
generator connecting $x$ and $y$, which implies that the saddle is located when $y$ is in the
geodesic connecting $x$ and $x'$. But unlike Eq.~\eqref{boundary_saddle}, we no longer have 
a one dimensional saddle. Instead, we have the point, $y_*$, where the ``velocities'' are the same,
\begin{equation}
    \frac{L(x,y_*)}{T_1} =
    \frac{L(x',y_*)}{T_2}\Rightarrow 
    \frac{\sigma(x,y_*)}{T_1}+
    \frac{\sigma(x',y_*)}{T_2} =
    \frac{\sigma(x,x')}{T_1+ T_2}.
\end{equation}
To evaluate the remaining bulk integral over the intermediate point $y$, we introduce normal coordinates around the localized saddle point $y_{*}$. Expanding the effective action to quadratic order, the path integral evaluates to a standard $(d+1)$-dimensional Gaussian, governed entirely by the determinant of the full $(d+1) \times (d+1)$ bulk Hessian matrix:
\begin{equation}
    \left.\det H \equiv \det\left[
        \nabla_a \nabla_b \left(
        \frac{\sigma(x,y)}{2 T_1} +
    \frac{\sigma(x',y)}{2 T_2}\right)\right|_{y=y_*}
    \right].
    \label{worldline hessian}
\end{equation}
In the strict heavy limit, executing this Gaussian integration over the bulk fluctuations simplifies the gluing relation to a purely multiplicative identity:
\begin{equation}
    \sqrt{\frac{\mathcal D(x,x')}{ (4\pi (T_1+T_2))^{d+1}}} = 
    \sqrt{\frac{\mathcal D(x,y_*)}{ (4\pi T_1)^{d+1}}} 
    \sqrt{\frac{\mathcal D(x',y_*)}{ (4\pi T_2)^{d+1}}} 
    \sqrt{\frac{(2\pi)^{d+1}}{\det H}}.
    \label{multiplicative identity}
\end{equation}
The physical role of the Van Vleck determinant is now manifest: it provides the exact gluing mechanism that ensures intermediate bulk integrations depend exclusively on the external boundary points.
Since, as we have shown, the propagator in the heavy limit is related to the worldline
quantum mechanical propagator by the saddle integral that fixes the Schwinger
time to be proportional to the proper time, this gluing property is preserved. 
 To check this explicitly,
let us first consider the transport equation, that relates the Van Vleck determinant
with the divergence of geodesics,
\begin{equation}
    \square \sigma(x,x') = d+1 - L(x,x')\derivative{\log(\mathcal D(x,x'))}{L(x,x')}.
    \label{eq:transport}
\end{equation}

Using the symmetries of the background, the covariant Hessian of Synge's world function can be decomposed into longitudinal and transverse components along the geodesic as
\begin{equation}
    \nabla_a \nabla_b \sigma = a(L)u_a u_b + b(L) (g_{ab}-u_a u_b), 
\end{equation}
where $u_a \equiv \partial_a L(x,x')$ is the unit tangent vector. By projecting along the longitudinal direction and using the definition of $\sigma$ one can derive that $a(L)=\frac{d^2}{dL^2}(\frac{1}{2}L^2)=1$. 
Similarly, the coefficient $b(L)$ is related to the transverse Hessian acting on the geodesic length
\begin{equation}
    \nabla_a \nabla_b \left(\frac{1}{2}L^2\right) = u_a u_b + L\nabla_a\nabla_b L.
\end{equation}
The second term, $\nabla_a \nabla_b L$, is purely spatial \cite{Wald:1984rg}, in the sense that 
$u^a\nabla_a\nabla_b L = 
u^b\nabla_a\nabla_b L = 0$. Replacing this decomposition into the transport equation establishes 
a direct relation between the Van Vleck determinant and the coefficient $b(L)$,
\begin{equation}
    L\derivative{}{L}\log \mathcal D = d(1-b(L)),
    \label{eq:VV-transport}
\end{equation}
which gives $b(L) = L \coth(L)$. Evaluating the determinant of the saddle-point Hessian $H$, Eq.~(\ref{worldline hessian}), with the previous results gives 
\begin{equation}
    \det H =\frac{1}{2^{d+1}}
        \left(\frac{1}{T_1}+\frac{1}{T_2}\right)
    \left(
        \frac{L(x,x')}{T_1 + T_2} \left(
            \coth L(x,y_*) + \coth L(x',y_*)
        \right)
    \right)^d.
\end{equation}
Using the hyperbolic identity $\coth(x) + \coth(y) = \sinh(x+y)/(\sinh(x)\sinh(y))$, this result precisely reproduces the multiplicative prefactor relation required by the gluing identity in Eq.~(\ref{multiplicative identity}).
As a consistency check, the determinant of the transverse Hessian $H_S$ governing the boundary-to-bulk spatial integration in the tree-level contact diagram (\ref{gaussian transverse integral}) can also be explicitly evaluated using this framework:
\begin{equation}
    \det H_S = \left(\coth L(X_1,X)  +\coth L(X_4,X)\right)^d \xrightarrow{X_i\rightarrow P_i} 2^d.
\end{equation}

We can now return to Eq.~\eqref{eq:bulkintegration} and use the Schwinger representation,
 \begin{equation}
 I[\mathcal R]
=
\int_{0}^{\infty} dT_1\, dT_2\,
e^{-m_H^2(T_1+T_2)}
\int_{\text{AdS}} dX\,
K(X_1,X;T_1)\,\mathcal R(X)\,K(X,X_2;T_2),
     \label{eq:schwinger rep bulk integrals}
 \end{equation}
and compute the AdS integral using Eq.~\eqref{multiplicative identity}. To compute the remaining integrals, introduce
 new variables, $T =T_1 + T_2$ and $\alpha = T_1/T$, such that $dT_1 dT_2 = TdTd\alpha$, and suppose the saddle, $X_*$, depended on the total Schwinger time $T$. Then a common rescaling of $T_1$ and $T_2$, which leaves both the endpoints and the ratio $\alpha$ unchanged, would move the saddle. But such a rescaling only changes the affine parametrization of the same geodesic, not its geometry\footnote{A constant rescaling of an affine parameter
is again an affine parametrization of the same geodesic.}. Therefore, $X_*$ cannot depend on $T$, and must be determined solely by $\alpha$. The integrals then factorize,
\begin{equation}
I[\mathcal R]
\simeq
\int_{0}^{\infty} dT\, T\, e^{-m_H^2 T}
K(X_1,X_2;T)
\int_{0}^{1} d\alpha\, \mathcal R\bigl(X_*(\alpha)\bigr).
\end{equation} 
The $T$ integral is again dominated by the saddle $T_* = L/2m_H$, which gives, at leading order,
\begin{equation}
    I[\mathcal R]
\simeq
\frac{L}{2m_H}
\Pi^{\rm B}_{\Delta_H}(X_1,X_2)
\int_{0}^{1} d\alpha\, \mathcal R\bigl(X_*(\alpha)\bigr).
\end{equation}
Writing $\tau=\alpha L$ for the length parameter along the geodesic $\gamma_{12}$ connecting the two endpoints, this is the composition rule Eq.~\eqref{eq: rule for bulk integrals} of the main text.

\subsection{Expansion of the geodesic congruence}\label{sec:congruence}
The coefficient $b(L)=L\coth L$, and through Eq.~\eqref{eq:VV-transport} the Van Vleck determinant, can also be obtained directly from the geodesic deviation equation. Consider the expansion $\vartheta\equiv h^{ab}\nabla_b u_a$ of the congruence of geodesics emanating from $x'$,
where $h_{ab}$ is the metric transverse to the
geodesic generator $u_a = \nabla_a L(x,x')$. The Jacobi field $\eta^a$ obeys the geodesic
deviation equation
\begin{equation}
    u^a\nabla_a(u^b\nabla_b \eta^c) = -R_{abd}^{\ \ \ \ c}\eta^bu^au^d,
\end{equation}
with the boundary condition $\eta^a(0) = 0$. This allows the
connection \cite{Wald:1984rg} between Jacobi fields, and deviation vectors for
the congruence passing through the fixed point, $x'$. The Riemann tensor for maximally
symmetric spacetimes characterized by a cosmological constant, $\Lambda$, is
\begin{equation}
    R_{abcd} =\frac{ 2\Lambda}{d(d-1)} \left(
        g_{ac}
        g_{bd}-
        g_{ad}
        g_{bc}
    \right),
\end{equation}
which implies the geodesic deviation equation is just
\begin{equation}
    \dv[2]{L}\eta^a = -\frac{ 2\Lambda}{d(d-1)} (u\cdot u) \eta^a,
\end{equation}
where $\dv{L} = u^a\nabla_a$. The differential equation
becomes oscillatory whenever $\Lambda \,(u\cdot u)>0$,
i.e., for Lorentzian AdS (with timelike geodesic) and for Euclidean de Sitter ($\mathbb S^{d+1}$),
where we have conjugate points. For Euclidean AdS (and Lorentzian de Sitter),
the equation becomes exponential, thus complementing the
argument used in section~\ref{Heavy limit of Witten diagrams}, showing the uniqueness of the saddle.
The solution is written linearly, in terms of the initial data,
\begin{equation}
    \eta^a(L) =  \sinh(L) \delta^a_b
    \dv{\eta^b}{L} \left(0\right)
    = A^a_{\ b }
    \dv{\eta^b}{L} \left(0\right).
\end{equation}
The expansion parameter $\vartheta$ can then be computed from the $A^a_{\ b}$ matrix \cite{Wald:1984rg},
\begin{equation}
    \vartheta = \dv{L} (\log\abs{ \det A}) = d\coth L,
\end{equation}
in agreement with the transport equation: since $\square\sigma=\nabla^a(L\nabla_a L)=1+L\vartheta$, Eq.~\eqref{eq:transport} and Eq.~\eqref{eq:VV-transport} give $\vartheta = d\,b(L)/L = d\coth L$.

\section{Scalar Yukawa theory in AdS}
\label{app:yukawa}
We now consider an exchange Witten diagram, by studying three scalar fields
$\phi$, $\chi$ and $\sigma$, dual to the scalar operators 
$\mathcal{O}_L$, $\mathcal{O}_H$ and $\mathcal{O}_\sigma$ with scaling dimensions 
$\Delta_L$, $\Delta_H$ and $\Delta_\sigma$ respectively. Consider the Yukawa interaction Lagrangian
\begin{equation}
    \mathcal L = \sqrt{\abs{g}}\lambda \left(
         \phi^2 \sigma + 
         \chi^2 \sigma  
    \right).
\end{equation}
We would like to consider the 4-point function 
$\langle 
\mathcal O_H(P_1)
\mathcal O_L(P_2)
\mathcal O_L(P_3)
\mathcal O_H(P_4)
\rangle$, and its correction from the tree-level $t$-channel exchange Witten diagram, $\mathcal A_t(P_i)$,
\begin{equation}
    \mathcal A_t(P_i) = \lambda^2\int_{\text{AdS}} dX dY \left(
        \Pi_{\Delta_H}^{\partial} (P_1,Y)
        \Pi_{\Delta_L}^{\partial} (P_2,X)
        \Pi_{\Delta_\sigma}^{B} (Y,X)
        \Pi_{\Delta_L}^{\partial} (P_3,X)
        \Pi_{\Delta_H}^{\partial} (P_4,Y)
    \right).
\end{equation}
After integrating out the heavy field through the procedure of section~\ref{Heavy limit of Witten diagrams},
we are left with the integral
\begin{equation}
    \mathcal A_t(P_i)    =-\frac{\lambda^2}{2\Delta_H} \int_{\text{AdS}} dX \int_{-\infty}^{\infty} dt \left(
        \Pi_{\Delta_L}^{\partial} (P_2,X)
        \Pi_{\Delta_\sigma}^{B} (Y_*(t),X)
        \Pi_{\Delta_L}^{\partial} (P_3,X)
    \right).
\end{equation}
To exhibit the Born series, we must then solve the integral
\begin{equation}
    V(X) \equiv -\frac{\lambda^2}{2\Delta_H} \int_{-\infty}^{\infty} dt \ 
        \Pi_{\Delta_\sigma}^{B} (Y_*(t),X),
\end{equation}
where the embedding-space invariant for this geodesic is just
\begin{equation}
    X\cdot Y_*(t) = -\sqrt{(1+r_1^2)} \cosh(t_1 - t) ,
\end{equation}
where $(r_1,t_1)$ parametrize the point $X$ in global AdS coordinates. The integral
is then
\begin{equation}
    V(X) = 
  -  \frac{\lambda^2 }{2\Delta_H} 
    \frac{2^{-\Delta_\sigma}\mathcal C_{\Delta_\sigma}}{(1+r_1^2)^{\frac{\Delta_\sigma}{2}}}
    \int_{-\infty}^{\infty}dt \frac{
        \ _2F_1 \left(
            \frac{\Delta_\sigma}{2},\frac{\Delta_\sigma+1}{2} ; 1+ \Delta_\sigma - \frac{d}{2}; 
            \frac{1}{(1+r_1^2)\cosh^2(t-t_1)}
        \right)
    }{
        \left(\cosh(t-t_1)\right)^{\Delta_\sigma}
    }.
\end{equation}
The procedure to compute this integral is similar to computing the Fourier transform of the
bulk-to-bulk propagator done in section~\ref{Loop diagrams}, except that the resummed hypergeometric function 
is no longer evaluated at 1. Using the series expansion, the integral we get is
\begin{equation}
    \int_{-\infty}^{\infty}  \frac{dt}{\left(\cosh^2(t-t_1)\right)^{
            \frac{\Delta_\sigma}{2} + n 
    }}
            = \sqrt{\pi}\frac{\Gamma\left(
                    \frac{\Delta_\sigma}{2} + n
            \right)}{
                \Gamma \left(
                    \frac{\Delta_\sigma + 1}{2} + n
                \right)
            }.
\end{equation}
The result is
\begin{equation}
    V(X) = 
    -\frac{\lambda^2 \mathcal C_{\Delta_\sigma}}{2^{1+\Delta_\sigma}\Delta_H}
        \frac{
            \Gamma \left(
                \frac{\Delta_\sigma}{2}
            \right)
    }{
            \Gamma \left(
                \frac{\Delta_\sigma+1}{2}
            \right)
    }
    \frac{\sqrt{\pi}}{(1+r_1^2)^{\frac{\Delta_\sigma}{2}}}
        \ _2F_1 \left(
            \frac{\Delta_\sigma}{2},\frac{\Delta_\sigma}{2} ; 1+ \Delta_\sigma - \frac{d}{2}; 
            \frac{1}{(1+r_1^2)}
        \right).
\end{equation}
This defines the potential that appears in the Born series for the scalar Yukawa theory.
Using a standard transformation of the hypergeometric function, this can be rewritten as
\begin{equation}
V(r) = \frac{c_{\Delta_\sigma}}{r^{\Delta_\sigma}}\;
{}_2F_1\!\left(1+\frac{\Delta_\sigma}{2}-\frac d2,\;\frac{\Delta_\sigma}{2};
\,1+\Delta_\sigma-\frac d2;\,-\frac{1}{r^2}\right),
\end{equation}
where we dropped the indices and gathered all the constants in $c_{\Delta_\sigma}$.
As an illustration, for a massless exchanged field in AdS$_4$, the potential is just
\begin{equation}
    V(r) = c_{\sigma} \left(\frac{1}{r} -  \arccot(r)\right).
\end{equation}
As a consistency check we can also consider the 1-loop correction
to the Born series coming from the box diagram plus the cross diagram.
Let $\mathcal A^{(2)}(P_i) = \mathcal A_{\square}(P_i) +
\mathcal A_{\cross}(P_i)$ denote the correction to the 4-point function coming from the two Witten
diagrams, and where
\begin{equation}
    \begin{aligned}
    \mathcal A_\square(P_i) = \lambda^4\int_{\text{AdS}} dX_1 dX_2 dY_1 dY_2 \Big(
        &\Pi_{\Delta_H}^{\partial} (P_1,Y_1)
        \Pi_{\Delta_H}^{B} (Y_1,Y_2)
        \Pi_{\Delta_H}^{\partial} (P_4,Y_2)\times \\
        &\Pi_{\Delta_L}^{\partial} (P_2,X_1)
        \Pi_{\Delta_L}^{B} (X_1,X_2)
        \Pi_{\Delta_L}^{\partial} (P_3,X_2)\times \\
        &\Pi_{\Delta_\sigma}^{B} (Y_1,X_1)
        \Pi_{\Delta_\sigma}^{B} (Y_2,X_2)
    \Big),
\end{aligned}
\end{equation}
\begin{equation}
    \begin{aligned}
        \mathcal A_{\cross}(P_i) = \lambda^4 \int_{\text{AdS}} dX_1 dX_2 dY_1 dY_2 \Big(
        &\Pi_{\Delta_H}^{\partial} (P_1,Y_1)
        \Pi_{\Delta_H}^{B} (Y_1,Y_2)
        \Pi_{\Delta_H}^{\partial} (P_4,Y_2)\times \\
        &\Pi_{\Delta_L}^{\partial} (P_2,X_1)
        \Pi_{\Delta_L}^{B} (X_1,X_2)
        \Pi_{\Delta_L}^{\partial} (P_3,X_2)\times \\
        &\Pi_{\Delta_\sigma}^{B} (Y_1,X_2)
        \Pi_{\Delta_\sigma}^{B} (Y_2,X_1)
    \Big).
\end{aligned}
\end{equation}
Repeating the steps of section~\ref{Loop diagrams}, we are left with the two integrals
\begin{equation}
    \mathcal I_{\square}=
    \left(\frac{ \lambda^2}{2\Delta_H}\right)^2
    \int dt_1 dt_2 \left(
        \Pi^B_{\Delta_\sigma} (X_1,Y_*(t_1))
        \Pi^B_{\Delta_\sigma} (X_2,Y_*(t_2))
    \right),
\end{equation}
\begin{equation}
    \mathcal I_{\cross}=
    \left(\frac{ \lambda^2}{2\Delta_H}\right)^2
    \int dt_1 dt_2 \left(
        \Pi^B_{\Delta_\sigma} (X_2,Y_*(t_1))
        \Pi^B_{\Delta_\sigma} (X_1,Y_*(t_2))
    \right),
\end{equation}
where the time ordering $t_2>t_1$ is implied. 
Just like in section~\ref{Loop diagrams},
the difference between the two integrals is just the ordering. This means that the 
sum $\mathcal I_M = \mathcal I_{\square} + \mathcal I_{\cross}$ can be written in
terms of a master integral where we just relax the ordering and allow both integrals
to run over the real numbers,
\begin{equation}
    \mathcal I_M = 
    \left(\frac{ \lambda^2}{2\Delta_H}\right)^2
    \int_{-\infty}^{\infty} dt_1\int_{-\infty}^{\infty} dt_2 \left(
        \Pi^B_{\Delta_\sigma} (X_1,Y_*(t_1))
        \Pi^B_{\Delta_\sigma} (X_2,Y_*(t_2))
    \right).
\end{equation}
More importantly, this just means that the integrals are completely factorized and 
we just recover the definition of the potential,
\begin{equation}
    \mathcal I_M = V(X_1) V(X_2),
\end{equation}
which means the sum of the box and cross Witten diagram, in the heavy limit, can be written as
\begin{equation}
    \mathcal A^{(2)}(P_i)=
        \int_{\text{AdS}} dX_1 dX_2 \left(
            \Pi^{\partial}_{\Delta_L}(P_2,X_1)
            V(X_1)
            \Pi^{B}_{\Delta_L}(X_1,X_2)
            V(X_2)
            \Pi^{\partial}_{\Delta_L}(P_3,X_2)
    \right),
\end{equation}
thus exhibiting the second order Born correction for the scalar exchange potential.

\section{CFT data via QM perturbation theory}
\label{app:anom2}
In this appendix, we compute the correction to the anomalous
dimension up to second order in the Born approximation for a contact interaction with derivatives associated with the potential
\begin{equation}
  V = \sum_{\ell = 0}^\infty f_\ell \frac{(-1)^\ell   }{\ell!} D^L \delta^{(d)}(\mathbf{x}) D_L 
  \equiv \sum_{\ell=0}^\infty f_\ell  V_\ell,
\end{equation}
with $f_\ell\propto\bar\mu^{3-d}F_\ell$ for a static response, cf.~ Eq.~\eqref{eq:frequency-space-worldline-potential}. The important quantity we need is the matrix element
\begin{equation}
    \bra{n' \ell'}V_\ell\ket{n,\ell} = 
    \left( D^L \psi_{n',\ell'}^* (0)Y_{\ell',\vec{m}'}(\Omega)\right)
    \left( D_L \psi_{n,\ell} (0)Y_{\ell,\vec{m}}(\Omega)\right),
\end{equation}
where the symmetric trace-free derivatives $D_L$ act diagonally and just extract the overall factor of the wavefunction at the origin. The behavior of the wavefunction near the defect is
\begin{equation}
\psi_{n,\ell} (r) \xrightarrow[r\rightarrow 0]{} N_{n,\ell}^{-1}\,  r^\ell, \qquad N^{-1}_{n,\ell}  = \left[
\frac{
\Gamma(n+\ell+\Delta_L)\,
\Gamma\left(n+\ell+\frac{d}{2}\right)
}{
n!\,
\Gamma\left(n+\Delta_L+1-\frac{d}{2}\right)\,
\Gamma\left(\ell+\frac{d}{2}\right)^2
}
\right]^{1/2},
\end{equation}
in agreement with  Eq.~\eqref{eq:Nnl}. The matrix element is therefore diagonal in $\ell$ and factorizes,
\begin{equation}
    \bra{n',\ell'}V_\ell\ket{n,\ell} =
   -\Lambda_\ell \,
    N^{-1}_{n,\ell}
    N^{-1}_{n',\ell} \,
    \delta_{\ell,\ell'},
\end{equation}
where $\Lambda_\ell$ is proportional to $f_\ell$, the constant of proportionality being fixed by the STF contraction  Eq.~\eqref{eq:STF-angular-momentum-projection} and the addition theorem  Eq.~\eqref{addition theorem}.
We can now compute the CFT data explicitly. The trivial case is the anomalous dimension at first order, which gives
\begin{equation}
    \gamma_{n,\ell}^{(1)} = \Lambda_\ell N^{-2}_{n,\ell}.
    \label{eq:gamma1Lambda}
\end{equation}
Comparing with  Eq.~\eqref{eq:gamma1} fixes the normalization of $\Lambda_\ell$: the $n$-dependence cancels identically and one finds 
\begin{equation}\Lambda_\ell = \frac{2^{\ell-1}\Gamma\left(
\ell + \frac{d}{2}\right)}{\pi^{d/2}}\,\bar\mu^{3-d}F_\ell(\omega_{n,\ell}),
\end{equation}
where the sign convention for $\Lambda_\ell$ follows from  Eq.~\eqref{eq:gammaV}.

The first non-trivial correction is the first order correction to the OPE coefficient. 
In order to compute the series, one first needs to analytically continue in $n$. Consider a continuous parameter, $q$, such that
the series can be written as
\begin{equation}
\sum_{n'\neq n}\frac{2\omega_{n',\ell}k_{n',\ell}N^{-1}_{n',\ell}}{\omega^2_{n,\ell} - \omega^2_{n',\ell}} 
= \lim_{q\rightarrow n} \left[
\mathcal B_\ell (q)
-\frac{2\omega_{n,\ell} k_{n,\ell}N^{-1}_{n,\ell}}{\omega^2(q) - \omega_{n,\ell}^2}
\right],
\end{equation}
where,
\begin{equation}
\mathcal B_\ell(q) \equiv
\sum_{n=0}^{\infty}\frac{2\omega_{n,\ell}k_{n,\ell}N^{-1}_{n,\ell}}{\omega^2(q) - \omega^2_{n,\ell}},
\end{equation}
is reminiscent of a boundary to origin radial Green's function.
Adding and subtracting the divergent $n'=n$ term, we resum the full series in closed form and extract the finite part. Importantly, the divergent piece we added also cancels the first term in  Eq.~\eqref{first order OPE formula}.
This is easy to see by expanding the ratio in partial fractions
\begin{equation}
\frac{2\omega_{n,\ell} }{\omega^2(q) - \omega_{n,\ell}^2}= 
\left(
\frac{1}{\omega(q) - \omega_{n,\ell}} - 
\frac{1}{\omega(q) + \omega_{n,\ell}} 
\right),
\end{equation}
where the latter term admits a smooth limit when $q\rightarrow n$.
It is also useful to write $\omega (q) = \Delta_L+2q + \ell$. This isolates the $n$-dependence in Pochhammer symbols,
\begin{align}
\frac{2\omega_{n,\ell}}{\omega^2(q) - \omega_{n,\ell}^2} &= 
\frac{1}{2}\frac{\Delta_L+\ell+2n}{(q-n)(\Delta_L+\ell+q+n)}  \nonumber \\
&=\frac{1}{2}\frac{\Delta_L+\ell}{q(\Delta_L+\ell+q)}
\frac{
\left(1+\frac{\Delta_L+\ell}{2}\right)_n
}{
\left(\frac{\Delta_L+\ell}{2}\right)_n
}
\frac{
(-q)_n
}{
(1-q)_n
}
\frac{
(\Delta_L+\ell+q)_n
}{
(\Delta_L+\ell+q+1)_n
}.
\end{align}
Using the Pochhammer representation of the ratio of Gamma functions appearing in $N_{n,\ell}$, the full $n$-dependence can be reorganized as a ratio of Pochhammer symbols (with a $(-1)^n/n!$).
The series then resums into a hypergeometric function at argument $-1$,
\begin{equation}
\mathcal B_\ell(q)
=
\frac{c_\ell}{2}\,
\frac{\Delta_L+\ell}{q\,(\Delta_L+\ell+q)}\;
{}_4F_3\!\left(
\begin{matrix}
\Delta_L+\ell,\;\; 1+\frac{\Delta_L+\ell}{2},\;\; -q,\;\; \Delta_L+\ell+q
\\[3pt]
\frac{\Delta_L+\ell}{2},\;\; 1-q,\;\; \Delta_L+\ell+q+1
\end{matrix}
\,;\,-1\right),
\label{eq:B4F3}
\end{equation}
where the prefactor is 
\begin{equation}
  c_\ell =    \frac{\Gamma(\Delta_L+\ell)}
         {\Gamma\left(\ell+\frac d2\right)\Gamma\left(\Delta_L+1-\frac d2\right)}.
\end{equation}
This is a very-well-poised ${}_4F_3$, and can be written as a ratio of Gamma functions
\begin{equation}
\mathcal B_\ell(q)
=
-\frac{1}{2}
\frac{
\Gamma(\Delta_L+\ell+q)\Gamma(-q)
}{
\Gamma\left(\ell+\frac{d}{2}\right)
\Gamma\left(\Delta_L+1-\frac{d}{2}\right)
}.
\end{equation}
The behavior near the poles is
\begin{equation}
    \mathcal B_\ell (n+\delta) =
    \frac{(-1)^n}{2}
    \frac{\Gamma(\Delta_L+\ell + n)}{
    \Gamma\left(\Delta_L+1-\frac{d}{2}\right)
    \Gamma\left(\ell+\frac{d}{2}\right)
    n!
    }\left(
    \frac{1}{\delta} + \psi(\Delta_L+\ell+n) -\psi(n+1)
    \right),
\end{equation}
leading to the result
\begin{equation}
    a^{(1)}_{n,\ell} = 
    a^{(0)}_{n,\ell} 
    \gamma^{(1)}_{n,\ell} \left[
    \psi(\Delta_L+\ell+n) -\psi(n+1) 
    \right].
\end{equation}
Importantly, this can be recovered from the derivative relation \cite{Heemskerk:2009pn},
\begin{equation}
    a^{(1)}_{n,\ell} =  \frac{1}{2} \partial_n\left( 
    a^{(0)}_{n,\ell} 
    \gamma^{(1)}_{n,\ell}
    \right).
    \label{eq:derivrel}
\end{equation}
This equation follows from the meromorphic structure of the correlator, but in this formalism it emerges as a non-trivial relation
between off-diagonal and diagonal matrix elements of the potential in AdS.

The second-order anomalous dimension is given by
\begin{equation}
    \gamma_{n,\ell}^{(2)} = \Lambda^2_\ell N^{-2}_{n,\ell}
    \sum_{k\neq n}\frac{2\omega_{k,\ell}N^{-2}_{k,\ell}}{\omega^2_{n,\ell} - \omega^2_{k,\ell}} 
    -\Lambda_\ell^2 \frac{N^{-4}_{n,\ell}}{2\omega_{n,\ell}}.
    \label{second order pt}
\end{equation}
The derivation follows the same strategy as the one used for the first correction to the OPE coefficient.
We begin by defining the resolvent
\begin{equation}
    \mathcal R_\ell(q) = 
    \sum_{k=0}^\infty\frac{2\omega_{k,\ell}N^{-2}_{k,\ell}}{\omega^2(q) - \omega^2_{k,\ell}} ,
\end{equation}
which is a hypergeometric series,
\begin{align}
\mathcal R_\ell(q) &=  \mathcal M_{\ell}(q)
\sum_{k=0}^{\infty}
\frac{
(\Delta_L+\ell)_k
\left(1+\frac{\Delta_L+\ell}{2}\right)_k
\left(\ell + \frac{d}{2}\right)_k
(-q)_k
(\Delta_L+\ell+q)_k
}{
\left(\frac{\Delta_L+\ell}{2}\right)_k
\left(\Delta_L+1-\frac{d}{2}\right)_k
(1-q)_k
(\Delta_L+\ell+q+1)_k
}
\frac{1}{k!} \nonumber \\
&=\mathcal M_{\ell}(q)
{}_5F_4\!\left(
\begin{matrix}
\Delta_L+\ell,\;\; 1+\frac{\Delta_L+\ell}{2},\;\; \ell+\frac d2,\;\; -q,\;\; \Delta_L+\ell+q
\\[3pt]
\frac{\Delta_L+\ell}{2},\;\; \Delta_L+1-\frac d2,\;\; 1-q,\;\; \Delta_L+\ell+q+1
\end{matrix}
\,;\,1\right),
\end{align}
with the prefactor
\begin{equation}
    \mathcal M_{\ell}(q)
=
\frac{1}{2}\,
\frac{\Delta_L+\ell}{q(\Delta_L+\ell+q)}
\frac{
\Gamma(\Delta_L+\ell)
}{
\Gamma\left(\Delta_L+1-\frac{d}{2}\right)
\Gamma\left(\ell+\frac{d}{2}\right)
}.
\end{equation}
This is a very-well-poised ${}_5F_4$, and can be written as a ratio of Gamma functions
\begin{equation}
{}_5F_4(1) 
=
\frac{
\Gamma\!\left(\Delta_L+1-\frac{d}{2}\right)
\Gamma\!\left(1+\Delta_L+\ell+q\right)
\Gamma(1-q)
\Gamma\!\left(1-\ell-\frac{d}{2}\right)
}{
\Gamma(1+\Delta_L+\ell)
\Gamma\!\left(\Delta_L+1-\frac{d}{2}+q\right)
\Gamma\!\left(1-\ell-\frac{d}{2}-q\right)
}.
\end{equation}
We note that the domain of convergence is $\operatorname{Re}\left(\ell+\frac{d}{2}\right)<1$. We again stress that dimensional regularization is essential here. Let now $q= n+ \delta$ and
expand around $\delta =0$. The pole in $\delta$ is the $k=n$ term excluded from the sum in  Eq.~\eqref{second order pt}. Its finite remainder, $-N^{-2}_{n,\ell}/(2\omega_{n,\ell})$, cancels the last term of  Eq.~\eqref{second order pt}, leaving a finite result. To see this explicitly, we first simplify the result using the properties of Gamma functions to write
\begin{equation}
    \mathcal R_\ell(q) = -\frac{1}{2}
\frac{
\Gamma\left(1-\ell-\frac{d}{2}\right)
}{
\Gamma\left(\ell+\frac{d}{2}\right)
}
\frac{
\Gamma(\Delta_L+\ell+q)\Gamma(-q)
}{
\Gamma\left(\Delta_L+1-\frac{d}{2}+q\right)
\Gamma\left(1-\ell-\frac{d}{2}-q\right)
}.
\end{equation}
Expanding then gives
\begin{equation}
\mathcal R_\ell(n+\delta)=
\frac{N_{n,\ell}^{-2}}{2\delta}
+
\frac{N_{n,\ell}^{-2}}{2}\,
\mathcal S_{n,\ell}
+
O(\delta),
\end{equation}
where the finite part is
\begin{align}
\mathcal S_{n,\ell}
&\equiv
\psi(\Delta_L+\ell+n)
-\psi\left(\Delta_L+1-\frac{d}{2}+n\right)
\nonumber\\
&
+\psi\left(1-\ell-\frac{d}{2}-n\right)
-\psi(n+1).
\end{align}
Rewriting the digamma functions in terms of (analytically continued) harmonic numbers, we get 
\begin{equation}
\gamma_{n,\ell}^{(2)}=
\frac{
\left(\gamma_{n,\ell}^{(1)}\right)^2}{2}
\left[
H_{\Delta_L+\ell+n-1}
+
H_{\frac d2+\ell+n-1}
-
H_n
-
H_{\Delta_L-\frac d2+n}
+
\pi \cot\!\left(\frac{\pi d}{2}\right)
\right],
\end{equation}
in agreement with  Eq.~\eqref{second energy correction}.

\section{Relation to non-relativistic two-body limit}
\label{app:nrel}
In the non-relativistic limit, defined by taking both scaling dimensions to be large (with fixed ratio, and an inverse suppression of the cross ratios), the two-body problem is turned into an effective one-body problem with reduced mass $m_{\rm red} = m_1 m_2/(m_1+m_2)$ by choosing center-of-mass coordinates. This also induces a factorization of the two-particle space of primaries,
and we can achieve a factorization in terms of an auxiliary Verma module of a scalar with dimension $\Delta_{\text{rel}} = m_{\rm red} + \frac{d}{2}$, and a COM primary with dimension $\Delta_\text{COM} = \Delta_1+\Delta_2-m_{\rm red}-\frac{d}{2}$.
It is important to note that, in this limit, the Hamiltonian factorizes into
a reduced mass term,
plus a non-relativistic Hamiltonian of the harmonic oscillator furnishing the quantized states with
energies, $E = \frac{d}{2} + 2n + \ell$. 
The interpretation that imposing the center of mass state to be a primary equates to choosing center-of-mass coordinates is maintained in the heavy--light setup. It becomes a simple statement that, in the heavy limit, the center-of-mass frame is the rest frame of the heavy particle.

Indeed, one can generically ask when the two-particle space of primaries can be written as
\begin{equation}
 \ket{\left[\mathcal O_1 \mathcal O_2\right]_{n,\ell}} \approx
 \ket{\mathcal O_\text{COM}} \otimes 
 \left( P_{\langle\mu_1} \cdots P_{\mu_\ell\rangle} \left(P^2)^n\right)\ket{\mathcal O_\text{rel}}\right),
\end{equation}
for some effective operators $\mathcal O_{\text{COM}}$ and $\mathcal O_{\text{rel}}$. The idea is that,
in this new basis, the COM momenta, $P_{\text{COM}}$, generates the descendants for a given primary, $(n,\ell)$, and the relative momenta, $P_{\text{rel}}$, jumps between the Verma modules,
\begin{equation}
    P^{\text{rel}}_{\mu}
 \ket{\left[\mathcal O_1 \mathcal O_2\right]_{n,\ell}} =
 \ket{\left[\mathcal O_1 \mathcal O_2\right]_{n,\ell+1}}.
\end{equation}
To see this explicitly we consider the $n=0$ case, for which the primaries are known in closed form for generic $\ell$,
\begin{equation}
 \ket{\left[\mathcal O_1 \mathcal O_2\right]_{0,\ell}} =
\sum_{k=0}^{\ell} b_\ell (k) 
\left(P_{\langle\mu_1}\cdots P_{\mu_k}  \ket{\mathcal O_1}\right)
\otimes
\left(P_{\mu_{k+1}}\cdots P_{\mu_\ell\rangle}  \ket{\mathcal O_2}\right),
\end{equation}
where the coefficient $b_\ell(k)$ is fixed up to a normalization
\begin{equation}
    b_\ell(k) = \frac{(-1)^k}{\Gamma(k+1) \Gamma(\ell-k+1)\Gamma(\Delta_1+k)\Gamma(\Delta_2+\ell-k)}.
\end{equation}
 Taking the non-relativistic limit amounts to applying Stirling's formula to the gamma functions (with large $\Delta_i$),
 \begin{equation}
     b_\ell(k) \xrightarrow[\Delta_i\rightarrow\infty]{}
     \frac{1}{
    \Gamma( \Delta_1)
    \Gamma( \Delta_2)
     }
\binom{\ell}{k} 
     \frac{(-1)^k}{
( \Delta_1)^k
    ( \Delta_2)^{\ell-k}
     },
 \end{equation}
 where we multiplied the coefficient by $\ell!$ to make the combinatorics manifest. Defining the
 one-particle-momenta $P^{(i)}$ as $P^{(1)} \equiv P\otimes\mathbb I$ (and vice versa), we can recognize the sum as the binomial expansion
\begin{equation}
 \ket{\left[\mathcal O_1 \mathcal O_2\right]_{0,\ell}} 
 = \mathcal N_{1,2,\ell} 
 \left(\frac{P^{(2)}}{\Delta_2}- \frac{P^{(1)}}{\Delta_1}\right)^\ell 
 \ket{\left[\mathcal O_1 \mathcal O_2\right]_{0,0}} +\mathcal O(\Delta_i^{-1}),
\end{equation}
for some normalization factor, $\mathcal N_{1,2,\ell}$. The powers of $\ell$ are to be interpreted as products of momentum with different indices (and taking into account the trace subtraction so as to furnish an irreducible representation of $SO(d)$).
It is therefore natural to define the center-of-mass and relative momenta,
\begin{equation}
    P_{\mathrm{COM}}
    \equiv P^{(1)}+P^{(2)},
\qquad
    P_{\mathrm{rel}}
    \equiv
    \frac{\Delta_2 P^{(1)}-\Delta_1 P^{(2)}}
    {\Delta_1+\Delta_2},
\end{equation}
with inverse transformation,
\begin{equation}
    P^{(1)}
    =
    \frac{\Delta_1}{\Delta_1+\Delta_2}P_{\mathrm{COM}}
    +
    P_{\mathrm{rel}},
    \qquad
    P^{(2)}
    =
    \frac{\Delta_2}{\Delta_1+\Delta_2}P_{\mathrm{COM}}
    -
    P_{\mathrm{rel}},
\end{equation}
very much like the standard two-body change of variables, with the scaling dimensions playing the role of masses.
The dimensions of the primaries in the new basis are not separately fixed by the original CFT data. The total spectrum should be the same,
\begin{equation}
    \Delta_{\rm COM}
    =
    \Delta_1+\Delta_2-\Delta_{\rm rel},
\end{equation}
and the scaling dimensions of the relative descendants are
\begin{equation}
    \Delta_{\rm rel}(n,\ell)
    =
    \Delta_{\rm rel}+2n+\ell,
\end{equation}
so that the full state has
\begin{equation}
    \Delta_{\rm COM}+\Delta_{\rm rel}(n,\ell)
    =
    \Delta_1+\Delta_2+2n+\ell,
\end{equation}
as required by the MFT double-trace spectrum. The space of primaries can then be identified with the Verma module of the relative state, thus proving the factorization.

The Born regime can then be defined purely in terms of CFT arguments. In a holographic CFT, it is exactly the limit where the $s$--channel contribution is dominated by the double-twist operators, which admit a factorization in terms of a new basis of COM primaries and relative descendants. If we choose $\Delta_{\text{rel}}$ dual to the reduced mass, $m_{\rm red}^2 = \Delta_{\text{rel}}(\Delta_{\text{rel}}-d)$, we recover both the non-relativistic two-body problem, when both operators have large dimensions,
$\Delta_\text{rel}= m_{\rm red} + \frac{d}{2}+\mathcal O(\Delta_i^{-1})$,
and also the heavy--light setup $\Delta_{\text{rel}} = \Delta_L + \mathcal O(\Delta_H^{-1})$.
We interpret this as an analogue of the effective-one-body formalism defined in the CFT.

\section{Heavy limit for single graviton exchange}
\label{sec:graviton_heavy}

In this section we discuss the heavy limit of the graviton-exchange Mellin amplitude
computed in \cite{Costa:2014kfa}. To translate their conventions to ours,
denote their quantities by a subscript $C$ and identify the boundary
points as
\begin{equation}
(P_1^C,P_2^C,P_3^C,P_4^C)=(P_1,P_4,P_2,P_3)\,.
\end{equation}
Their cross ratios are therefore
\begin{equation}
u_C=\frac{P_{14}P_{23}}{P_{12}P_{34}}=\frac vu\,,
\qquad
v_C=\frac{P_{13}P_{24}}{P_{12}P_{34}}=\frac1u\,,
\end{equation}
and their reduced correlator is defined by
\begin{equation}
\left\langle \mathcal O_H(P_1)\mathcal O_L(P_2)\mathcal O_L(P_3)\mathcal O_H(P_4)\right\rangle
=\frac{\mathcal G_C(u_C,v_C)}{P_{14}^{\Delta_H}P_{23}^{\Delta_L}}\,.
\end{equation}
Comparing with our prefactor in Eq.~\eqref{eq:redGuv}, we obtain
\begin{equation}
\mathcal G(u,v)
=\frac{u^{(\Delta_H+\Delta_L)/2}}{v^{\Delta_L}}\,
\mathcal G_C\!\left(\frac vu,\frac1u\right),
\label{eq:CGP_correlator_map}
\end{equation}
and matching the powers of the cross ratios in the two Mellin
representations gives
\begin{equation}
t_C=t\,,\qquad s_C=s-\Delta_H-\Delta_L=\sigma-\Delta_L\,.
\label{eq:CGP_mellin_map}
\end{equation}
Writing $h=d/2$, the graviton-exchange Mellin amplitude consists of a
sum over stress-tensor exchange poles, $t_C=d-2+2m$, and a term
$R(s_C)$ regular in $t_C$,
\begin{equation}
M^{C}_{\rm grav}(s_C,t_C)
=C_{\mathcal O_H\mathcal O_HT}\,C_{\mathcal O_L\mathcal O_LT}
\sum_{m=0}^{\infty}\frac{Q_{2,m}(s_C)}{t_C-2h+2-2m}
+R(s_C)\,.
\label{eq:graviton_mellin_CGP}
\end{equation}
At fixed $s_C$ and descendant level $m$, the dependence of $Q_{2,m}$ on
$\Delta_H$ is confined to a single gamma function,
\begin{equation}
Q_{2,m}(s_C)=\frac{\widehat Q_{2,m}(s_C)}{\Gamma(\Delta_H+1-h-m)}\,,
\end{equation}
where
\begin{equation}
\widehat Q_{2,m}(s_C)=
\frac{(1-2h)\,h\,\Gamma(2h+2)\,P_{ih,2}(s_C,2h-2+2m)}
{4\,m!\,\Gamma^4(h+1)\,(h+1)_m\,\Gamma(\Delta_L+1-h-m)}\,,
\label{eq:Q_heavy_scaling}
\end{equation}
and $P_{ih,2}$ is the Mack polynomial defined in \cite{Costa:2014kfa},
so that $Q_{2,m}\simeq\widehat Q_{2,m}\,\Delta_H^{\,h+m-1}/\Gamma(\Delta_H)$
at large $\Delta_H$. The regular contribution scales as
\begin{equation}
R(s_C)=
-\frac{2\pi^{1-h}G_N\,(s_C+\Delta_L)}
{\Gamma(\Delta_L)\,\Gamma(\Delta_L+1-h)}\,
\frac{\Delta_H^{\Delta_L}}{\Gamma(\Delta_H)}
\left[1+\mathcal O(\Delta_H^{-1})\right].
\label{eq:R_heavy_scaling}
\end{equation}
These scalings should not be compared directly since these two terms contribute
through different poles of the full Mellin integrand and must be combined
with the gamma function $\Gamma(\Delta_H-t/2)$, where $t$ is evaluated at some poles. These poles are related to
the operators appearing in the OPE of the external operators. In the
$t$-channel there are three families: the stress-tensor poles,
light double-twist operators,
$[\mathcal O_L\mathcal O_L]_{n,\ell}$, and heavy
double twists $[\mathcal O_H\mathcal O_H]_{n,\ell}$, which are pushed to
infinity in the heavy limit and play no role in what follows.

At a stress-tensor pole, $t=d-2+2m$, the dependence on $\Delta_H$ cancels
exactly,
\begin{equation}
\Gamma\!\left(\Delta_H-\frac t2\right)Q_{2,m}(s_C)\Big|_{t=d-2+2m}
=\widehat Q_{2,m}(s_C)\,,
\label{eq:Tpole_exact}
\end{equation}
and the remaining OPE-coefficient product scales as
$C_{\mathcal O_H\mathcal O_HT}C_{\mathcal O_L\mathcal O_LT}\sim G_N\Delta_H$,
so every descendant level of the stress tensor survives, at order
$\mu\sim G_N\Delta_H$, when $\mu$ is held fixed. This is the manifestation of the fact the conformal block in the t-channel does not depend on the external dimensions, as discussed in section \ref{sec:CB decomp}.

The second family of poles, at $t=2\Delta_L+2n$, is where the light
double-twist operators enter, and here the power counting must be read
with care, since they may coincide with descendants of the stress tensor. Multiplying Eq.~\eqref{eq:Q_heavy_scaling} and
\eqref{eq:R_heavy_scaling} by the gamma at the pole,
$\Gamma(\Delta_H-\Delta_L-n)\simeq\Gamma(\Delta_H)\Delta_H^{-\Delta_L-n}$,
and writing $N\equiv\Delta_L-h$,
\begin{equation}
C_{\mathcal O_H\mathcal O_HT}C_{\mathcal O_L\mathcal O_LT}\,
\Gamma(\Delta_H-\Delta_L-n)\,Q_{2,m}\sim\mu\,\Delta_H^{\,m-N-n-1},
\label{eq:LL_naive}
\end{equation}
\begin{equation}
\Gamma(\Delta_H-\Delta_L-n)\,R\sim\mu\,\Delta_H^{-n-1},
\end{equation}
from which we conclude that the regular part, $R$, is suppressed for every $n$. If we were to take the heavy limit before computing the series, each term in Eq.~(\ref{eq:graviton_mellin_CGP}), given in Eq.~(\ref{eq:LL_naive}),
grows in $\Delta_H$ with increasing powers in the descendant label, $m$, implying the series does not commute with the limit. 

When $N$ is a non-negative integer the sum over $m$
terminates at $m=N$, because $1/\Gamma(\Delta_L+1-h-m)$, in Eq.~(\ref{eq:Q_heavy_scaling}), vanishes beyond
it, and the residues at $t=2\Delta_L+2n$ are of order
$\mu\Delta_H^{-n-1}$. For those poles the twist $2\Delta_L+2n$ is
also that of the stress-tensor descendants at level $m=N+n+1$, so the
residue at that pole is the sum of the two contributions. The stress-tensor
block is independent of $\Delta_H$ and carries the coefficient
$C_{\mathcal O_H\mathcal O_HT}C_{\mathcal O_L\mathcal O_LT}\propto\mu$,
so its descendants at this twist contribute at order $\mu$.
In order for the residue to be of order $\mu\Delta_H^{-n-1}$,
the
double-twist contribution must cancel them to that order. The light
double twists are therefore present with coefficients of order $\mu$,
equal to minus those of the stress-tensor descendants beyond the
truncation up to corrections of order $\mu \Delta_H^{-n-1}$, and
fixed entirely by the stress-tensor data (since the stress tensor OPE coefficient is the only input).
When $N$ is not an integer, there is no truncation. We conjecture that the contribution of the light double twist operators is still leading in $\mu$.

\bibliographystyle{JHEP}
\bibliography{bib}

@article{Correia:2026utp,
    author = "Correia, Miguel and Isabella, Giulia and Wolz, Anna M.",
    title = "{Gravitational Compton Amplitude to All Orders in Perturbation Theory}",
    eprint = "2608.26284",
    archivePrefix = "arXiv",
    primaryClass = "hep-th",
    month = "8",
    year = "2026"
}

@article{Dodelson:2022yvn,
    author = "Dodelson, Matthew and Grassi, Alba and Iossa, Cristoforo and Panea Lichtig, Daniel and Zhiboedov, Alexander",
    title = "{Holographic thermal correlators from supersymmetric instantons}",
    eprint = "2206.07720",
    archivePrefix = "arXiv",
    primaryClass = "hep-th",
    reportNumber = "CERN-TH-2022-095",
    doi = "10.21468/SciPostPhys.14.5.116",
    journal = "SciPost Phys.",
    volume = "14",
    number = "5",
    pages = "116",
    year = "2023"
}

@article{Ivanov:2024sds,
    author = "Ivanov, Mikhail M. and Li, Yue-Zhou and Parra-Martinez, Julio and Zhou, Zihan",
    title = "{Gravitational Raman Scattering in Effective Field Theory: A Scalar Tidal Matching at O(G3)}",
    eprint = "2401.08752",
    archivePrefix = "arXiv",
    primaryClass = "hep-th",
    reportNumber = "MIT-CTP/5664",
    doi = "10.1103/PhysRevLett.132.131401",
    journal = "Phys. Rev. Lett.",
    volume = "132",
    number = "13",
    pages = "131401",
    year = "2024",
    note = "[Erratum: Phys.Rev.Lett. 134, 159901 (2025)]"
}

@article{Apostolidis:2026qsg,
    author = "Apostolidis, Thomas and De Luca, Valerio and Gualtieri, Leonardo and Katagiri, Takuya and Pani, Paolo and Santoni, Luca",
    title = "{Dynamical tidal response of neutron stars: From effective field theory to gravitational waveforms}",
    eprint = "2606.19446",
    archivePrefix = "arXiv",
    primaryClass = "gr-qc",
    month = "6",
    year = "2026"
}

@article{Combaluzier--Szteinsznaider:2025eoc,
    author = "Combaluzier--Szteinsznaider, Oscar and Glazer, Daniel and Joyce, Austin and Rodriguez, Maria J. and Santoni, Luca",
    title = "{Dynamical tidal response of Schwarzschild Black Holes}",
    eprint = "2511.02372",
    archivePrefix = "arXiv",
    primaryClass = "gr-qc",
    doi = "10.1007/JHEP06(2026)032",
    journal = "JHEP",
    volume = "06",
    pages = "032",
    year = "2026"
}

@article{Chang:2026eti,
    author = "Chang, Chih-Hao and Shen, Chia-Hsien and Zhou, Zihan",
    title = "{Gravitational Sommerfeld Effects: Formalism, Renormalization, and Perturbation to $O(G^{10})$}",
    eprint = "2604.14112",
    archivePrefix = "arXiv",
    primaryClass = "hep-th",
    month = "4",
    year = "2026"
}

@article{Karlsson:2019dbd,
    author = "Karlsson, Robin and Kulaxizi, Manuela and Parnachev, Andrei and Tadi{\'c}, Petar",
    title = "{Leading Multi-Stress Tensors and Conformal Bootstrap}",
    eprint = "1909.05775",
    archivePrefix = "arXiv",
    primaryClass = "hep-th",
    doi = "10.1007/JHEP01(2020)076",
    journal = "JHEP",
    volume = "01",
    pages = "076",
    year = "2020"
}

@article{Fitzpatrick:2010zm,
    author = "Fitzpatrick, A. Liam and Katz, Emanuel and Poland, David and Simmons-Duffin, David",
    title = "{Effective Conformal Theory and the Flat-Space Limit of AdS}",
    eprint = "1007.2412",
    archivePrefix = "arXiv",
    primaryClass = "hep-th",
    reportNumber = "BUHET-07-14-10",
    doi = "10.1007/JHEP07(2011)023",
    journal = "JHEP",
    volume = "07",
    pages = "023",
    year = "2011"
}

@article{Gubser:1998bc,
    author = "Gubser, S. S. and Klebanov, Igor R. and Polyakov, Alexander M.",
    title = "{Gauge theory correlators from noncritical string theory}",
    eprint = "hep-th/9802109",
    archivePrefix = "arXiv",
    reportNumber = "PUPT-1767",
    doi = "10.1016/S0370-2693(98)00377-3",
    journal = "Phys. Lett. B",
    volume = "428",
    pages = "105--114",
    year = "1998"
}

@article{Fitzpatrick:2015zha,
    author = "Fitzpatrick, A. Liam and Kaplan, Jared and Walters, Matthew T.",
    title = "{Virasoro Conformal Blocks and Thermality from Classical Background Fields}",
    eprint = "1501.05315",
    archivePrefix = "arXiv",
    primaryClass = "hep-th",
    doi = "10.1007/JHEP11(2015)200",
    journal = "JHEP",
    volume = "11",
    pages = "200",
    year = "2015"
}

@article{Goldberger:2005cd,
    author = "Goldberger, Walter D. and Rothstein, Ira Z.",
    title = "{Dissipative effects in the worldline approach to black hole dynamics}",
    eprint = "hep-th/0511133",
    archivePrefix = "arXiv",
    doi = "10.1103/PhysRevD.73.104030",
    journal = "Phys. Rev. D",
    volume = "73",
    pages = "104030",
    year = "2006"
}

@article{Goldberger:2020fot,
    author = "Goldberger, Walter D. and Li, Jingping and Rothstein, Ira Z.",
    title = "{Non-conservative effects on spinning black holes from world-line effective field theory}",
    eprint = "2012.14869",
    archivePrefix = "arXiv",
    primaryClass = "hep-th",
    doi = "10.1007/JHEP06(2021)053",
    journal = "JHEP",
    volume = "06",
    pages = "053",
    year = "2021"
}

@book{Wald:1984rg,
    author = "Wald, Robert M.",
    title = "{General Relativity}",
    doi = "10.7208/chicago/9780226870373.001.0001",
    publisher = "Chicago Univ. Pr.",
    address = "Chicago, USA",
    year = "1984"
}

@article{Cornalba:2007zb,
    author = "Cornalba, Lorenzo and Costa, Miguel S. and Penedones, Joao",
    title = "{Eikonal approximation in AdS/CFT: Resumming the gravitational loop expansion}",
    eprint = "0707.0120",
    archivePrefix = "arXiv",
    primaryClass = "hep-th",
    reportNumber = "ROM2F-2007-11, LPTENS-07-27",
    doi = "10.1088/1126-6708/2007/09/037",
    journal = "JHEP",
    volume = "09",
    pages = "037",
    year = "2007"
}

@article{Cornalba:2006xk,
    author = "Cornalba, Lorenzo and Costa, Miguel S. and Penedones, Joao and Schiappa, Ricardo",
    title = "{Eikonal Approximation in AdS/CFT: From Shock Waves to Four-Point Functions}",
    eprint = "hep-th/0611122",
    archivePrefix = "arXiv",
    reportNumber = "ROM2F-2006-25, LPTENS-06-49, CERN-PH-TH-2006-233",
    doi = "10.1088/1126-6708/2007/08/019",
    journal = "JHEP",
    volume = "08",
    pages = "019",
    year = "2007"
}

@article{Jafferis:2017zna,
    author = "Jafferis, Daniel and Mukhametzhanov, Baur and Zhiboedov, Alexander",
    title = "{Conformal Bootstrap At Large Charge}",
    eprint = "1710.11161",
    archivePrefix = "arXiv",
    primaryClass = "hep-th",
    doi = "10.1007/JHEP05(2018)043",
    journal = "JHEP",
    volume = "05",
    pages = "043",
    year = "2018"
}

@article{Maxfield:2022hkd,
    author = "Maxfield, Henry and Zahraee, Zahra",
    title = "{Holographic solar systems and hydrogen atoms: non-relativistic physics in AdS and its CFT dual}",
    eprint = "2207.00606",
    archivePrefix = "arXiv",
    primaryClass = "hep-th",
    doi = "10.1007/JHEP11(2022)093",
    journal = "JHEP",
    volume = "11",
    pages = "093",
    year = "2022"
}

@article{Maxfield:2022nat,
    author = "Maxfield, Henry and Zahraee, Zahra",
    title = "{Lorentzian inversion in non-relativistic and classical limits}",
    eprint = "2210.12147",
    archivePrefix = "arXiv",
    primaryClass = "hep-th",
    month = "10",
    year = "2022"
}

@book{Gradshteyn:1943cpj,
    author = "Gradshteyn, I. S. and Ryzhik, I. M.",
    title = "{Table of Integrals, Series, and Products}",
    isbn = "978-0-12-294757-5, 978-0-12-294757-5",
    year = "1943"
}

@article{Maxfield:2017rkn,
    author = "Maxfield, Henry",
    title = "{A view of the bulk from the worldline}",
    eprint = "1712.00885",
    archivePrefix = "arXiv",
    primaryClass = "hep-th",
    month = "12",
    year = "2017"
}

@article{Nastase:2007kj,
    author = "Nastase, Horatiu",
    title = "{Introduction to AdS-CFT}",
    eprint = "0712.0689",
    archivePrefix = "arXiv",
    primaryClass = "hep-th",
    reportNumber = "TIT-HEP-578",
    month = "12",
    year = "2007"
}

@article{Mack:2009mi,
    author = "Mack, Gerhard",
    title = "{D-independent representation of Conformal Field Theories in D dimensions via transformation to auxiliary Dual Resonance Models. Scalar amplitudes}",
    eprint = "0907.2407",
    archivePrefix = "arXiv",
    primaryClass = "hep-th",
    month = "7",
    year = "2009"
}

@misc{OsbornCFTNotes,
  author       = {Osborn, Hugh},
  title        = {Lectures on Conformal Field Theories in more than two dimensions},
  year         = {2025},
  month        = feb,
  howpublished = {Lecture notes},
  note         = {Available at \url{https://www.damtp.cam.ac.uk/user/ho10/CFTNotes.pdf}},
}

@article{IossaKarlssonZhiboedov:toappear,
    author  = "Iossa, Cristoforo and Karlsson, Robin and Zhiboedov, Alexander",
    title   = "{AdS black holes and CFT Love numbers}",
    journal = "to appear",
    year    = "2026"
}

@article{Maldacena:1997re,
    author = "Maldacena, Juan Martin",
    title = "{The Large $N$ limit of superconformal field theories and supergravity}",
    eprint = "hep-th/9711200",
    archivePrefix = "arXiv",
    reportNumber = "HUTP-97-A097, HUTP-98-A097",
    doi = "10.4310/ATMP.1998.v2.n2.a1",
    journal = "Adv. Theor. Math. Phys.",
    volume = "2",
    pages = "231--252",
    year = "1998"
}

@article{Witten:1998qj,
    author = "Witten, Edward",
    title = "{Anti de Sitter space and holography}",
    eprint = "hep-th/9802150",
    archivePrefix = "arXiv",
    reportNumber = "IASSNS-HEP-98-15",
    doi = "10.4310/ATMP.1998.v2.n2.a2",
    journal = "Adv. Theor. Math. Phys.",
    volume = "2",
    pages = "253--291",
    year = "1998"
}

@article{Kulaxizi:2018dxo,
    author = "Kulaxizi, Manuela and Ng, Gim Seng and Parnachev, Andrei",
    title = "{Black Holes, Heavy States, Phase Shift and Anomalous Dimensions}",
    eprint = "1812.03120",
    archivePrefix = "arXiv",
    primaryClass = "hep-th",
    doi = "10.21468/SciPostPhys.6.6.065",
    journal = "SciPost Phys.",
    volume = "6",
    number = "6",
    pages = "065",
    year = "2019"
}

@article{Dodelson:2022eiz,
    author = "Dodelson, Matthew and Zhiboedov, Alexander",
    title = "{Gravitational orbits, double-twist mirage, and many-body scars}",
    eprint = "2204.09749",
    archivePrefix = "arXiv",
    primaryClass = "hep-th",
    reportNumber = "CERN-TH-2022-065",
    doi = "10.1007/JHEP12(2022)163",
    journal = "JHEP",
    volume = "12",
    pages = "163",
    year = "2022"
}

@article{Huang:2024wbq,
    author = "Huang, Kuo-Wei",
    title = "{Resummation of multistress tensors in higher dimensions}",
    eprint = "2406.07458",
    archivePrefix = "arXiv",
    primaryClass = "hep-th",
    doi = "10.1103/PhysRevD.111.046016",
    journal = "Phys. Rev. D",
    volume = "111",
    number = "4",
    pages = "046016",
    year = "2025"
}

@article{Alday:2020eua,
    author = "Alday, Luis F. and Kologlu, Murat and Zhiboedov, Alexander",
    title = "{Holographic correlators at finite temperature}",
    eprint = "2009.10062",
    archivePrefix = "arXiv",
    primaryClass = "hep-th",
    reportNumber = "CERN-TH-2020-155",
    doi = "10.1007/JHEP06(2021)082",
    journal = "JHEP",
    volume = "06",
    pages = "082",
    year = "2021"
}

@article{Dodelson:2023vrw,
    author = "Dodelson, Matthew and Iossa, Cristoforo and Karlsson, Robin and Zhiboedov, Alexander",
    title = "{A thermal product formula}",
    eprint = "2304.12339",
    archivePrefix = "arXiv",
    primaryClass = "hep-th",
    reportNumber = "CERN-TH-2023-062",
    doi = "10.1007/JHEP01(2024)036",
    journal = "JHEP",
    volume = "01",
    pages = "036",
    year = "2024"
}

@article{Dodelson:2023nnr,
    author = "Dodelson, Matthew and Iossa, Cristoforo and Karlsson, Robin and Lupsasca, Alexandru and Zhiboedov, Alexander",
    title = "{Black hole bulk-cone singularities}",
    eprint = "2310.15236",
    archivePrefix = "arXiv",
    primaryClass = "hep-th",
    reportNumber = "CERN-TH-2023-192",
    doi = "10.1007/JHEP07(2024)046",
    journal = "JHEP",
    volume = "07",
    pages = "046",
    year = "2024"
}

@article{Hijano:2015rla,
    author = "Hijano, Eliot and Kraus, Per and Snively, River",
    title = "{Worldline approach to semi-classical conformal blocks}",
    eprint = "1501.02260",
    archivePrefix = "arXiv",
    primaryClass = "hep-th",
    doi = "10.1007/JHEP07(2015)131",
    journal = "JHEP",
    volume = "07",
    pages = "131",
    year = "2015"
}

@article{Kraus:2017kyl,
    author = "Kraus, Per and Sivaramakrishnan, Allic and Snively, River",
    title = "{Black holes from CFT: Universality of correlators at large c}",
    eprint = "1706.00771",
    archivePrefix = "arXiv",
    primaryClass = "hep-th",
    doi = "10.1007/JHEP08(2017)084",
    journal = "JHEP",
    volume = "08",
    pages = "084",
    year = "2017"
}

@article{Goldberger:2004jt,
    author = "Goldberger, Walter D. and Rothstein, Ira Z.",
    title = "{An Effective field theory of gravity for extended objects}",
    eprint = "hep-th/0409156",
    archivePrefix = "arXiv",
    reportNumber = "UCSD-PTH-04-17, CMU-HEP-04-06",
    doi = "10.1103/PhysRevD.73.104029",
    journal = "Phys. Rev. D",
    volume = "73",
    pages = "104029",
    year = "2006"
}

@article{Kol:2011vg,
    author = "Kol, Barak and Smolkin, Michael",
    title = "{Black hole stereotyping: Induced gravito-static polarization}",
    eprint = "1110.3764",
    archivePrefix = "arXiv",
    primaryClass = "hep-th",
    doi = "10.1007/JHEP02(2012)010",
    journal = "JHEP",
    volume = "02",
    pages = "010",
    year = "2012"
}

@article{Giusto:2018ovt,
    author = "Giusto, Stefano and Russo, Rodolfo and Wen, Congkao",
    title = "{Holographic correlators in AdS$_{3}$}",
    eprint = "1812.06479",
    archivePrefix = "arXiv",
    primaryClass = "hep-th",
    reportNumber = "QMUL-PH-18-31",
    doi = "10.1007/JHEP03(2019)096",
    journal = "JHEP",
    volume = "03",
    pages = "096",
    year = "2019"
}

@article{Giusto:2019pxc,
    author = "Giusto, Stefano and Russo, Rodolfo and Tyukov, Alexander and Wen, Congkao",
    title = "{Holographic correlators in AdS$_3$ without Witten diagrams}",
    eprint = "1905.12314",
    archivePrefix = "arXiv",
    primaryClass = "hep-th",
    reportNumber = "QMUL-PH-19-13",
    doi = "10.1007/JHEP09(2019)030",
    journal = "JHEP",
    volume = "09",
    pages = "030",
    year = "2019"
}

@article{Giusto:2023awo,
    author = "Giusto, Stefano and Iossa, Cristoforo and Russo, Rodolfo",
    title = "{The black hole behind the cut}",
    eprint = "2306.15305",
    archivePrefix = "arXiv",
    primaryClass = "hep-th",
    doi = "10.1007/JHEP10(2023)050",
    journal = "JHEP",
    volume = "10",
    pages = "050",
    year = "2023"
}

@article{McGreevy:2000cw,
    author = "McGreevy, John and Susskind, Leonard and Toumbas, Nicolaos",
    title = "{Invasion of the giant gravitons from Anti-de Sitter space}",
    eprint = "hep-th/0003075",
    archivePrefix = "arXiv",
    reportNumber = "SU-ITP-00-09",
    doi = "10.1088/1126-6708/2000/06/008",
    journal = "JHEP",
    volume = "06",
    pages = "008",
    year = "2000"
}

@article{Corley:2001zk,
    author = "Corley, Steve and Jevicki, Antal and Ramgoolam, Sanjaye",
    title = "{Exact correlators of giant gravitons from dual N=4 SYM theory}",
    eprint = "hep-th/0111222",
    archivePrefix = "arXiv",
    reportNumber = "BROWN-HET-1292",
    doi = "10.4310/ATMP.2001.v5.n4.a6",
    journal = "Adv. Theor. Math. Phys.",
    volume = "5",
    pages = "809--839",
    year = "2002"
}

@article{Lin:2004nb,
    author = "Lin, Hai and Lunin, Oleg and Maldacena, Juan Martin",
    title = "{Bubbling AdS space and 1/2 BPS geometries}",
    eprint = "hep-th/0409174",
    archivePrefix = "arXiv",
    reportNumber = "PUPT-2136",
    doi = "10.1088/1126-6708/2004/10/025",
    journal = "JHEP",
    volume = "10",
    pages = "025",
    year = "2004"
}

@article{Giusto:2024trt,
    author = "Giusto, Stefano and Rosso, Alessandro",
    title = "{The geometry of large charge multi-traces in $ \mathcal{N} $ = 4 SYM}",
    eprint = "2401.01254",
    archivePrefix = "arXiv",
    primaryClass = "hep-th",
    doi = "10.1007/JHEP10(2024)200",
    journal = "JHEP",
    volume = "10",
    pages = "200",
    year = "2024"
}

@article{Turton:2024afd,
    author = "Turton, David and Tyukov, Alexander",
    title = "{Four-point correlators in $ \mathcal{N} $ = 4 SYM from AdS$_{5}$ bubbling geometries}",
    eprint = "2408.16834",
    archivePrefix = "arXiv",
    primaryClass = "hep-th",
    doi = "10.1007/JHEP10(2024)244",
    journal = "JHEP",
    volume = "10",
    pages = "244",
    year = "2024"
}

@article{Aprile:2024lwy,
    author = "Aprile, Francesco and Giusto, Stefano and Russo, Rodolfo",
    title = "{Holographic correlators with BPS bound states in $\mathcal{N} = 4$ SYM}",
    eprint = "2409.12911",
    archivePrefix = "arXiv",
    primaryClass = "hep-th",
    doi = "10.1103/PhysRevLett.134.091602",
    journal = "Phys. Rev. Lett.",
    volume = "134",
    number = "9",
    pages = "091602",
    year = "2025"
}

@article{Fitzpatrick:2019zqz,
    author = "Fitzpatrick, A. Liam and Huang, Kuo-Wei",
    title = "{Universal Lowest-Twist in CFTs from Holography}",
    eprint = "1903.05306",
    archivePrefix = "arXiv",
    primaryClass = "hep-th",
    doi = "10.1007/JHEP08(2019)138",
    journal = "JHEP",
    volume = "08",
    pages = "138",
    year = "2019"
}

@article{Li:2019zba,
    author = "Li, Yue-Zhou",
    title = "{Heavy-light Bootstrap from Lorentzian Inversion Formula}",
    eprint = "1910.06357",
    archivePrefix = "arXiv",
    primaryClass = "hep-th",
    doi = "10.1007/JHEP07(2020)046",
    journal = "JHEP",
    volume = "07",
    pages = "046",
    year = "2020"
}

@article{Li:2020dqm,
    author = "Li, Yue-Zhou and Zhang, Hao-Yu",
    title = "{More on heavy-light bootstrap up to double-stress-tensor}",
    eprint = "2004.04758",
    archivePrefix = "arXiv",
    primaryClass = "hep-th",
    doi = "10.1007/JHEP10(2020)055",
    journal = "JHEP",
    volume = "10",
    pages = "055",
    year = "2020"
}

@article{Parnachev:2020fna,
    author = "Parnachev, Andrei",
    title = "{Near Lightcone Thermal Conformal Correlators and Holography}",
    eprint = "2005.06877",
    archivePrefix = "arXiv",
    primaryClass = "hep-th",
    doi = "10.1088/1751-8121/abec16",
    journal = "J. Phys. A",
    volume = "54",
    number = "15",
    pages = "155401",
    year = "2021"
}

@article{Ceplak:2024bja,
    author = "{\v{C}}eplak, Nejc and Liu, Hong and Parnachev, Andrei and Valach, Samuel",
    title = "{Black hole singularity from OPE}",
    eprint = "2404.17286",
    archivePrefix = "arXiv",
    primaryClass = "hep-th",
    doi = "10.1007/JHEP10(2024)105",
    journal = "JHEP",
    volume = "10",
    pages = "105",
    year = "2024"
}

@article{Buric:2025anb,
    author = "Buri{\'c}, Ilija and Gusev, Ivan and Parnachev, Andrei",
    title = "{Thermal holographic correlators and KMS condition}",
    eprint = "2505.10277",
    archivePrefix = "arXiv",
    primaryClass = "hep-th",
    doi = "10.1007/JHEP09(2025)053",
    journal = "JHEP",
    volume = "09",
    pages = "053",
    year = "2025"
}

@article{Buric:2025fye,
    author = "Buri{\'c}, Ilija and Gusev, Ivan and Parnachev, Andrei",
    title = "{Holographic correlators from thermal bootstrap}",
    eprint = "2508.08373",
    archivePrefix = "arXiv",
    primaryClass = "hep-th",
    doi = "10.1007/JHEP05(2026)059",
    journal = "JHEP",
    volume = "05",
    pages = "059",
    year = "2026"
}

@article{Barrat:2025nvu,
    author = "Barrat, Julien and Bozkurt, Deniz N. and Marchetto, Enrico and Miscioscia, Alessio and Pomoni, Elli",
    title = "{The analytic bootstrap at finite temperature}",
    eprint = "2506.06422",
    archivePrefix = "arXiv",
    primaryClass = "hep-th",
    reportNumber = "DESY-25-078",
    doi = "10.1007/JHEP05(2026)104",
    journal = "JHEP",
    volume = "05",
    pages = "104",
    year = "2026"
}

@article{Barrat:2025twb,
    author = "Barrat, Julien and Bozkurt, Deniz N. and Marchetto, Enrico and Miscioscia, Alessio and Pomoni, Elli",
    title = "{Analytic thermal bootstrap meets holography}",
    eprint = "2510.20894",
    archivePrefix = "arXiv",
    primaryClass = "hep-th",
    reportNumber = "DESY-25-139 , YITP-SB-2025-16",
    doi = "10.1007/JHEP05(2026)180",
    journal = "JHEP",
    volume = "05",
    pages = "180",
    year = "2026"
}

@article{Afkhami-Jeddi:2025wra,
    author = "Afkhami-Jeddi, Nima and Caron-Huot, Simon and Chakravarty, Joydeep and Maloney, Alexander",
    title = "{Imprint of the black hole singularity on thermal two-point functions}",
    eprint = "2510.21673",
    archivePrefix = "arXiv",
    primaryClass = "hep-th",
    doi = "10.1103/1j3s-fzjl",
    journal = "Phys. Rev. D",
    volume = "114",
    number = "2",
    pages = "025005",
    year = "2026"
}

@article{Giombi:2026kdz,
    author = "Giombi, Simone and Li, Yue-Zhou and Shan, Jieru",
    title = "{Bouncing singularities and thermal correlators on line defects}",
    eprint = "2603.11012",
    archivePrefix = "arXiv",
    primaryClass = "hep-th",
    doi = "10.1007/JHEP07(2026)170",
    journal = "JHEP",
    volume = "07",
    pages = "170",
    year = "2026"
}

@article{Arnaudo:2026der,
    author = "Arnaudo, Paolo and Withers, Benjamin",
    title = "{Analytic structure of holographic thermal correlators from Fourier series}",
    eprint = "2603.13469",
    archivePrefix = "arXiv",
    primaryClass = "hep-th",
    doi = "10.1007/JHEP06(2026)205",
    journal = "JHEP",
    volume = "06",
    pages = "205",
    year = "2026"
}

@article{Jia:2026ryl,
    author = "Jia, Hewei Frederic and Rangamani, Mukund",
    title = "{Exact holographic thermal spectral functions: OPE, non-perturbative corrections, and black hole singularity}",
    eprint = "2604.10803",
    archivePrefix = "arXiv",
    primaryClass = "hep-th",
    month = "4",
    year = "2026"
}

@article{Fitzpatrick:2012yx,
    author = "Fitzpatrick, A. Liam and Kaplan, Jared and Poland, David and Simmons-Duffin, David",
    title = "{The Analytic Bootstrap and AdS Superhorizon Locality}",
    eprint = "1212.3616",
    archivePrefix = "arXiv",
    primaryClass = "hep-th",
    doi = "10.1007/JHEP12(2013)004",
    journal = "JHEP",
    volume = "12",
    pages = "004",
    year = "2013"
}

@article{Komargodski:2012ek,
    author = "Komargodski, Zohar and Zhiboedov, Alexander",
    title = "{Convexity and Liberation at Large Spin}",
    eprint = "1212.4103",
    archivePrefix = "arXiv",
    primaryClass = "hep-th",
    doi = "10.1007/JHEP11(2013)140",
    journal = "JHEP",
    volume = "11",
    pages = "140",
    year = "2013"
}

@article{Fitzpatrick:2014vua,
    author = "Fitzpatrick, A. Liam and Kaplan, Jared and Walters, Matthew T.",
    title = "{Universality of Long-Distance AdS Physics from the CFT Bootstrap}",
    eprint = "1403.6829",
    archivePrefix = "arXiv",
    primaryClass = "hep-th",
    doi = "10.1007/JHEP08(2014)145",
    journal = "JHEP",
    volume = "08",
    pages = "145",
    year = "2014"
}

@article{Caron-Huot:2017vep,
    author = "Caron-Huot, Simon",
    title = "{Analyticity in Spin in Conformal Theories}",
    eprint = "1703.00278",
    archivePrefix = "arXiv",
    primaryClass = "hep-th",
    doi = "10.1007/JHEP09(2017)078",
    journal = "JHEP",
    volume = "09",
    pages = "078",
    year = "2017"
}

@article{Heemskerk:2009pn,
    author = "Heemskerk, Idse and Penedones, Joao and Polchinski, Joseph and Sully, James",
    title = "{Holography from Conformal Field Theory}",
    eprint = "0907.0151",
    archivePrefix = "arXiv",
    primaryClass = "hep-th",
    reportNumber = "NSF-KITP-09-110",
    doi = "10.1088/1126-6708/2009/10/079",
    journal = "JHEP",
    volume = "10",
    pages = "079",
    year = "2009"
}

@article{Albayrak:2019gnz,
    author = "Albayrak, Soner and Meltzer, David and Poland, David",
    title = "{More Analytic Bootstrap: Nonperturbative Effects and Fermions}",
    eprint = "1904.00032",
    archivePrefix = "arXiv",
    primaryClass = "hep-th",
    doi = "10.1007/JHEP08(2019)040",
    journal = "JHEP",
    volume = "08",
    pages = "040",
    year = "2019"
}

@article{Porto:2016pyg,
    author = "Porto, Rafael A.",
    title = "{The effective field theorist{\textquoteright}s approach to gravitational dynamics}",
    eprint = "1601.04914",
    archivePrefix = "arXiv",
    primaryClass = "hep-th",
    doi = "10.1016/j.physrep.2016.04.003",
    journal = "Phys. Rept.",
    volume = "633",
    pages = "1--104",
    year = "2016"
}

@article{Burgess:2016lal,
    author = "Burgess, C. P. and Hayman, Peter and Williams, Matt and Zalavari, Laszlo",
    title = "{Point-Particle Effective Field Theory I: Classical Renormalization and the Inverse-Square Potential}",
    eprint = "1612.07313",
    archivePrefix = "arXiv",
    primaryClass = "hep-ph",
    doi = "10.1007/JHEP04(2017)106",
    journal = "JHEP",
    volume = "04",
    pages = "106",
    year = "2017"
}

@article{Chakrabarti:2013xza,
    author = "Chakrabarti, Sayan and Delsate, T{\'e}rence and Steinhoff, Jan",
    title = "{Effective action and linear response of compact objects in Newtonian gravity}",
    eprint = "1306.5820",
    archivePrefix = "arXiv",
    primaryClass = "gr-qc",
    doi = "10.1103/PhysRevD.88.084038",
    journal = "Phys. Rev. D",
    volume = "88",
    pages = "084038",
    year = "2013"
}

@article{Steinhoff:2016rfi,
    author = "Steinhoff, Jan and Hinderer, Tanja and Buonanno, Alessandra and Taracchini, Andrea",
    title = "{Dynamical Tides in General Relativity: Effective Action and Effective-One-Body Hamiltonian}",
    eprint = "1608.01907",
    archivePrefix = "arXiv",
    primaryClass = "gr-qc",
    doi = "10.1103/PhysRevD.94.104028",
    journal = "Phys. Rev. D",
    volume = "94",
    number = "10",
    pages = "104028",
    year = "2016"
}

@article{Galley:2012hx,
    author = "Galley, Chad R.",
    title = "{Classical Mechanics of Nonconservative Systems}",
    eprint = "1210.2745",
    archivePrefix = "arXiv",
    primaryClass = "gr-qc",
    doi = "10.1103/PhysRevLett.110.174301",
    journal = "Phys. Rev. Lett.",
    volume = "110",
    number = "17",
    pages = "174301",
    year = "2013"
}

@article{Correia:2024jgr,
    author = "Correia, Miguel and Isabella, Giulia",
    title = "{The Born regime of gravitational amplitudes}",
    eprint = "2406.13737",
    archivePrefix = "arXiv",
    primaryClass = "hep-th",
    doi = "10.1007/JHEP03(2025)144",
    journal = "JHEP",
    volume = "03",
    pages = "144",
    year = "2025"
}

@article{Caron-Huot:2025tlq,
    author = "Caron-Huot, Simon and Correia, Miguel and Isabella, Giulia and Solon, Mikhail",
    title = "{Gravitational Wave Scattering via the Born Series: Scalar Tidal Matching to O(G7) and Beyond}",
    eprint = "2503.13593",
    archivePrefix = "arXiv",
    primaryClass = "hep-th",
    doi = "10.1103/qd3c-nfz6",
    journal = "Phys. Rev. Lett.",
    volume = "135",
    number = "19",
    pages = "191601",
    year = "2025"
}

@article{Regge:1957td,
    author = "Regge, Tullio and Wheeler, John A.",
    title = "{Stability of a Schwarzschild singularity}",
    doi = "10.1103/PhysRev.108.1063",
    journal = "Phys. Rev.",
    volume = "108",
    pages = "1063--1069",
    year = "1957"
}

@article{Zerilli:1970se,
    author = "Zerilli, Frank J.",
    title = "{Effective potential for even parity Regge-Wheeler gravitational perturbation equations}",
    doi = "10.1103/PhysRevLett.24.737",
    journal = "Phys. Rev. Lett.",
    volume = "24",
    pages = "737--738",
    year = "1970"
}

@article{Teukolsky:1972my,
    author = "Teukolsky, S. A.",
    title = "{Rotating black holes - separable wave equations for gravitational and electromagnetic perturbations}",
    reportNumber = "OAP-291",
    doi = "10.1103/PhysRevLett.29.1114",
    journal = "Phys. Rev. Lett.",
    volume = "29",
    pages = "1114--1118",
    year = "1972"
}

@article{Teukolsky:1973ha,
    author = "Teukolsky, Saul A.",
    title = "{Perturbations of a rotating black hole. 1. Fundamental equations for gravitational electromagnetic and neutrino field perturbations}",
    doi = "10.1086/152444",
    journal = "Astrophys. J.",
    volume = "185",
    pages = "635--647",
    year = "1973"
}

@article{Pound:2021qin,
    author = "Pound, Adam and Wardell, Barry",
    title = "{Black hole perturbation theory and gravitational self-force}",
    eprint = "2101.04592",
    archivePrefix = "arXiv",
    primaryClass = "gr-qc",
    doi = "10.1007/978-981-15-4702-7_38-1",
    month = "1",
    year = "2021"
}

@article{Ivanov:2022hlo,
    author = "Ivanov, Mikhail M. and Zhou, Zihan",
    title = "{Revisiting the matching of black hole tidal responses: A systematic study of relativistic and logarithmic corrections}",
    eprint = "2208.08459",
    archivePrefix = "arXiv",
    primaryClass = "hep-th",
    doi = "10.1103/PhysRevD.107.084030",
    journal = "Phys. Rev. D",
    volume = "107",
    number = "8",
    pages = "084030",
    year = "2023"
}

@article{Saketh:2023bul,
    author = "Saketh, M. V. S. and Zhou, Zihan and Ivanov, Mikhail M.",
    title = "{Dynamical tidal response of Kerr black holes from scattering amplitudes}",
    eprint = "2307.10391",
    archivePrefix = "arXiv",
    primaryClass = "hep-th",
    doi = "10.1103/PhysRevD.109.064058",
    journal = "Phys. Rev. D",
    volume = "109",
    number = "6",
    pages = "064058",
    year = "2024"
}

@article{Aharony:1999ti,
    author = "Aharony, Ofer and Gubser, Steven S. and Maldacena, Juan Martin and Ooguri, Hirosi and Oz, Yaron",
    title = "{Large N field theories, string theory and gravity}",
    eprint = "hep-th/9905111",
    archivePrefix = "arXiv",
    reportNumber = "CERN-TH-99-122, HUTP-99-A027, LBNL-43113, RU-99-18, UCB-PTH-99-16, LBL-43113",
    doi = "10.1016/S0370-1573(99)00083-6",
    journal = "Phys. Rept.",
    volume = "323",
    pages = "183--386",
    year = "2000"
}

@article{Costa:2011mg,
    author = "Costa, Miguel S. and Penedones, Joao and Poland, David and Rychkov, Slava",
    title = "{Spinning Conformal Correlators}",
    eprint = "1107.3554",
    archivePrefix = "arXiv",
    primaryClass = "hep-th",
    reportNumber = "LPTENS-11-22, NSF-KITP-11-128",
    doi = "10.1007/JHEP11(2011)071",
    journal = "JHEP",
    volume = "11",
    pages = "071",
    year = "2011"
}

@article{Costa:2014kfa,
    author = "Costa, Miguel S. and Gon{\c{c}}alves, Vasco and Penedones, Jo{\~a}o",
    title = "{Spinning AdS Propagators}",
    eprint = "1404.5625",
    archivePrefix = "arXiv",
    primaryClass = "hep-th",
    doi = "10.1007/JHEP09(2014)064",
    journal = "JHEP",
    volume = "09",
    pages = "064",
    year = "2014"
}

@article{Minahan:2012fh,
    author = "Minahan, Joseph A.",
    title = "{Holographic three-point functions for short operators}",
    eprint = "1206.3129",
    archivePrefix = "arXiv",
    primaryClass = "hep-th",
    reportNumber = "UUITP-02-12",
    doi = "10.1007/JHEP07(2012)187",
    journal = "JHEP",
    volume = "07",
    pages = "187",
    year = "2012"
}

@article{Hijano:2015zsa,
    author = "Hijano, Eliot and Kraus, Per and Perlmutter, Eric and Snively, River",
    title = "{Witten Diagrams Revisited: The AdS Geometry of Conformal Blocks}",
    eprint = "1508.00501",
    archivePrefix = "arXiv",
    primaryClass = "hep-th",
    doi = "10.1007/JHEP01(2016)146",
    journal = "JHEP",
    volume = "01",
    pages = "146",
    year = "2016"
}

@article{DHoker:1999kzh,
    author = "D'Hoker, Eric and Freedman, Daniel Z. and Mathur, Samir D. and Matusis, Alec and Rastelli, Leonardo",
    title = "{Graviton exchange and complete four point functions in the AdS / CFT correspondence}",
    eprint = "hep-th/9903196",
    archivePrefix = "arXiv",
    reportNumber = "MIT-CTP-2843, UCLA-99-TEP-2",
    doi = "10.1016/S0550-3213(99)00525-8",
    journal = "Nucl. Phys. B",
    volume = "562",
    pages = "353--394",
    year = "1999"
}

@book{Rychkov:2016iqz,
    author = "Rychkov, Slava",
    title = "{EPFL Lectures on Conformal Field Theory in D{\ensuremath{>}}= 3 Dimensions}",
    eprint = "1601.05000",
    archivePrefix = "arXiv",
    primaryClass = "hep-th",
    reportNumber = "CERN-TH-2016-012",
    doi = "10.1007/978-3-319-43626-5",
    isbn = "978-3-319-43625-8, 978-3-319-43626-5",
    series = "SpringerBriefs in Physics",
    month = "1",
    year = "2016"
}

@article{Penedones:2010ue,
    author = "Penedones, Joao",
    title = "{Writing CFT correlation functions as AdS scattering amplitudes}",
    eprint = "1011.1485",
    archivePrefix = "arXiv",
    primaryClass = "hep-th",
    doi = "10.1007/JHEP03(2011)025",
    journal = "JHEP",
    volume = "03",
    pages = "025",
    year = "2011"
}

@article{Schwinger:1960qe,
    author = "Schwinger, Julian S.",
    title = "{Brownian motion of a quantum oscillator}",
    doi = "10.1063/1.1703727",
    journal = "J. Math. Phys.",
    volume = "2",
    pages = "407--432",
    year = "1961"
}

@article{Keldysh:1964ud,
    author = "Keldysh, L. V.",
    title = "{Diagram Technique for Nonequilibrium Processes}",
    doi = "10.1142/9789811279461_0007",
    journal = "Sov. Phys. JETP",
    volume = "20",
    pages = "1018--1026",
    year = "1965"
}

@article{Simmons-Duffin:2012juh,
    author = "Simmons-Duffin, David",
    title = "{Projectors, Shadows, and Conformal Blocks}",
    eprint = "1204.3894",
    archivePrefix = "arXiv",
    primaryClass = "hep-th",
    doi = "10.1007/JHEP04(2014)146",
    journal = "JHEP",
    volume = "04",
    pages = "146",
    year = "2014"
}

@article{DeWitt:1964mxt,
    author = "DeWitt, Bryce S.",
    editor = "DeWitt, C. and DeWitt, B.",
    title = "{Dynamical theory of groups and fields}",
    journal = "Conf. Proc. C",
    volume = "630701",
    pages = "585--820",
    year = "1964"
}

@article{Chen:2026ium,
    author = "Chen, Junding and Jiang, Yunfeng and Zhou, Xinan",
    title = "{Defect approach to giant graviton dynamics}",
    eprint = "2602.13570",
    archivePrefix = "arXiv",
    primaryClass = "hep-th",
    reportNumber = "USTC-ICTS/PCFT-26-14",
    doi = "10.1088/1361-6633/ae79a4",
    journal = "Rept. Prog. Phys.",
    volume = "89",
    number = "6",
    pages = "067801",
    year = "2026"
}

@article{Chen:2025yxg,
    author = "Chen, Junding and Jiang, Yunfeng and Zhou, Xinan",
    title = "{Giant Graviton Correlators as Defect Systems}",
    eprint = "2503.22987",
    archivePrefix = "arXiv",
    primaryClass = "hep-th",
    reportNumber = "USTC-ICTS/PCFT-25-14",
    doi = "10.1103/hg9p-hblr",
    journal = "Phys. Rev. Lett.",
    volume = "135",
    number = "8",
    pages = "081602",
    year = "2025"
}

\end{document}